\documentclass[twocolumn]{aastex63}
\usepackage{graphicx}
\usepackage{amssymb,amsmath}
\usepackage{epstopdf}
\usepackage{textcomp,txfonts} 

\usepackage{rotating,multirow,threeparttablex,topcapt,booktabs}

\usepackage{natbib}
\setcitestyle{aysep={}}

\newcommand{\lbol}{$L_{\mathrm{bol}}$}
\newcommand{\teff}{$T_{\mathrm{eff}}$}

\shortauthors{Pineda et al.}
\shorttitle{MUSS}

\begin{document}

\title{The M-dwarf Ultraviolet Spectroscopic Sample I. Determining Stellar Parameters for Field Stars}

\author[0000-0002-4489-0135]{   J.~Sebastian Pineda}\altaffiliation{\href{mailto:sebastian.pineda@lasp.colorado.edu}{sebastian.pineda@lasp.colorado.edu}}
\affiliation{University of Colorado Boulder, Laboratoy for Atmospheric and Space Physics, 3665 Discovery Drive, Boulder CO, 80303, USA}

\author{Allison Youngblood}
\affiliation{University of Colorado Boulder, Laboratoy for Atmospheric and Space Physics, 3665 Discovery Drive, Boulder CO, 80303, USA}
\affiliation{NASA Goddard Space Flight Center, 8800 Greenbelt Rd, Greenbelt, MD 20771, USA}

\author{Kevin France}
\affiliation{University of Colorado Boulder, Laboratoy for Atmospheric and Space Physics, 3665 Discovery Drive, Boulder CO, 80303, USA}

\begin{abstract}
	Accurate stellar properties are essential for precise stellar astrophysics and exoplanetary science. In the M-dwarf regime, much effort has gone into defining empirical relations that can use readily accessible observables to assess physical stellar properties. Often, these relations for the quantity of interest are cast as a non-linear function of available data; however, in Bayesian modeling the reverse is needed. In this article, we introduce a new Bayesian framework to self-consistently and simultaneously apply multiple empirical calibrations to fully characterize the mass, luminosity, radius, and effective temperature of a field age M-dwarf. This framework includes a new M-dwarf mass-radius relation with a scatter of 3.1\% at fixed mass. We further introduce the M-dwarf Ultraviolet Spectroscopic Sample (MUSS), and apply our methodology to provide consistent stellar parameters for these nearby low-mass stars, selected as having available spectroscopic data in the ultraviolet. These targets are of interest largely as either exoplanet hosts or benchmarks in multi-wavelength stellar activity. We use the field MUSS stars to define a low-mass main sequence in the solar neighborhood through Gaussian Process (GP) regression. These results enable us to empirically measure a feature in the GP derivative at $\mathcal{M}_{b} = 0.337 \pm ^{0.013}_{0.026}$ $M_{\odot}$ that indicates where the M-dwarf Ultraviolet Spectroscopic Sample transitions from fully to partly convective interiors.
\end{abstract}

\section{Introduction}

Understanding the physics of low-mass stars, M-dwarfs, is of critical importance to current astronomy research. These objects represent the most numerous stars in our Galaxy \citep{Bochanski2010} and the majority of our nearest stellar neighbors are M-dwarfs \citep[e.g.,][]{Henry2018AJ....155..265H}. They are frequent targets of exoplanet searches, revealing an abundant population of rocky planets \citep[e.g.,][]{Dressing2015}, with ongoing searches often dedicated to M-dwarf hosts \citep[e.g.,][]{Ricker2015, Reiners2018A&A...612A..49R}. The fundamental stellar properties, mass, radius, and luminosity, are essential parameters in assessing stellar structure, evolution, and their interplay with other observable phenomena like magnetic activity. Additionally, these are the stellar parameters necessary to accurately characterize the properties and atmospheres of exoplanetary systems. 

Unlike warmer stars, M dwarf properties have been difficult to model, revealing large discrepancies between different models and observed properties; for example, field M-dwarfs are known to display larger radii than predicted by evolutionary models \citep[e.g.,][]{LopezMorales2007ApJ...660..732L, Kraus2011ApJ...728...48K}, including across the fully convective boundary \citep{Kesseli2018, Jackson2018MNRAS.476.3245J}. In the cool M-dwarf atmospheres displaying a large number of molecular lines, model spectra can sometimes fail to reproduce the observed features. This is especially the case at blue optical wavelengths, a discrepancy often attributed to missing line opacities \citep[e.g.,][]{Allard2000ApJ...540.1005A,Allard2012}. This evidence over the years has led to a focus on utilizing empirical methods to estimate M-dwarf properties, including mass-luminosity relations \citep[e.g.,][]{Delfosse2000,Benedict2016,Mann2019}, mass-radius relations \citep[e.g.,][]{Boyajian2012, Schweitzer2019, Rabus2019}, and ever more precise photometric bolometric corrrections \citep[e.g.,][]{Tinney1993AJ....105.1045T,Mann2015}. These often rely on eclipsing binaries, or interferometrically observed stars with precise bolometric fluxes, and benefit from improved distance and stellar metallicity estimates. The statistical scatter in these relationships can be calculated and is the basis for accurate values with reliable uncertainties in the estimates of M-dwarf fundamental properties.

Because of the varied development of such relations, literature estimates of M-dwarf properties contain an assortment of underlying assumptions, in aggregate producing populations that incorporate the uncertainties inherent to the different methods. This approach makes it difficult to assess systematic effects from the property determinations when analyzing populations of stars and exoplanets. To reveal potential biases introduced by the chosen set of stellar parameters, it is important to understand the underlying assumptions using consistently determined stellar fundamental properties, and accurately characterize their often correlated statistical uncertainties. Analysis of samples with consistently determined properties can thus point to the underlying causes of observed trends or which of the known sources of error need to be improved to reveal new findings. The statistically correlated uncertainties between properties is a feature often ignored in the literature. Regression analyses relying on such estimates are likely to yield biased uncertainties that are enlarged relative to the underlying data, potentially obscuring important astrophysical relationships.

In this paper, we demonstrate how to self-consistently apply multiple empirical relations to simultaneously determine the fundamental stellar properties of field M-dwarfs, with reliable uncertainties. Our Monte Carlo methods readily enable the use of our results in additional analyses to account for the statistical correlations in the parameter uncertainties. These methods are flexible to observational inputs, allowing multiple and distinct measurements to inform the stellar properties. Our methods also avoid the approximations inherent to standard error propagation through nonlinear empirical relationships, which can fail to reproduce the distribution tails --- the basis for statistically significant results. As part of this effort we define a new field M-dwarf mass-radius relation, using literature results of eclipsing binaries.

To illustrate our methods we introduce the M-dwarf Ultraviolet Spectroscopic Sample (MUSS), a compilation of low-mass stars generally defined as being previously targeted for ultraviolet spectra by the \textit{Hubble Space Telescope}, but including a few other M dwarfs of interest and M dwarfs with interferometric radii. We treat the field stars of this sample in this work, and address how to assess the properties of young stars in our companion paper (Pineda et al.\ 2021c). Applying our methodology to this sample, we create a set of M dwarfs with consistently determined properties that are already of interest to the astronomy community, as planet hosts, or as interesting/useful objects in probing stellar magnetic activity. We discuss this sample in Section~\ref{sec:MUSS}. We define the new field mass-radius relation in Section~\ref{sec:relationMR}, demonstrate and use our unified Bayesian framework for field M-dwarfs in Section~\ref{sec:empfield}, with applications for the MUSS ensemble stellar properties, discuss our conclusions in Section~\ref{sec:conc}, and summarize our findings in Section~\ref{sec:summary}.

\startlongtable
\begin{deluxetable*}{l c c c c c}
	\tablecaption{ MUSS Sample Stars.\tablenotemark{a}
		\label{tab:sample} }
	\tablehead{
		\colhead{Name} & \colhead{RA} & \colhead{DEC} & \colhead{Spectral Type\tablenotemark{b}} & \colhead{UV Data} & \colhead{Program\tablenotemark{c}} }
	\startdata
GJ 15A &  00 18 22.8849  &  +44 01 22.637  & M2V & \checkmark & Mega-MUSCLES \\ 
LHS 1140 &  00 44 59.3314  &  -15 16 17.543  & M4.5 & \checkmark & Berta-Thompson \\ 
GJ 49 &  01 02 38.8679  &  +62 20 42.172  & M1.5V & \checkmark & FUMES  \\ 
G 75-55 &  02 58 20.0853  &  -00 59 33.435  & M0.5 & \checkmark & HAZMAT \\ 
LP 247-13 &  03 15 37.8542  &  +37 24 14.233  & M2.7V & \checkmark &  FUMES \\ 
GJ 1061 &  03 35 59.6996  &  -44 30 45.725  & M5.5V & \checkmark & MUSCLES \\ 
GJ 163 &  04 09 15.6684  &  -53 22 25.289  & M3.5 & \checkmark & Mega-MUSCLES  \\ 
GJ 173 &  04 37 41.8658  &  -11 02 19.970  & M1V & \checkmark & HAZMAT \\ 
GJ 176 &  04 42 55.7753  &  +18 57 29.394  & M2.5V & \checkmark  & MUSCLES  \\ 
LP 55-41 &  05 02 05.0677  &  +68 27 25.290  & M3 & \checkmark  & FUMES  \\ 
GJ 3325 &  05 03 20.0842  &  -17 22 24.722  & M3V & \checkmark & Redfield  \\ 
G 249-11 &  05 30 40.3586  &  +68 54 07.968  & M4 & \checkmark & FUMES \\ 
GJ 205 &  05 31 27.3958  &  -03 40 38.021  & M1.5Ve & \checkmark & Wood \\ 
GJ 213 &  05 42 09.2672  &  +12 29 21.611  & M4V & \checkmark & Guinan \\ 
GJ 3378 &  06 01 11.0454  &  +59 35 49.885  & M4.0Ve & $\times$ & --- \\ 
GJ 273 &  07 27 24.4997  &  +05 13 32.833  & M3.5V & \checkmark & Wood \\ 
YZ CMi &  07 44 40.1726  &  +03 33 08.877  & M4.0Ve & \checkmark & Hawley, Wood \\ 
GJ 3470 &  07 59 05.8395  &  +15 23 29.240  & M2.0Ve & \checkmark & PanCET \\ 
LHS 2065 &  08 53 36.1607  &  -03 29 32.201  & M9Ve & \checkmark & Osten \\ 
GJ 338 A &  09 14 22.7754  &  +52 41 11.792  & K7V & \checkmark & Wood \\ 
TOI-1235 &  10 08 51.8069  &  +69 16 35.564  & M0.5Ve & $\times$ & --- \\ 
GJ 1132 &  10 14 51.7783  &  -47 09 24.189  & M4 & \checkmark & Berta-Thompson, Mega-MUSCLES \\ 
AD Leo &  10 19 36.2808  &  +19 52 12.014  & dM3 & \checkmark  & Brown, Hawley \\ 
GJ 402 &  10 50 52.0312  &  +06 48 29.263  & M4V & $\times$ & --- \\ 
GJ 403 &  10 52 04.2435  &  +13 59 51.304  & M4V & $\times$ & --- \\ 
GJ 406 &  10 56 28.8262  &  +07 00 52.344  & dM6 & \checkmark & Johns-Krull  \\ 
GJ 410 &  11 02 38.3419  &  +21 58 01.704  & M1.0V & \checkmark & FUMES \\ 
GJ 411 &  11 03 20.1940  &  +35 58 11.568  & M2V & \checkmark  & Youngblood  \\ 
K2-3 &  11 29 20.3917  &  -01 27 17.279  & M1V & \checkmark & Kreidberg  \\ 
GJ 436 &  11 42 11.0933  &  +26 42 23.658  & M3V & \checkmark & e.g, MUSCLES, PanCET \\ 
GJ 447 &  11 47 44.3968  &  +00 48 16.404  & dM4 & $\times$ & ---   \\ 
LHS 2686 &  13 10 12.6294  &  +47 45 18.671  & M5.0V & \checkmark & Mega-MUSCLES \\ 
GJ 526 &  13 45 43.7754  &  +14 53 29.473  & M2V & $\times$ & --- \\ 
GJ 545 &  14 20 07.3705  &  -09 37 13.391  & M4 & $\times$ &  --- \\ 
L 980-5 &  14 21 15.1250  &  -01 07 19.813  & M4 & \checkmark & Mega-MUSCLES  \\ 
Proxima Centauri &  14 29 42.9451  &  -62 40 46.170  & M5.5Ve & \checkmark & e.g., Linsky, MacGregor \\ 
LHS 3003 &  14 56 38.2643  &  -28 09 48.622  & M7.0Ve & \checkmark & Osten \\ 
GJ 581 &  15 19 26.8271  &  -07 43 20.190  & M3V & \checkmark & MUSCLES \\ 
GJ 588 &  15 32 12.9326  &  -41 16 32.131  & M2.5V & \checkmark  & Wood  \\ 
GJ 628 &  16 30 18.0582  &  -12 39 45.323  & M3V & \checkmark  & MUSCLES \\ 
VB 8 &  16 55 35.2561  &  -08 23 40.752  & M7Ve & \checkmark & e.g., Hawley, Johns-Krull \\ 
GJ 1207 &  16 57 05.7363  &  -04 20 56.317  & dM4 & $\times$ & ---  \\ 
GJ 649 &  16 58 08.8497  &  +25 44 38.971  & M2V & \checkmark & Mega-MUSCLES \\ 
GJ 3991 &  17 09 31.5428  &  +43 40 52.734  & M3.5V & $\times$ & --- \\ 
GJ 1214 &  17 15 18.9337  &  +04 57 50.064  & M4.5V & \checkmark & MUSCLES \\ 
GJ 667C &  17 18 58.8271  &  -34 59 48.612  & M1.5V & \checkmark & MUSCLES \\ 
GJ 674 &  17 28 39.9455  &  -46 53 42.693  & M3V & \checkmark & Mega-MUSCLES \\ 
GJ 676A &  17 30 11.2030  &  -51 38 13.139  & M0V & \checkmark & Mega-MUSCLES \\ 
GJ 687 &  17 36 25.8991  &  +68 20 20.904  & M3.0V & $\times$ & ---  \\ 
GJ 699 &  17 57 48.4997  &  +04 41 36.111  & M4V & \checkmark & Mega-MUSCLES \\ 
GJ 725A &  18 42 46.7048  &  +59 37 49.411  & M3V & \checkmark & MUSCLES \\ 
GJ 729 &  18 49 49.3638  &  -23 50 10.453  & M3.5Ve & \checkmark & Mega-MUSCLES \\ 
VB 10 &  19 16 57.6103  &  +05 09 01.588  & M8V & \checkmark & Hawley \\ 
Kepler-138 &  19 21 31.5681  &  +43 17 34.680  & M1V & $\times$ & --- \\ 
GJ 1243 &  19 51 09.3205  &  +46 29 00.208  & M4.0V & \checkmark & Kowalski \\ 
GJ 809 &  20 53 19.7890  &  +62 09 15.813  & M1.0Ve &  $\times$ & ---  \\ 
LP 756-18 &  20 55 37.1152  &  -14 03 54.880  & M4.5V & \checkmark & Mega-MUSCLES \\ 
GJ 821 &  21 09 17.4274  &  -13 18 09.019  & M1V & \checkmark & Guinan  \\ 
GJ 832 &  21 33 33.9749  &  -49 00 32.403  & M2/3V & \checkmark & MUSCLES \\ 
GJ 849 &  22 09 40.3443  &  -04 38 26.651  & M3.5V & \checkmark & Mega-MUSCLES  \\ 
EV Lac &  22 46 49.7311  &  +44 20 02.372  & M4.0V & \checkmark & Osten, Redfield \\ 
GJ 876 &  22 53 16.7323  &  -14 15 49.303  & M3.5V & \checkmark & MUSCLES  \\ 
GJ 880 &  22 56 34.8047  &  +16 33 12.353  & M1.5Ve & $\times$ & --- \\ 
GJ 887 &  23 05 52.0354  &  -35 51 11.058  & M2V & \checkmark & MUSCLES \\ 
TRAPPIST-1 &  23 06 29.3684  &  -05 02 29.031  & M7.5e & \checkmark & e.g, Bourrier, Mega-MUSCLES  \\ 
GJ 4334 &  23 25 40.2972  &  +53 08 05.900  & M5V & \checkmark  & FUMES \\ 
GJ 4367 &  23 50 31.6427  &  -09 33 32.655  & dM4.0 & $\times$ & ---  \\ 
	\hline
	\hline
BPS CS 22188-0067 &  00 39 35.7994  &  -38 16 58.554  & M1.4 & \checkmark & HAZMAT \\ 
UCAC3 61-3820 &  01 52 18.2985  &  -59 50 16.769  & M2.0Ve & \checkmark & HAZMAT \\ 
UCAC3 163-5082 &  02 00 12.7784  &  -08 40 51.915  & M2.5 & \checkmark & HAZMAT \\ 
2MASS J02125819-5851182 &  02 12 58.1893  &  -58 51 18.150  & M2.0Ve &\checkmark & HAZMAT \\ 
GSC 8056-0482 &  02 36 51.7015  &  -52 03 03.601  & M2Ve & \checkmark & HAZMAT \\ 
GSC 08057-00342 &  02 54 33.1649  &  -51 08 31.391  & M1.5Ve & \checkmark &HAZMAT  \\ 
CD-44 1173 &  03 31 55.6414  &  -43 59 13.561  & K6Ve &\checkmark  & HAZMAT \\ 
HIP17695 &  03 47 23.3412  &  -01 58 19.945  & M3.0V & \checkmark & FUMES \\ 
HG 7-162 &  04 18 47.0366  &  +13 21 58.714  & M1V~ & \checkmark & HAZMAT \\ 
LP 415-27 &  04 22 39.5612  &  +18 16 09.590  & M0.5 & \checkmark & HAZMAT \\ 
LP5-282 &  04 22 59.9094  &  +13 18 58.533  & M1 & \checkmark & HAZMAT \\ 
LP 475-68 &  04 26 04.7064  &  +15 02 28.987  & M1V & \checkmark & HAZMAT  \\ 
GJ 3290 &  04 27 16.6394  &  +17 14 30.664  & M1.5 & \checkmark & HAZMAT \\ 
LP 415-1619 &  04 36 38.9455  &  +18 36 56.715  & M2.0V & \checkmark & HAZMAT \\ 
HIP23309 &  05 00 47.1300  &  -57 15 25.453  & M0Ve & \checkmark & FUMES \\ 
CD-35 2722 &  06 09 19.2082  &  -35 49 31.062  & M1Ve & \checkmark & FUMES \\ 
TWA 7A &  10 42 30.1021  &  -33 40 16.226  & M2Ve & \checkmark & Herczeg \\ 
TWA 13A &  11 21 17.2193  &  -34 46 45.500  & M1Ve & \checkmark & Brown, Redfield  \\ 
TWA 13B &  11 21 17.4429  &  -34 46 49.774  & M1Ve & \checkmark & Brown, Redfield \\ 
AU Mic &  20 45 09.5323  &  -31 20 27.241  & M1VeBa1 & \checkmark & Linsky, Newton  \\ 
BPS CS 22956-0074 &  22 02 54.5032  &  -64 40 44.219  & M1.8 & \checkmark & HAZMAT \\ 
HIP112312 &  22 44 57.9628  &  -33 15 01.745  & M4IVe & \checkmark & FUMES \\ 
UCAC3 33-129092 &  22 46 34.7012  &  -73 53 50.426  & M2.3 & \checkmark & HAZMAT \\ 
TYC 9344-293-1 &  23 26 10.6971  &  -73 23 49.856  & M0Ve & \checkmark & HAZMAT \\ 
UCAC4 110-129613 &  23 28 57.6374  &  -68 02 34.025  & M2.5Ve & \checkmark & HAZMAT \\ 
	\enddata
	\tablenotetext{a}{Stars above the double line divide in the table are treated here as pertaining to the field, and below it are treated in Pineda et al. (2021c), as young with defined ages.}
	\tablenotetext{b}{Spectral Types are as listed in SIMBAD, except for LP 55-41 and G 249-11, which we took from \citet{Pineda2021arXiv210212485P}.}
	\tablenotetext{c}{ References for programs: \emph{Berta-Thompson}: HST-14462, 14757, \citet{Waalkes2019AJ....158...50W}; \emph{Bourrier}: HST-14493, \citet{Bourrier2017b}; \emph{Brown}: HST-9271, 12361, \citet{France2012ApJ...756..171F}; \emph{FUMES}: HST-14640, \citet{Pineda2021arXiv210212485P}; \emph{Guinan}: HST-13020, \citet{Guinan2016}; \emph{Hawley}: HST-8129, 8613, 9090, \citet{Hawley2003,Hawley2003b,Hawley2007}; \emph{HAZMAT}: HST-14784, \citet{Loyd2021ApJ...907...91L}; \emph{Herzceg}: HST-11616, \citet{France2012ApJ...756..171F}; \emph{Johns-Krull}: HST-15660; \emph{Kowalski}: HST-13323, \citet{Kowalski2019ApJ...871..167K}; \emph{Kriedberg}: HST-15110; \emph{Linsky}: HST-7556, 8040, \citet{Pagano2000ApJ...532..497P,Wood2001ApJ...547L..49W}; \emph{MacGregor}: HST-15651; \emph{Mega-MUSCLES}: HST-15071, \citet{Froning2019}; \emph{MUSCLES}: HST-12464, 13650, \citet{France2013,France2016}; \emph{Newton}: HST-15836; \emph{PanCET}: HST-14767, \citet{Bourrier2018AA...620A.147B}; \emph{Osten}: HST-8880, 12011, \citet{Osten2006ApJ...647.1349O}; \emph{Redfield}: HST-13332, 14084, 16225; \emph{Wood}: HST-15326; \emph{Youngblood}: HST-15190, \citet{Melbourne2020AJ....160..269M}. }
\end{deluxetable*}

\section{M-dwarf Ultraviolet Spectroscopic Sample}\label{sec:MUSS}

The \textit{Hubble Space Telescope} is the only current observatory capable of spectroscopically observing across the entire ultraviolet regime. This unique capability has been used to study low-mass stars in a variety of ways with two major themes in recent years: 1) the high-energy spectral energy distribution and its impact on planetary atmospheres \citep[e.g.,][]{France2013, France2016, Bourrier2017a, Froning2019, France2020AJ....160..237F}, and 2) the nature of these emissions, and temporal variability on long and short time-scales as probes of magnetic activity and stellar upper atmospheric structure \citep[e.g.,][]{Hawley2003b, Osten2006ApJ...647.1349O, Hawley2007,Loyd2018ApJ...867...70L,Loyd2018, Kowalski2019ApJ...871..167K, Loyd2021ApJ...907...91L,Pineda2021arXiv210212485P}. 

We chose these HST targets across a variety of programs as the focus of our study of M-dwarf stellar properties. These programs include MUSCLES (HST-12464, 13650, PI-France), Mega-MUSCLES (HST-15071, PI - Froning), FUMES (HST-14640, PI - Pineda), and HAZMAT (HST-14784, PI-Shkolnik), among miscellaneous others cataloged in Table~\ref{tab:sample}. These objects are of interest to the community, and future analyses will benefit from consistently determined stellar properties.

\begin{figure}[tbp]
	\centering
	\includegraphics{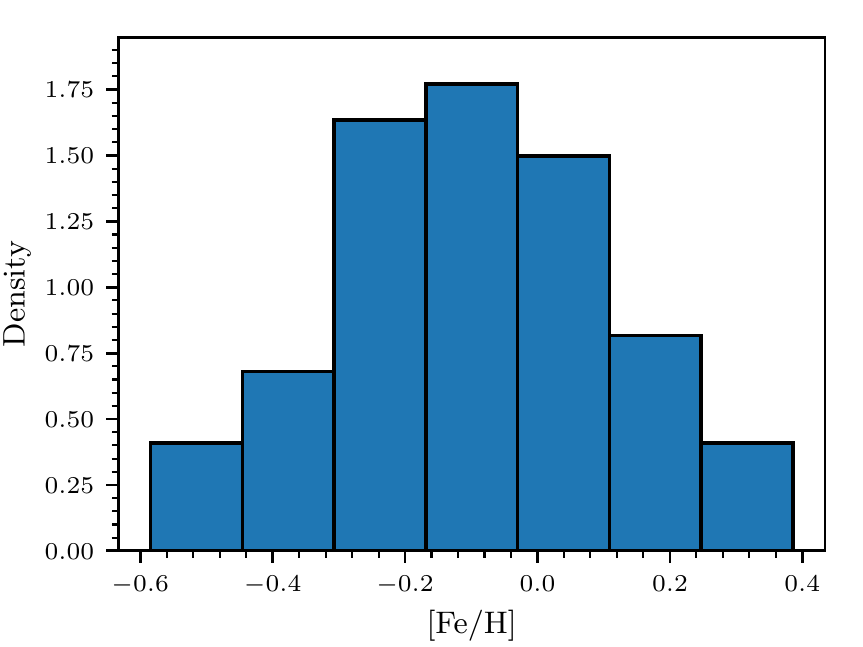} 
	\caption{The MUSS field stars are well distributed around the solar metallicity value with a mean of -0.006 dex and standard deviation 0.22 dex. We only include here the stars (55/67) which had consistent literature metallicity estimates on the scale used by \citet{Mann2015}. Typical [Fe/H] uncertainties are of order 0.1 dex.}
	\label{fig:fedist}
\end{figure}

We dub the compilation of these objects the M-dwarf Ultraviolet Spectroscopic Sample (MUSS). The targets are enumerated in Table~\ref{tab:sample}. The sample fundamental properties and their determination are described in Section~\ref{sec:empfield} for field stars and we defer the discussion of young stars to Pineda et al.\ (2021c). Many of these targets host exoplanetary systems for which we provide consistently determined stellar properties. With uniformly determined stellar properties, this work enables new consistent ensemble statistical examinations of planet property trends, for example, the planet radius valley around low-mass stars \citep[e.g.,][]{Cloutier2020AJ....159..211C}. Similarly, the methods presented here have already informed the activity analyses of \citet{Pineda2021arXiv210212485P} and  \citet{Youngblood2021arXiv210212504Y}. We also show the metallicity distribution of the MUSS field stars used in this paper in Figure~\ref{fig:fedist}. Those metallicities are taken from the literature when available on the [Fe/H] scale defined by the work of \citet{Mann2013AJ....145...52M}, or converted to that scale as necessary \citep{Neves2014,Gaidos2014ApJ...791...54G,Mann2015,Terrien2015}.

Although the focus for the MUSS objects is targets with spectroscopic UV data, we included additional stars that may be of interest to the community (exoplanet hosts with close-in or habitable zone planets), as well as a sample of M-dwarfs with interferometric angular diameter measurements to provide an additional consistency check for the accuracy of our methods, with certain objects selected from existing catalogues with measured bolometric fluxes to help fill out the entire M-dwarf mass range. Notes on individual stars can be found in Appendix~\ref{sec:ap_starspec}, with the observational properties of the sample compiled in the Appendix, Table~\ref{tab:field_obsdata}. The MUSS stars include 67 targets treated as field objects, and 25 targets with well determined ages. The latter are treated in our companion paper Pineda et al.\ (2021c).

\section{A Mass-Radius Relation for Field M-dwarfs}\label{sec:relationMR}

Empirical mass-radius relationships have been used considerably in the literature to both assess stellar properties and test stellar models in the low-mass regime \citep[e.g.,][]{Chen2014}. We build on previous work to derive a new mass-radius relation for low-mass stars, utilizing an expanded data set and a thorough statistical treatment. This sample is distinct from the MUSS stars, but we will use the results discussed in this section to analyze the MUSS stars in Section~\ref{sec:empfield}. 

\begin{figure}[tbp]
	\centering
	\includegraphics{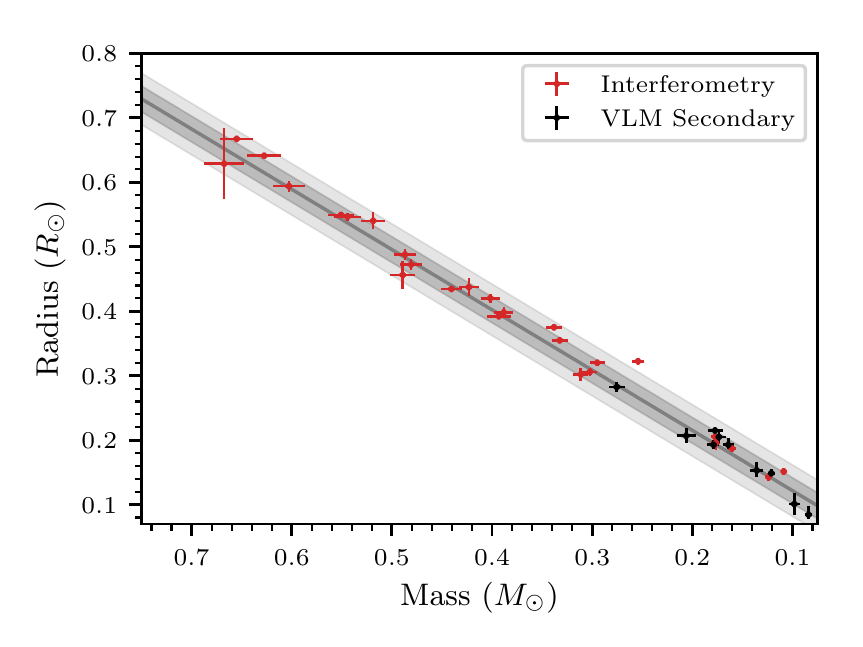} 
	\caption{The M-dwarf mass-radius relationship of \cite{Schweitzer2019} derived from eclipsing binaries (gray line) is consistent with single star measurements (red points) from interferometry \citep{Kervella2008,Demory2009, Boyajian2012,vonBraun2012,vonBraun2014,Rabus2019} and very-low mass eclipsing dwarfs (black points) that are secondaries to FG primaries \citep{vonBoetticher2019}. None of the individual points plotted here were used in the construction of the mass-radius relationship (gray). The light and dark gray shaded regions denote the 2$\sigma$ and 1$\sigma$ scatter about the relation of \cite{Schweitzer2019}, respectively.}
	\label{fig:MRlit}
\end{figure}

Recently, \cite{Schweitzer2019} published an empirical relationship based on a sample of 55 eclipsing binaries spanning 0.092-0.73 $M_{\odot}$. We show their best fit relation for radius as a function of mass and their reported scatter (0.02 $R_{\odot}$, i.e., $>$3\% for $\mathcal{R} < 0.6$ $R_{\odot}$) in Figure~\ref{fig:MRlit} against interferometric radius measurements and eclipsing spectroscopic binaries at the end of the main sequence from \cite{vonBoetticher2019}. Their relation is visually consistent with available interferometric measurements of single stars and the eclipsing binaries at the lowest stellar masses, where there are few targets in the \citet{Schweitzer2019} sample (only 5 stars of $\mathcal{M}<$$0.2$ $M_{\odot}$). This supports previous findings that there is little difference in the observed masses and radii of single stars versus those in binary systems. \cite{Kesseli2018} stated this conclusion as both single stars and binary stars showing the same extent of radius inflation ($\sim$10\%) relative to model predictions. These result demonstrate confidence in the use of such an empirical relation for single stars in the field as well as binaries.

We derive a new $\mathcal{M}$-$\mathcal{R}$ relationship by supplementing the sample used in \cite{Schweitzer2019}, the measurements as compiled in their Table~B.2, with the 10 additional targets from \cite{vonBoetticher2019} to fill in the parameter space for $\mathcal{M}<$0.2 $M_{\odot}$. Although these additional measurements depend on the model based masses of the FG primaries \citep{vonBoetticher2019}, those models do not suffer from large discrepancies relative to observations for sun-like stars, and because they are consistent with the extant $\mathcal{M}$-$\mathcal{R}$ relationships, we conclude that the addition of these points improves the reliability of our relation at the lowest masses. We do not include the interferometrically observed single stars, as we will use them as a check on our results, and to confine underlying systematics strictly to assumptions inherent to the eclipsing binary measurements. 

The need for a new $\mathcal{M}$-$\mathcal{R}$ relation is motivated by our desire to assess potential correlations, which were not reported in \cite{Schweitzer2019}, in the best fit parameters, and model the scatter about that fit relationship with the expanded data set. We followed \cite{Kelly2007} in our regression model for stellar radius as a function of mass, accounting for uncertainty in both the mass and radius measurements. That is, each observed data point in mass and radius is jointly sampled from a distribution defined by their observed measurements and uncertainties. The underlying true masses and radii, $\mathcal{M}_{i}$ and $\mathcal{R}_{i}$, are then constrained to follow our model relationship 

\begin{eqnarray}
\mathcal{R}_{m}  (\mathcal{M}_{i} ) = \sum_{j=0}^{n} b_{j} \mathcal{M}_{i}^{j} & = & b_{0} + b_{1}\mathcal{M}_{i} + ... + b_{n} \mathcal{M}_{i}^{n}  \; ,
\label{eq:mod_MR2}  \\
\mathcal{R}_{i} \, | \, \mathcal{M}_{i}, \,  \{ b \}, \, \sigma_{\mathcal{R}} \; & \sim & \; \mathcal{N}( \mathcal{R}_{m}(\mathcal{M}_{i}), \sigma_{\mathcal{R}}) \; , \label{eq:mod_MR1} 
\end{eqnarray}

\noindent where $i$ corresponds to each measured data pair, with predicted radius, $\mathcal{R}_{m}$, at a given mass. The notation of Equation~\ref{eq:mod_MR1} indicates that the random variable $\mathcal{R}_{i}$ follows the distribution $\mathcal{N}(\mu,\sigma)$, a normal distribution of mean $\mu$ and standard deviation $\sigma$, $\{ b \}$ denotes the polynomial coefficients of the empirical relation, and $\sigma_{\mathcal{R}}$ is the intrinsic scatter for the regression model. The polynomial coefficients are treated with uniform priors and $\sigma_{\mathcal{R}}$ uses a log-uniform prior. The likelihood function is thus the product of Equation~\ref{eq:mod_MR1} over the data set, indexed by $i$, conditioned on the fit parameters $\{ b \}$ and $\sigma_{\mathcal{R}}$. We tested linear and quadratic relations, finding like \citet{Schweitzer2019}, that the linear relationship well characterized the data set, and that the extra freedom in the higher order polynomial was not warranted by the data using BIC-like comparisons (see below). We thus focus our discussion on comparing two linear models, one with a constant scatter value, and one with a fractional scatter relative to the predicted radius, i.e., a scatter at a fixed fraction of $\mathcal{R}_{m}$.

For the literature eclipsing M-dwarf binary measurements, we treated the masses and radii as uncorrelated bivariate normal distributions consistent with their measured errors.\footnote{The individual mass and radius measurements are likely somewhat correlated in many studies. Our assumption of independence is consistent with the literature treatment of these measurements in currently available $\mathcal{M}$-$\mathcal{R}$ relations.} Incorporating the measurements of \cite{vonBoetticher2019}, for which many of their reported masses and radii show asymmetric confidence intervals about the modal value, we treated those results within our model as skew normal distributions following Appendix~\ref{sec:ap_prior}, to better capture the shapes of the asymmetric posterior distributions. Without the additional 10 data points from \citet{vonBoetticher2019}, our Bayesian regression model in the constant scatter case gave results consistent within errors to the values determined by \cite{Schweitzer2019}.

\begin{deluxetable*}{l c c c c c c c c}
	\tablecaption{ $\mathcal{M}$-$\mathcal{R}$ Relation Model Parameters and Comparison.\tablenotemark{a}
		\label{tab:MRbest} }
	\tablehead{
		\colhead{Model \tablenotemark{b}} & \colhead{$b_{0}$} & \colhead{$b_{1}$} &  \colhead{$\sigma_{\mathcal{R}}$} & \colhead{WAIC} & \colhead{pWAIC}& \colhead{dWAIC} &  \colhead{SE}}
	\startdata
	Fractional Scatter & 0.0375$\pm$0.0038  & 0.918$\pm$ 0.012 &    $\mathcal{R}_{m} \times $ 0.031  $\pm^{0.006}_{0.003}$ &-374.43  &21.19&      0 &  10.7\\
	Constant Scatter & 0.0315$\pm$ 0.0052&  0.932$\pm$0.013&   0.0143$\pm^{0.0023}_{0.0015}$ &-341.19 & 19.04 & 33.23 &   9.16
	\enddata
	\tablenotetext{a}{Best model parameters for our $\mathcal{M}$-$\mathcal{R}$ relation are listed as medians when symmetric and modes when asymmetric with marginalized 68\% confidence intervals. The two coefficients are highly correlated (see Figure~\ref{fig:MRpost}), and when using our linear relationship should be sampled from their joint posterior to account for the uncertainty in the relation.}
	\tablenotetext{b}{We also tested a quadratic model for both fractional and constant scatter models, which yielded WAIC values only marginally better than the linear model, consistent well within their standard errors (SE). We choose the simpler linear models as more representative of the astrophysical $\mathcal{M}$-$\mathcal{R}$ relation.}
\end{deluxetable*}

\begin{figure}[tbp]
	\centering
	\includegraphics{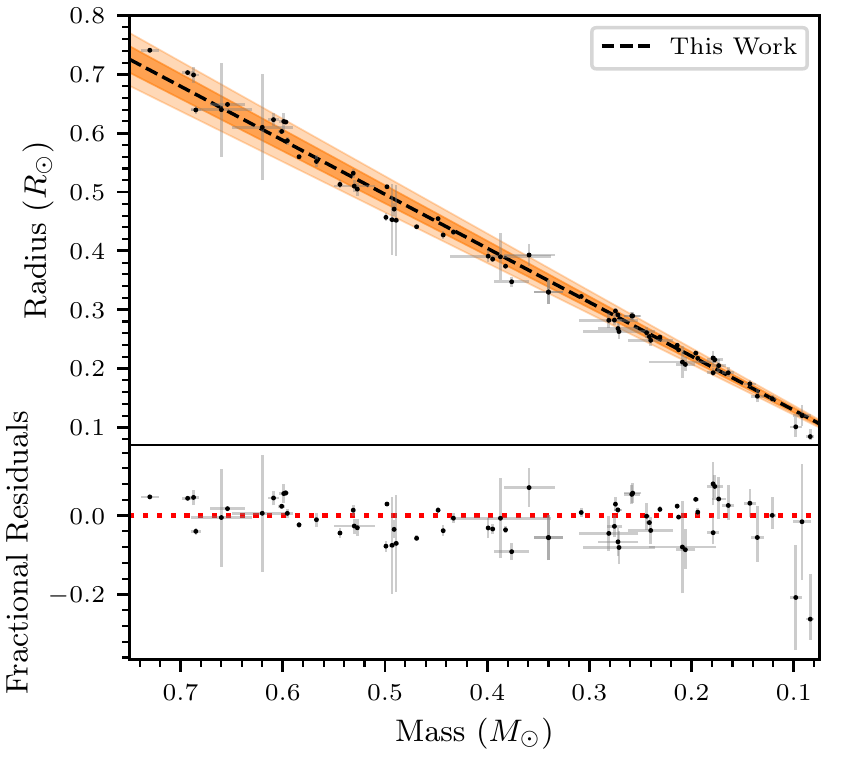} 
	\caption{Combining the eclipsing binary data used in \cite{Schweitzer2019} with additional points at the end of the main sequence from \cite{vonBoetticher2019}, we derive a new $\mathcal{M}$-$\mathcal{R}$ relationship shown as a dashed line with a best estimate of 3.1\% scatter in stellar radius at a fixed mass. The light and dark orange shaded regions denote, respectively, the 2$\sigma$ and 1$\sigma$ scatter about the best fit relationship. Data points with uncertainties show the combined eclipsing binary data set used in our regression analysis.}
	\label{fig:MRdata}
\end{figure}

\begin{figure}[tbp]
	\centering
	\includegraphics[width=0.5\textwidth]{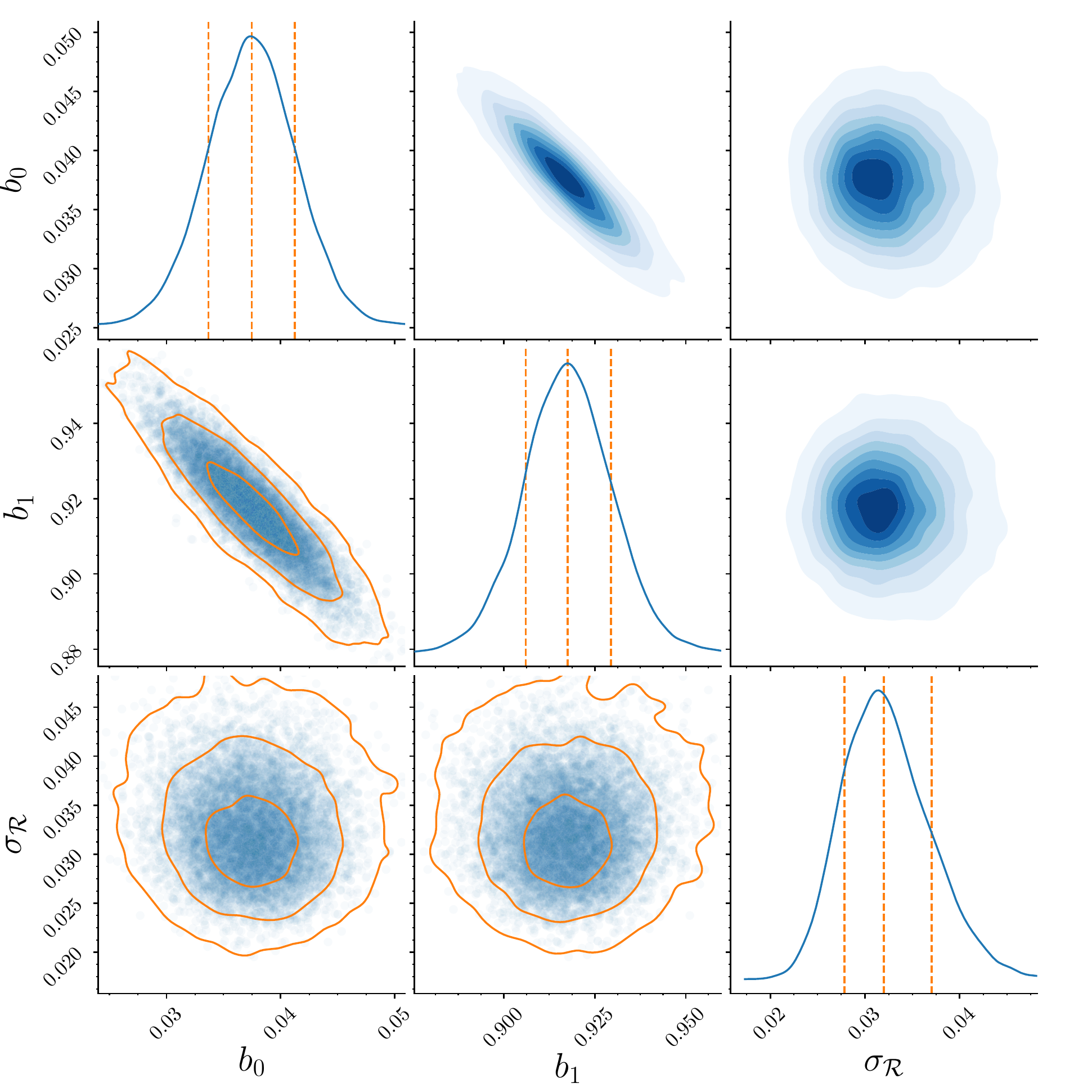} 
	\caption{Joint posterior distributions for our best fit fractional model for the mass-radius relationship of M-dwarf systems using eclipsing binary data. The best fit slope and intercept of the linear model are strongly correlated, as expected, with a best fit scatter of about 3.1\%. The lines in the lower left corner show the 1$\sigma$, 2$\sigma$ and 3$\sigma$ contours on top of semi-transparent points from our MCMC sampling. The diagonal shows the marginalized distributions for each property with the distribution visualized from kernel density estimation. Along the diagonal the dashed lines indicate the median and middle 68\% confidence intervals. The upper right corner visualizes the detailed shape of the joint posteriors through kernel density estimation, showing well defined peaks.}
	\label{fig:MRpost}
\end{figure}

We sampled the posterior distribution for the fit parameters of our regression model within PyMC3 \citep{Salvatier2016} using the Hamiltonian NUTS sampler. We tested convergence by examining the trace for each variable, making sure the auto correlation functions vanish and computing the Gelman-Rubin statistic, which should be very close to unity. We report our best fit parameters (median/modes) in Table~\ref{tab:MRbest} for both the fractional and constant scatter models, with their 68\% confidence intervals. We plot our preferred fractional scatter model against the data in Figure~\ref{fig:MRdata}.

\begin{figure}[tbp]
	\centering
	\includegraphics{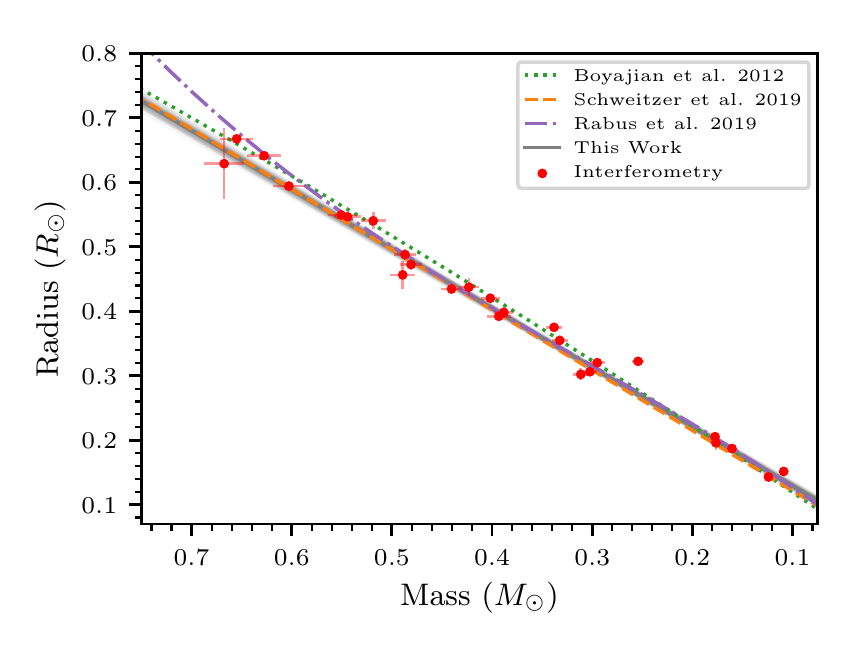} 
	\caption{Our M-dwarf mass-radius relationship (gray lines), shown as 100 samples from the joint posterior distribution of the best model parameters, is consistent with the masses and radii of the sample of interferometrically observed low-mass stars \citep{Kervella2008,Demory2009, Boyajian2012,vonBraun2012,vonBraun2014, Rabus2019} with masses estimated from the mass-luminosity relationship of \cite{Mann2019}. The 3.1\% scatter about our best fit line is not shown here. For comparison the eclipsing binary $\mathcal{M}$-$\mathcal{R}$ relations of \cite{Boyajian2012} and \cite{Schweitzer2019} are included as dotted and dashed lines, respectively, with the interferometric relation of \cite{Rabus2019} also shown as the dash-dotted line.}
	\label{fig:MRcomp}
\end{figure}

We used the Watanabe-Akaike Information Criteria \citep[WAIC;][]{Watanabe2010,Gelman2014}, for model comparison. WAIC is useful for assessing how well the models can predict data not included within the regression analysis, accounting for the effective number of free parameters in each model.\footnote{As compared to a Bayesian Information Criterion (BIC), which considers the relative likelihood of different models, WAIC is more suited to our question at hand: which model will better predict the properties of stars not in our sample? See discussion in \cite{Gelman2014} for comparison of WAIC and BIC.} We show this comparison in Table~\ref{tab:MRbest}, including the WAIC score, pWAIC which reflects the number of parameters (higher means more flexible model), dWAIC as the difference relative to the best model, and SE as the standard error on the WAIC assessment. These scores are also consistent with leave-one-out cross validation comparison of the models \citep[see][]{Gelman2014}. Our results show that the fractional scatter model is preferred. In Figure~\ref{fig:MRpost}, we show the pair-wise joint posteriors of this fractional scatter model fit. We use this relation, which has a scatter of 3.1\% in radius at fixed stellar mass, for our parameter estimations in Section~\ref{sec:empfield}. We compare our new relation to currently available relations in the literature from \cite{Boyajian2012}, \cite{Schweitzer2019} and \cite{Rabus2019} in Figure~\ref{fig:MRcomp}. The relations largely agree, and are consistent with the interferometric radius measurements.

In both cases the best fit parameters $b_{0}$ and $b_{1}$ show correlations (see Figure~\ref{fig:MRpost}) and thus in practice when accounting for the uncertainty in our $\mathcal{M}$-$\mathcal{R}$ relations these parameters should be drawn from their joint distributions (see Appendix~\ref{sec:ap_corr }). We have made these posterior distributions publically available for both the linear and fractional scatter models with accompanying code to facilitate their use.\footnote{ \url{https://github.com/jspineda/stellarprop}} The fitted sample spans masses of 0.084-0.73 $M_{\odot}$, but with few points below 0.09 $M_{\odot}$ and above 0.7 $M_{\odot}$. We thus recommend 0.09-0.7 $M_{\odot}$ as the secure range of applicability for input masses in the field (also see residuals in Figure~\ref{fig:MRdata}).

\begin{figure}[tbp]
	\centering
	\includegraphics{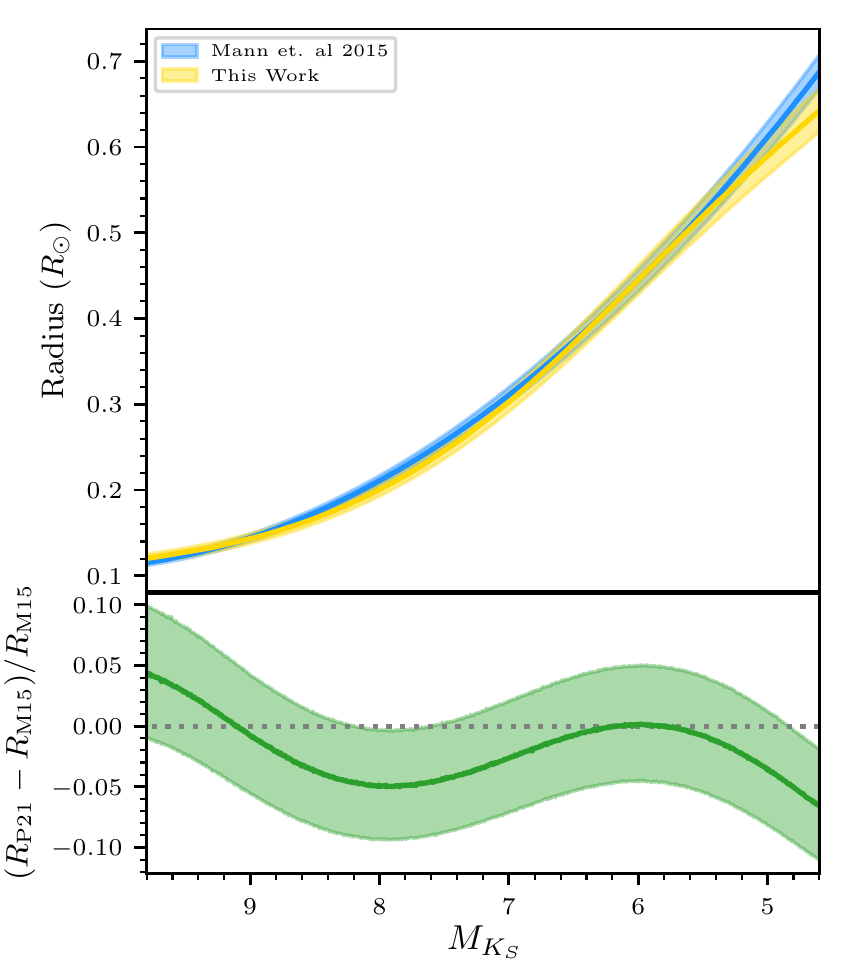} 
	\caption{Combining our mass-radius result with the mass-absolute $K_{S}$ relation of \cite{Mann2019}, enables a comparison to the work of \cite{Mann2015}, illustrating a very close match, typically within 5\%. In each panel the shaded bands indicate the central 68\% confidence intervals.}
	\label{fig:manncomp}
\end{figure}

While our result agrees with the recent literature, our mass-radius relation has more rigorously characterized scatter, and, with the publicly available code, allows for readers to estimate stellar radii from masses properly accounting for the fit correlations. As a final comparison, we consider how our result compares to the semi-empirical radius relation of \citet[][]{Mann2015}. While that work relies on effective temperatures based on atmospheric models, their calibration between absolute $K_{S}$-band magnitude and radius is observationally accessible, and uses methods tuned to match empirical radii. It has moreover been used widely for radius estimates in the literature \citep[e.g.,][]{Feinstein2019AJ....157...40F,Gilbert2020AJ....160..116G, Waalkes2021AJ....161...13W}. To compare, we first converted the absolute magnitudes to masses, and then used our mass-radius relation to plot $\mathcal{R}(M_{K_{S}})$. That comparison is illustrated in Figure~\ref{fig:manncomp} with central 68\% confidence intervals accounting for our uncertainties on the $\mathcal{M}$-$\mathcal{R}$ relation and error propagation through the mass-absolute $K_{S}$ relation of \cite{Mann2019}. 

The curves match very closely, usually well within the uncertainty. The total errors in determining radii in this fashion are $\sim$4\% ($M_{K_{S}} \rightarrow \mathcal{M} \rightarrow \mathcal{R}$), slightly larger than the 2.89\% quoted by \cite{Mann2015} for just using their semi-empirical method ($M_{K_{S}} \rightarrow \mathcal{R}$), although that omits the impact of any uncertainty in their fit parameters, as it is not reported in their article.

These results can be used on their own, however, in concert with other expressions, we reveal a self-consistent and powerful means to fully characterize field M-dwarfs in the next section.

\section{Empirical Relations and Bayesian Inference}\label{sec:empfield}

\subsection{Framework}\label{sec:bayes}

Estimating stellar properties in the low-mass star regime has heavily relied on empirical relations connecting quantities of interest, like mass and radius, to observables like absolute magnitude (if the distance is known) or potentially spectroscopic features. For example, M-dwarf masses are often estimated from mass-luminosity relationships \citep[e.g.,][]{Delfosse2000} calibrated on binary systems with measured masses. These calibrations are designed to take the observed quantities (e.g., absolute $K$-band magnitude, $M_{K_{S}}$) and convert them to the physical attribute of interest (e.g., mass, $\mathcal{M}$).

However, in Bayesian modeling, the reverse is often desired --- we want to be able to predict an observable quantity from the underlying physical properties, for comparison through a likelihood function on the data. Generally, the empirical relations are not invertible, but show a scatter at fixed observable. Even if the relation is invertible, 
the resulting probability density function deviates from normal. We demonstrate here a method (and use it in Section~\ref{sec:field}) to incorporate these empirical relationships into Bayesian modeling of stellar properties, which can be readily folded into any likelihood function. 

To illustrate our methods we consider a generic empirical mass-radius relationship which gives radius ($\mathcal{R}$) as a polynomial function of mass ($\mathcal{M}$), like Equation~\ref{eq:mod_MR2}, however we rewrite it as 

\begin{equation}
f(\mathcal{M},\mathcal{R}) = \mathcal{R}  - \sum_{j=0}^{n} b_{j} \mathcal{M}^{j} = 0 \; . 
\label{eq:ex_empf}
\end{equation}

\noindent The empirical relationship is then taken to have a measured scatter, $\sigma_{\mathcal{R}}$. This scatter can itself be a function of $\mathcal{R}$ or $\mathcal{M}$, as it would be for a fractional scatter. Equation~\ref{eq:ex_empf} states that empirically this combination of $\mathcal{M}$ and $\mathcal{R}$, $f(\mathcal{M},\mathcal{R})$, is observed to vanish. Thus $f$ can be taken to follow the probability density function,

\begin{equation}
f(\mathcal{M},\mathcal{R}) \sim \mathcal{N}(\mu=0,\sigma = \sigma_{\mathcal{R}}) \; ,
\label{eq:ex_pdf}
\end{equation}

\begin{figure}[tbp]
	\centering
	\includegraphics{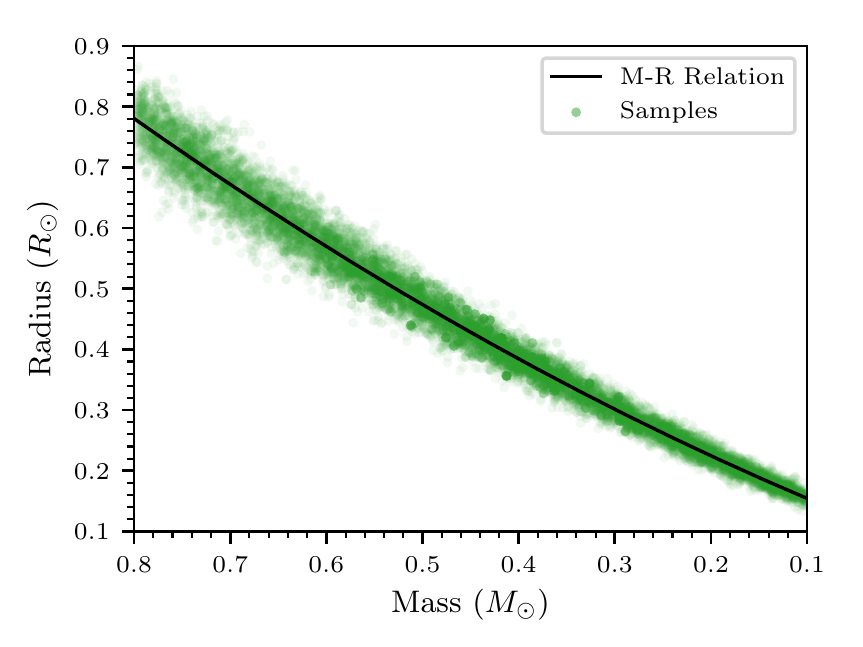} 
	\caption{Monte Carlo sampling of empirical relationships enables use of these constraints in Bayesian inference modeling, including intrinsic scatter. The example here uses the mass-radius relationship from \cite{Boyajian2012} for single stars and a 5\% fractional scatter in predicted radius at constant mass.}
	\label{fig:MRsample}
\end{figure}

\noindent where $\mathcal{N}$ denotes a normal distribution of mean $\mu$ and standard deviation $\sigma$. We are thus assuming that $f(\mathcal{M},\mathcal{R})$ vanishes probabilistically within a known scatter, $\sigma_{\mathcal{R}}$. This distribution (Equation~\ref{eq:ex_pdf}) can then be treated as a product within a likelihood function (see below). If uncertainties in the fit coefficients $b_{j}$ are also known then these parameters can be drawn from their own respective distributions when sampling $\mathcal{M}$ and $\mathcal{R}$. In practice the $b_{j}$ are likely highly correlated and should be drawn from their joint posterior. This requires having the output posterior, as we provide for the mass-radius relation we derive in Section~\ref{sec:relationMR}.

 In Figure~\ref{fig:MRsample}, we show an example of sampling from the single star mass-radius relationship of \cite{Boyajian2012} using Equation~\ref{eq:ex_pdf}, with a 5\% fractional scatter. The constant fractional scatter of the assumed relation is seen in the narrowing of the sample distribution at lower masses. Combined with an additional observational constraint, for example, the stellar density from a planetary transit light curve, the mass and radius can be constrained individually, accounting for all the observational uncertainty and the scatter of the empirical relation. Although we considered in particular a mass-radius relation, any useful empirical relationship can be recast in the form of Equation~\ref{eq:ex_empf} and a corresponding probability density function (Equation~\ref{eq:ex_pdf}) written down if an uncertainty about the best empirical relationship is well defined. This treatment also considers the two quantities, $\mathcal{M}$ and $\mathcal{R}$, as separate random variables constrained within the measured empirical relation, not treated deterministically, so that each may vary within the intrinsic scatter of these relationships, e.g., as a result of unaccounted for effects like magnetic activity or metallicity. Recast as in Equation~\ref{eq:ex_pdf}, multiple empirical relations can be multiplied within a single likelihood function, assuming they are independent, to simultaneously apply the available empirical constraints; we take this approach when determining the properties of MUSS field stars in the next section.

\subsection{Determining MUSS Field Properties}\label{sec:field}

We are interested in estimating the stellar properties, mass ($\mathcal{M}$), radius ($\mathcal{R}$), bolometric luminosity ($L_{\mathrm{bol}}$), and effective temperature ($T_{\mathrm{eff}}$) of field M-dwarfs in a consistent and uniform way. In general, the most widely available data is photometric, and we thus focus our approach to those relations most accessible in their application to any field M-dwarf. Additional spectroscopic relations exist in the literature \citep[e.g.,][]{Newton2015,Terrien2015}, which when independent, can readily supplement the methods used here. As these data are not always available, we proceed with largely photometric empirical relations.

We determined the physical properties within a Bayesian framework through Markov Chain Monte Carlo (MCMC) sampling by using a variety of empirical relations for low-mass M-dwarfs, outlined below. This allows us to jointly and accurately constrain the stellar $\mathcal{M}$, $\mathcal{R}$, and \lbol. We then determine the \teff\ from the combination of $\mathcal{R}$ and \lbol.\footnote{$T_{\mathrm{eff}} = [ L_{\mathrm{bol}} / 4\pi \sigma_{\mathrm{sb}}\mathcal{R}^{2} ]^{1/4}$, with $\sigma_{\mathrm{sb}}$ being the Stefan–Boltzmann constant.} Although \teff\ can be estimated spectroscopically, such estimates are often tied to atmospheric models which potentially introduce new systematics. Our approach for \teff\ is grounded in empirical measurements, and directly applies the definition of effective temperature. We used the methods outlined in Section~\ref{sec:bayes} to combine multiple empirical relations to jointly constrain stellar properties.

\textit{Mass-Luminosity:} To constrain the mass of our stellar sample we relied upon the empirical mass-luminosity relationship of \cite{Mann2019}, which defines $\mathcal{M}$ as a polynomial function of absolute $K_{S}$-band magnitude, $M_{K_{S}}$, based on resolved photometry and the measured orbits of a large sample of low-mass binary systems determined from accumulated community efforts. As has been discussed much in the literature, the choice of $K$-band, is due to its relative insensitivity to metal content, more specifically, C and O abundances \citep{Veyette2016ApJ...828...95V}, among stars in the solar neighborhood, providing a tight correlation for $\mathcal{M}$-$M_{K_{S}}$ \citep[e.g.,][]{Delfosse2000, Benedict2016, Mann2019}.

Although \cite{Mann2019} also developed a metallicity based calibration, they do not find the extra variable to significantly improve the fit to their data, and following their recommendations, their fifth order polynomial is favored for near solar metallicity main sequence stars in the range of 0.08-0.7 $M_{\odot}$, yielding $\sim$2\% errors on mass at fixed $M_{K_{S}}$. Following Section~\ref{sec:bayes}, we rearranged the results of \cite{Mann2019} to define a normally distributed variable $f_{\mathcal{M}}$ as a function $\mathcal{M}$ and $M_{K_{S}}$, which is empirically observed to vanish with a measured dispersion. We can thus use the probability distribution function for $f_{\mathcal{M}}$ within our Bayesian framework:

\begin{eqnarray}
\log_{10} m(M_{K_{S}}) &=&   \sum_{i=0}^{5} a_{i} \, (M_{K_{S}} - 7.5)^{i} \, , \label{eq:mann_mk} \\ 
f_{\mathcal{M}}(\mathcal{M},M_{K_{S}}) &=& \mathcal{M} - m \, ,  \\ 
f_{\mathcal{M}}(\mathcal{M},M_{K_{S}}) &\sim& \mathcal{N}_{\mathcal{M}}( 0, m \, \sigma_{m}  ) \,  , \label{eq:fM}
\end{eqnarray} 

\noindent where $m(M_{K_{S}})$ is the stellar mass in solar units implied by the empirical relationship, and $\sigma_{m} = 0.02$. We incorporated the probability distribution for $f_{\mathcal{M}}$ as a product within our total likelihood function (see below). We use this empirical relation from \citet{Mann2019} as it is precise, and allows us to account for fit coefficient uncertainties and their correlations. Relative to other $\mathcal{M}$-$M_{K_{S}}$ relations from \cite{Delfosse2000} or \cite{Benedict2016}, for early M-dwarfs, the \cite{Mann2019} relation yields smaller masses at fixed $M_{K_{S}}$. The effect is small ($\sim$5\%) but systematic, see discussion within \cite{Mann2019}.

\textit{Mass-Radius:} For typical isolated stars in the field, we cannot determine their radii unless they are sufficiently close for interferometric measurements \citep[e.g.,][]{Boyajian2012,vonBraun2014}. As such, to estimate radii we employed the empirical mass-radius relationship developed in Section~\ref{sec:relationMR} using the fraction scatter model. Like the mass-luminosity relation for $f_{\mathcal{M}}$ above, we summarize the mass-radius results (see Section~\ref{sec:relationMR}) defining the normally distributed random variable $f_{\mathcal{R}}$ as a function of $\mathcal{M}$ and $\mathcal{R}$:

\begin{eqnarray}
r_{m}(\mathcal{M}) &=&   b_{0} +  b_{1} \mathcal{M} \, , \label{eq:lineMR}\\ 
f_{\mathcal{R}}(\mathcal{M},\mathcal{R}) &=& \mathcal{R} - r_{m}(\mathcal{M})  \, , \\
f_{\mathcal{R}}(\mathcal{M},\mathcal{R}) & \sim& \mathcal{N}_{\mathcal{R}}( 0, \sigma_{\mathcal{R}}(\mathcal{M}) ) \,  , \label{eq:fR} 
\end{eqnarray} 

\noindent where $r_{m}(\mathcal{M})$ is the stellar radius implied by the empirical relations at fixed mass, and $\sigma_{\mathcal{R}} = 0.031 \times r_{m}(\mathcal{M})$, using the best value of the scatter parameter. We also sample the coefficients $b_{0}$ and $b_{1}$ from their joint posterior to account for the correlated fit parameter uncertainties.\footnote{We note here as well that from Figure~\ref{fig:MRpost}, $\sigma_{\mathcal{R}}$ is uncorrelated with the polynomial coefficients.} Although this relationship was determined from eclipsing binary measurements, as shown in Figure~\ref{fig:MRcomp} our result is consistent with the measured interferometric radii of single M-dwarfs.

\textit{Color-Bolometric Correction:} As an additional empirical result, we used bolometric correction as a function of color. This provides a way to constrain the stellar bolometric luminosity from widely available photometry, such as the Johnson $V$-band or SDSS $r$-band. We used the calibration published by \cite{Mann2015} for $J$-band bolometric correction as a function of $V-J$ or $r-J$, the two expressions from their work that provide the best precision.\footnote{To reproduce the plot for $J$-band bolometric correction as a function of color, Figure~12 of \cite{Mann2015}, the $X$ of their Table~3 has to be $J-V$ ($J-r$) instead of $V-J$ ($r-J$), or see erratum \cite{Mann2016}.} Their results are expressed, following Section~\ref{sec:bayes}, as 

\begin{eqnarray}
\mathcal{C} &=& (V-J)  \; \mathrm{or} \;  (r-J) \, , \\
\mathcal{B}(\mathcal{C}) &=&  \sum_{i=0}^{2 \; \mathrm{or} \; 3} c_{i} \, \mathcal{C}^{i} \, , \label{eq:BC} \\  
f_{L}(L_{\mathrm{bol}},M_{J},\mathcal{C}) &=& -2.5 \log [L_{\mathrm{bol}}/L_{0}]-  M_{J} - \mathcal{B}(\mathcal{C}) \\ 
f_{L}(L_{\mathrm{bol}},M_{J},\mathcal{C}) &\sim& \mathcal{N}_{L}( 0, \sigma_{L}  ) \,  , \label{eq:fL}
\end{eqnarray} 

\noindent where $\sigma_{L}$ = 0.016, and $L_{0} = 3.0128\times10^{35}$ erg s$^{-1}$ is the luminosity at zero bolometric magnitude. \cite{Mann2015} also demonstrated systemic residuals as a function of metallicity, [Fe/H]. Thus for objects that may deviate significantly from solar metallicity, we included a term in Equation~\ref{eq:BC} that is dependent on [Fe/H], which reduces the scatter to $\sigma_{L}$ = 0.012 (see Table~3 of \citealt{Mann2015}).\footnote{To use the metallicity relation requires a [Fe/H] measurement on the same scale as defined by \cite{Mann2015}.} However, this improved precision is lost if the [Fe/H] uncertainty is large.

\textit{Likelihood:} In general, we expressed Bayes' theorem for the probability distribution describing our parameters of interest as 

\begin{equation}
\mathcal{P}({\alpha} \, | \, {\mathcal{D}}) \propto \mathcal{L}( {\mathcal{D}} \, | \,  {\alpha} ) \,  p({\alpha}) \, , \label{eq:posterior}
\end{equation}

\noindent where $\mathcal{P}$ is the posterior distribution for the set of parameters $\alpha = \{\mathcal{M}, \mathcal{R}, L_{\mathrm{bol}}, d, M_{K_{S}}, M_{J}, \mathrm{[Fe/H]}, \mathcal{C}\}$ given the observational constraints from the data, $\mathcal{D}$, for the likelihood function, $\mathcal{L}$. By sampling $\alpha$, we can generate the observable constraints, for example, inverting the distance, $d$, to obtain the observed parallax, or combining $d$ and $M_{J}$ to obtain the observed $J$ magnitude. $p(\alpha)$ is the prior on these parameters, taken to be uniformly uninformative within the range of properties for which the empirical relations are applicable. Although [Fe/H] and $\mathcal{C}$ are randomly sampled, they are used to connect the observables to the physical stellar properties through the empirical relations as needed. They are in effect inputs and are not further constrained by our likelihood function. For example, when used, the [Fe/H] or color, $\mathcal{C}$, are simply drawn from their observed distributions, and used in the calibration for bolometric correction. The metallicity thus presents an instance by which spectroscopic data can provide an informative constraint.

\begin{figure*}[tbp]
	\centering
	\includegraphics[width=\textwidth]{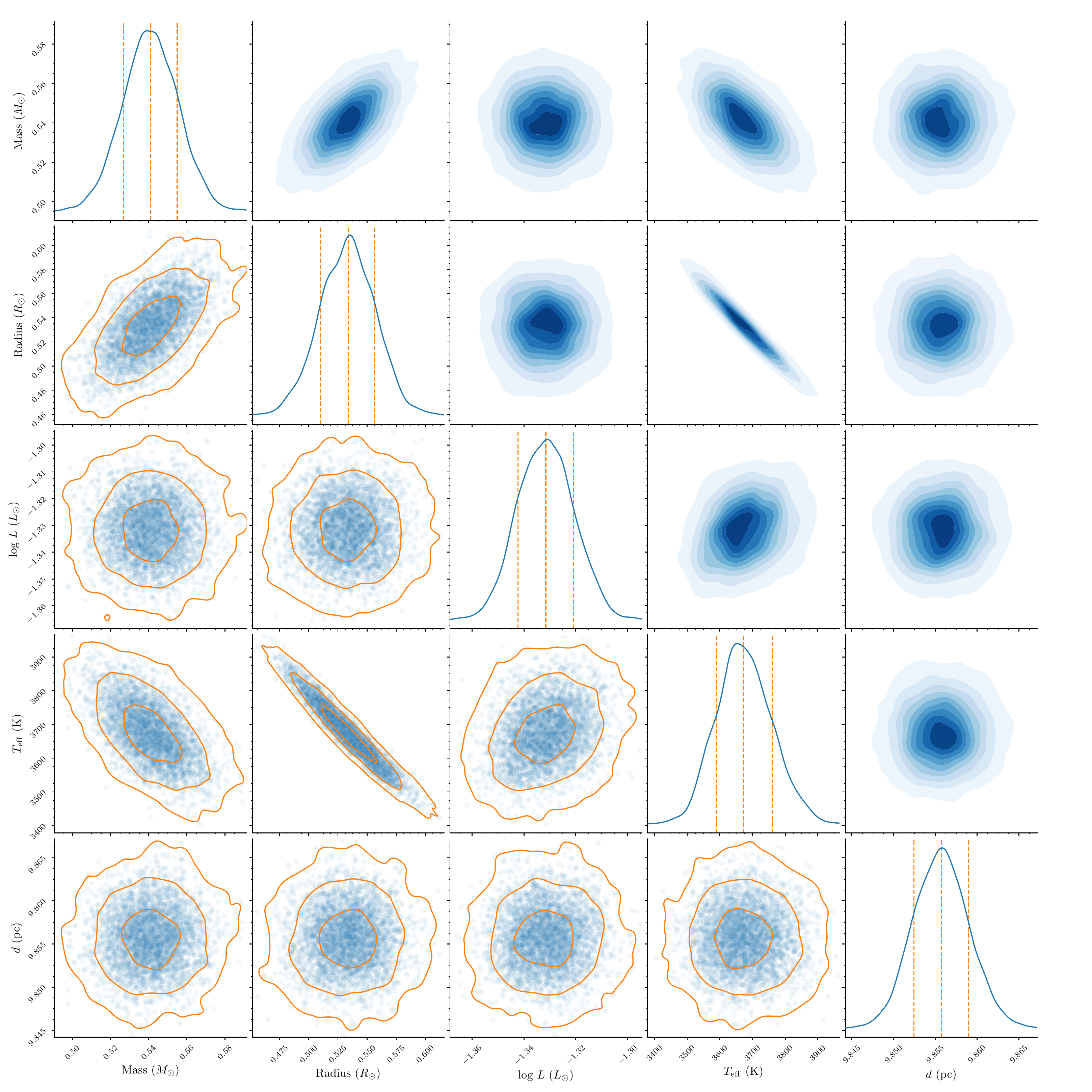} 
	\caption{This illustration of the joint posterior distributions for our parameter estimates of GJ~49, exemplifies our typical results for the sample of field stars, following the methods of Section~\ref{sec:field}. The lines in the lower left corner show the 1$\sigma$, 2$\sigma$ and 3$\sigma$ contours on top of semi-transparent points from our MCMC sampling. The diagonal shows the marginalized distributions for each property with the distribution visualized from kernel density estimation. Along the diagonal the dashed lines indicate the median and middle 68\% confidence intervals. The upper right corner visualizes the detailed shape of the joint posteriors through kernel density estimation, showing well defined peaks.}
	\label{fig:ex_pairs}
\end{figure*}

The likelihood for the data is the product of the observational constraints taken here to be normal distributions based on the observed values and uncertainties, 

\begin{eqnarray}
\pi_{\mathrm{obs}} \,| \, \sigma_{\pi}, d &\sim& \mathcal{N}_{\pi}(\pi(d), \sigma_{\pi}) \, , \\
J_{\mathrm{obs}} \,| \, \sigma_{J}, d, M_{J} &\sim& \mathcal{N}_{J}(J(d,M_{J}), \sigma_{J}) \, , \\
K_{\mathrm{obs}} \,| \, \sigma_{K_{S}}, d, M_{K_{S}} &\sim& \mathcal{N}_{K}(K(d,M_{K_{S}}), \sigma_{K_{S}}) \, , \\
\mathcal{L} (\mathcal{D} \, | \, \alpha) &=& \mathcal{N}_{\pi} \mathcal{N}_{J} \mathcal{N}_{K} \prod_{j \in \{ \mathcal{M}, \mathcal{R}, L \} } N_{j} \; , \label{eq:likelihood}
\end{eqnarray}

\noindent where the means of the distributions are determined from the set of random variables, $\alpha$. To these we multipy the distributions $N_{j}$ representing the empirical relations for $f_{\mathcal{M}}$, $f_{\mathcal{R}}$ and $f_{L}$ given in Equations~\ref{eq:fM}, \ref{eq:fR}, and \ref{eq:fL}. Additional empirical relations can be employed, assuming they are independent, as products to Equation~\ref{eq:likelihood}.We also sampled the distributions for the fit coefficients in Equations~\ref{eq:mann_mk} and \ref{eq:lineMR} (mass-luminosity and mass-radius relations, respectively) to account for that uncertainty. The former is made possible by the publicly available posterior from \citet[][]{Mann2019},\footnote{\url{https://github.com/awmann/M_-M_K-/tree/master/resources}} and for the latter we sample from the posterior reported in Section~\ref{sec:relationMR}. The bolometric correction relation does not have published coefficient uncertainties, and we can only include the scatter on the relation \citep{Mann2015}.

In the most general case, we employed all three empirical relations. To summarize this process: in effect the parallax and $K_{S}$-band magnitude constrain the mass with the $\mathcal{M}$-$M_{K_{S}}$ relation, which in turn through the $\mathcal{M}$-$\mathcal{R}$ relation constrains the radius. The color, [Fe/H], and $J$-band magnitude plus parallax constrain the bolometric luminosity, which combined with the radius estimate defines the effective temperature. By combining all the relations simultaneously we can account for correlations between the parameters and properly include all of the uncertainty in observables (e.g., the parallax measurement is used at multiple stages) and empirical relations in our final parameter estimates.

However, for some of the targets in the literature there are additional observational constraints that we can use. For example, certain targets have measurements of their bolometric flux, in which case we omitted Equation~\ref{eq:fL} for the bolometric correction calibration and simply used the observed normally distributed $F_{\mathrm{bol}} = L_{\mathrm{bol}} / 4\pi d^{2}$ to constrain the luminosity. Similarly, measured angular diameters, $\theta = 2R / d$, from interferometry \citep[e.g.,][]{Boyajian2012}, directly constrain the radius, and through the $\mathcal{M}$-$\mathcal{R}$ relation refine the mass estimate. In Table~\ref{tab:field_obsdata}, we display the set of observational constraints used for each field star. The use of our Bayesian framework allows us to flexibly incorporate these additional observational constraints. \footnote{We further note that all of our comparisons are done in the direct space of observables, where the assumption of normally distributed data is well justified.}

We sampled our posterior distribution (Equation~\ref{eq:posterior}) within PyMC3 \citep{Salvatier2016} using the Hamiltonian NUTS sampler, similar to the methods discussed in Section~\ref{sec:relationMR}. We then applied this process to the sample of field MUSS stars shown in Table~\ref{tab:sample}, yielding marginalized parameter distributions and uncertainties listed in Table~\ref{tab:field_sample}. We show an example for one of our targets, GJ~49, of the joint posterior distributions for our main stellar properties of interest in Figure~\ref{fig:ex_pairs}.\footnote{Posterior distributions for any of our fits are available upon request.} The results for each star are qualitatively identical, with more precise parameters available for those objects with observed bolometric fluxes and/or angular diameters.

\startlongtable
\begin{deluxetable*}{l c c c c }
	\tablecaption{ Physical Properties of MUSS Field Stars \tablenotemark{a}
		\label{tab:field_sample} }
	\tablehead{
		\colhead{Name\tablenotemark{b}}  & \colhead{$L_{\mathrm{bol}}$ (10$^{31}$ erg s$^{-1}$)}  & \colhead{Mass ($M_{\odot}$)} & \colhead{Radius ($R_{\odot}$)} & \colhead{$T_{\mathrm{eff}}$ (K)}
	}
	\startdata
		\multicolumn{5}{c}{MUSCLES} \\
	\hline
	Prox.\ Cen.\ * & $0.5997 \pm ^{0.0076}_{0.0075}$ & $0.123 \pm 0.003$ & $0.147 \pm 0.005$ & $2992 \pm ^{49}_{47}$ \\ 
	GJ 1061 & $0.628 \pm 0.014$ & $0.125 \pm 0.003$ & $0.152 \pm 0.007$ & $2977 \pm ^{72}_{69}$ \\ 
	GJ 1214 & $1.343 \pm ^{0.036}_{0.034}$ & $0.181 \pm 0.005$ & $0.204 \pm ^{0.0085}_{0.0084}$ & $3111 \pm ^{69}_{66}$ \\ 
	GJ 628 * & $4.209 \pm ^{0.037}_{0.038}$ & $0.304 \pm 0.007$ & $0.319 \pm 0.007 $ & $3307 \pm ^{38}_{36}$ \\ 
	GJ 581 * & $4.562 \pm 0.037$ & $0.307 \pm 0.007$ & $0.310 \pm 0.008$ & $3424 \pm ^{43}_{42}$ \\ 
	GJ 667C & $5.51 \pm 0.13$ & $0.327 \pm 0.008 $ & $0.337 \pm 0.014 $ & $3443 \pm ^{75}_{71}$ \\ 
	GJ 725A * & $5.783 \pm ^{0.069}_{0.068}$ & $0.336 \pm0.007$ & $0.354 \pm 0.003$ & $3401 \pm ^{18}_{17}$ \\ 
	GJ 876 * & $5.012 \pm 0.040$ & $0.346 \pm 0.007$ & $0.372 \pm 0.004$ & $3201 \pm ^{20}_{19}$ \\ 
	GJ 436 * & $9.43 \pm 0.11$ & $0.425 \pm 0.009$ & $0.432 \pm 0.011$ & $3477 \pm ^{46}_{44}$ \\ 
	GJ 832 & $10.56 \pm 0.34$ & $0.441 \pm 0.011$ & $0.442 \pm 0.018$ & $3539 \pm ^{79}_{74}$ \\ 
	GJ 887 * & $14.08 \pm 0.22$ & $0.479 \pm ^{0.011}_{0.010}$ & $0.474 \pm 0.008$ & $3672 \pm ^{36}_{34}$ \\ 
	GJ 176 * & $13.46 \pm 0.12$ & $0.485 \pm 0.012 $ & $0.474 \pm 0.015$ & $3632 \pm ^{58}_{56}$ \\ 
	\hline
	\multicolumn{5}{c}{Mega-MUSCLES} \\
	\hline
	TRAPPIST-1 & $0.234 \pm ^{0.009}_{0.008}$ & $0.090 \pm ^{0.003}_{0.002}$ & $0.120 \pm 0.006 $ & $2619 \pm ^{71}_{66}$ \\ 
	LP 756-18 & $0.993 \pm ^{0.027}_{0.026}$ & $0.145 \pm 0.004$ & $0.170 \pm 0.007$ & $3155 \pm ^{71}_{68}$ \\ 
	LHS 2686 & $1.078 \pm 0.024$ & $0.157 \pm 0.004$ & $0.182 \pm 0.008 $ & $3119 \pm ^{70}_{68}$ \\ 
	GJ 699 * & $1.302 \pm ^{0.024}_{0.023}$ & $0.1610 \pm ^{0.0036}_{0.0035} $ & $0.187 \pm 0.001$ & $3223 \pm 17 $ \\ 
	GJ 729 & $1.537 \pm 0.018$ & $0.177 \pm 0.004 $ & $0.200 \pm 0.008 $ & $3248 \pm ^{68}_{66}$ \\ 
	GJ 1132 & $1.668 \pm ^{0.049}_{0.047}$ & $0.194 \pm 0.005$ & $0.215 \pm 0.009$ & $3196 \pm ^{71}_{70}$ \\ 
	L 980-5 & $2.488 \pm ^{0.079}_{0.078}$ & $0.232 \pm 0.006$ & $0.250 \pm 0.010$ & $3278 \pm ^{74}_{70}$ \\ 
	GJ 674 * & $6.03 \pm 0.14 $ & $0.353 \pm 0.008$ & $0.361 \pm ^{0.012}_{0.011}$ & $3404 \pm ^{59}_{57}$ \\ 
	GJ 15A * & $8.608 \pm 0.069$ & $0.393 \pm ^{0.009}_{0.008}$ & $0.385 \pm0.002$ & $3601 \pm ^{12}_{11}$ \\ 
	GJ 163 & $8.28 \pm 0.24$ & $0.405 \pm 0.010 $ & $0.409 \pm ^{0.017}_{0.016}$ & $3460 \pm ^{76}_{74}$ \\ 
	GJ 849 & $11.051 \pm ^{0.095}_{0.094}$ & $0.465 \pm 0.011$ & $0.464 \pm 0.018$ & $3492 \pm ^{70}_{68}$ \\ 
	GJ 649 * & $16.74 \pm 0.170 $ & $0.524 \pm 0.012 $ & $0.531 \pm 0.012 $ & $3621 \pm ^{41}_{40}$ \\ 
	GJ 676A & $34.04 \pm ^{0.84}_{0.80}$ & $0.631 \pm 0.017$ & $0.617 \pm ^{0.028}_{0.027}$ & $4014 \pm ^{94}_{90}$ \\ 
	\hline
	\multicolumn{5}{c}{FUMES} \\
	\hline
	G 249-11 & $2.587 \pm 0.069$ & $0.237 \pm 0.006$ & $0.255 \pm 0.010$ & $3277 \pm ^{70}_{68}$ \\ 
	GJ 4334 & $3.680 \pm ^{0.090}_{0.088}$ & $0.294 \pm 0.007 $ & $0.307 \pm 0.012 $ & $3260 \pm ^{69}_{67}$ \\ 
	LP 55-41 & $8.10 \pm  0.22$ & $0.413 \pm ^{0.011}_{0.010}$ & $0.416 \pm 0.017 $ & $3412 \pm ^{75}_{73}$ \\ 
	LP 247-13 & $12.68 \pm 0.35$ & $0.495 \pm 0.013 $ & $0.492 \pm ^{0.021}_{0.020}$ & $3511 \pm ^{79}_{76}$ \\ 
	GJ 49 \tablenotemark{c} & $17.84 \pm 0.44$ & $0.541 \pm ^{0.014}_{0.015}$ & $0.534 \pm 0.023 $ &  $3670 \pm ^{86}_{80}$ \\ 
	GJ 410 & $21.34 \pm ^{0.52}_{0.51}$ & $0.557 \pm 0.015 $ & $0.549 \pm 0.024 $ & $3786 \pm ^{89}_{82}$ \\ 
	\hline
	\multicolumn{5}{c}{HAZMAT} \\
	\hline	
	GJ 173 & $12.57 \pm 0.12$ & $0.470 \pm ^{0.012}_{0.011}$ & $0.469 \pm 0.019 $ & $3589 \pm ^{75}_{71}$ \\
	G 75-55 & $22.28 \pm ^{0.58}_{0.56}$ & $0.560 \pm 0.015 $ & $0.552 \pm 0.024 $ & $3818 \pm ^{87}_{85}$ \\ 
	\hline
	\multicolumn{5}{c}{Living with Red Dwarf} \\
	\hline
	GJ 213 & $2.407 \pm ^{0.036}_{0.035}$ & $0.218 \pm 0.005 $ & $0.238 \pm 0.009 $ & $3334 \pm ^{67}_{63}$ \\ 
	GJ 821 & $6.977 \pm ^{0.089}_{0.087}$ & $0.355 \pm 0.009 $ & $0.363 \pm ^{0.015}_{0.014}$ & $3518 \pm ^{73}_{69}$ \\ 
	\hline
	\multicolumn{5}{c}{Miscellaneous} \\
	\hline
	LHS 2065 & $0.1271 \pm  0.0007$ & $0.082 \pm 0.002$ & $0.113 \pm 0.006 $ & $2317 \pm ^{61}_{56}$ \\ 
	VB 10 & $0.1911 \pm 0.0013$ & $0.0881 \pm ^{0.0026}_{0.0024}$ & $0.1183 \pm ^{0.0059}_{0.0057}$ & $2508 \pm ^{63}_{60}$ \\ 
	VB 8 & $0.2468 \pm 0.0016 $ & $0.0914 \pm ^{0.0026}_{0.0025}$ & $0.1214 \pm ^{0.0060}_{0.0057}$ & $2640 \pm ^{65}_{64}$ \\ 
	LHS 3003 & $0.230 \pm 0.0019$ & $0.0929 \pm ^{0.0027}_{0.0025}$ & $0.123 \pm 0.006 $ & $2580 \pm ^{64}_{61}$ \\ 
	GJ 406 * & $0.406 \pm ^{0.006}_{0.005}$ & $0.110 \pm 0.003$ & $0.144 \pm 0.004$ & $2749 \pm ^{44}_{41}$ \\ 
	LHS 1140 & $1.615 \pm ^{0.041}_{0.039}$ & $0.183 \pm 0.005$ & $0.205 \pm 0.008 $ & $3250 \pm ^{71}_{69}$ \\ 
	GJ 447 * & $1.401 \pm 0.019 $ & $0.176 \pm 0.004 $ & $0.198 \pm 0.007 $ & $3189 \pm ^{55}_{53}$ \\ 
	GJ 1207 & $2.586 \pm 0.042$ & $0.235 \pm 0.006 $ & $0.253 \pm 0.010$ & $3288 \pm ^{67}_{65}$ \\ 
	GJ 1243 & $2.729 \pm 0.090 $ & $0.240 \pm 0.006 $ & $0.258 \pm 0.010$ & $3303 \pm ^{72}_{68}$ \\ 	
	GJ 545 & $3.198 \pm ^{0.056}_{0.055}$ & $0.253 \pm 0.007$ & $0.269 \pm 0.011$ & $3362 \pm ^{70}_{68}$ \\ 
	GJ 3378 & $3.292 \pm 0.062 $ & $0.263 \pm 0.006 $ & $0.278 \pm ^{0.011}_{0.010}$ & $3332\pm ^{66}_{64}$ \\ 
	GJ 403 & $3.292 \pm ^{0.051}_{0.052}$ & $0.265 \pm 0.007$ & $0.280 \pm 0.011 $ & $3320 \pm ^{69}_{64}$ \\ 
	GJ 402 & $3.057 \pm 0.032 $ & $0.268 \pm ^{0.007}_{0.006}$ & $0.284 \pm 0.011 $ & $3240 \pm ^{65}_{60}$ \\ 
	GJ 3991 & $3.270 \pm ^{0.044}_{0.043}$ & $0.2714 \pm ^{0.0066}_{0.0065}$ & $0.286 \pm 0.011 $ & $3279 \pm ^{65}_{63}$ \\ 
	GJ 3325 & $3.595 \pm ^{0.042}_{0.043}$ & $0.274 \pm 0.007 $ & $0.289 \pm 0.011 $ & $3341 \pm ^{69}_{65}$ \\ 
	GJ 273 * & $4.143 \pm ^{0.058}_{0.057}$ & $0.2982 \pm ^{0.0067}_{0.0064}$ & $0.319 \pm 0.004 $ & $3297 \pm ^{22}_{21}$ \\ 
	GJ 4367 & $4.114 \pm ^{0.087}_{0.089}$ & $0.304 \pm 0.008$ & $0.316 \pm 0.012$ & $3305 \pm ^{69}_{65}$ \\ 
	YZ CMi & $4.35 \pm ^{0.15}_{0.14}$ & $0.316 \pm 0.008$ & $0.328 \pm 0.013 $ & $3293 \pm ^{74}_{71}$ \\ 
	EV Lac & $4.88 \pm0.15$ & $0.320 \pm 0.008 $ & $0.331 \pm 0.013 $ & $3370 \pm ^{75}_{71}$ \\ 
	GJ 411 * & $8.399 \pm 0.081$ & $0.389 \pm 0.008 $ & $0.392 \pm 0.004 $ & $3547 \pm 18 $ \\ 
	GJ 687 * & $8.330 \pm 0.070 $ & $0.411 \pm 0.009$ & $0.4187 \pm ^{0.0066}_{0.0063}$ & $3426 \pm ^{27}_{28}$ \\ 
	AD Leo & $8.67 \pm ^ 0.10 $ & $0.426 \pm 0.010$ & $0.428 \pm ^{0.017}_{0.016}$ & $3425 \pm 68$ \\ 
	GJ 588 & $10.95 \pm ^{0.32}_{0.31}$ & $0.461 \pm 0.012$ & $0.460 \pm 0.019 $ & $3499 \pm ^{79}_{73}$ \\ 
	GJ 526 * & $14.18 \pm 0.19$ & $0.488 \pm 0.011$ & $0.487 \pm 0.008$ & $3628 \pm ^{32}_{31}$ \\ 
	GJ 3470 & $15.20 \pm ^{0.43}_{0.41}$ & $0.502 \pm 0.013 $ & $0.499 \pm 0.021 $ & $3649 \pm ^{82}_{78}$ \\ 
	Kepler-138 & $21.78 \pm ^{0.56}_{0.53}$ & $0.545 \pm 0.014 $ & $0.537 \pm ^{0.023}_{0.022}$ & $3846 \pm ^{86}_{84}$ \\ 
	GJ 809 * & $19.85 \pm ^{0.22}_{0.21}$ & $0.547 \pm 0.012$ & $0.546 \pm 0.006 $ & $3729 \pm ^{22}_{23}$ \\ 
	K2-3 & $21.45 \pm ^{0.64}_{0.62}$ & $0.551 \pm 0.015 $ & $0.543 \pm ^{0.024}_{0.023}$ & $3810 \pm ^{89}_{84}$ \\ 
	GJ 880 * & $20.00 \pm ^{0.16}_{0.15}$ & $0.553 \pm 0.012 $ & $0.549 \pm 0.003 $ & $3724 \pm 12$ \\ 
	GJ 205 * & $24.65 \pm 0.22 $ & $0.590 \pm ^{0.014}_{0.013}$ & $0.5780 \pm ^{0.0024}_{0.0025}$ & $3824 \pm ^{12}_{11}$ \\ 
	GJ 338A & $29.22 \pm ^{0.27}_{0.28}$ & $0.599 \pm 0.016 $ & $0.587 \pm ^{0.025}_{0.024}$ & $3959 \pm ^{86}_{80}$ \\ 
	TOI-1235 & $33.51 \pm ^{0.77}_{0.78}$ & $0.617 \pm  0.016 $ & $0.604 \pm 0.025 $ & $4041 \pm ^{89}_{85}$ \\ 
	\enddata
	\tablenotetext{a}{Stellar properties are as determined in Sections~\ref{sec:empfield}, quoting medians and the central 68\% confidence interval. Joint posteriors for parameters are available upon request.}
	\tablenotetext{b}{Star names listed with an asterisk have measured angular diameters from interferometric measurements, see Table~\ref{tab:field_obsdata}.}
	\tablenotetext{c}{For GJ~49 we updated the input optical photometry used in the estimation relative to those used in \citet{Pineda2021arXiv210212485P}, leading to a slight change in luminosity and temperature.}
	
\end{deluxetable*}

\begin{figure*}[tbp]
	\centering
	\includegraphics{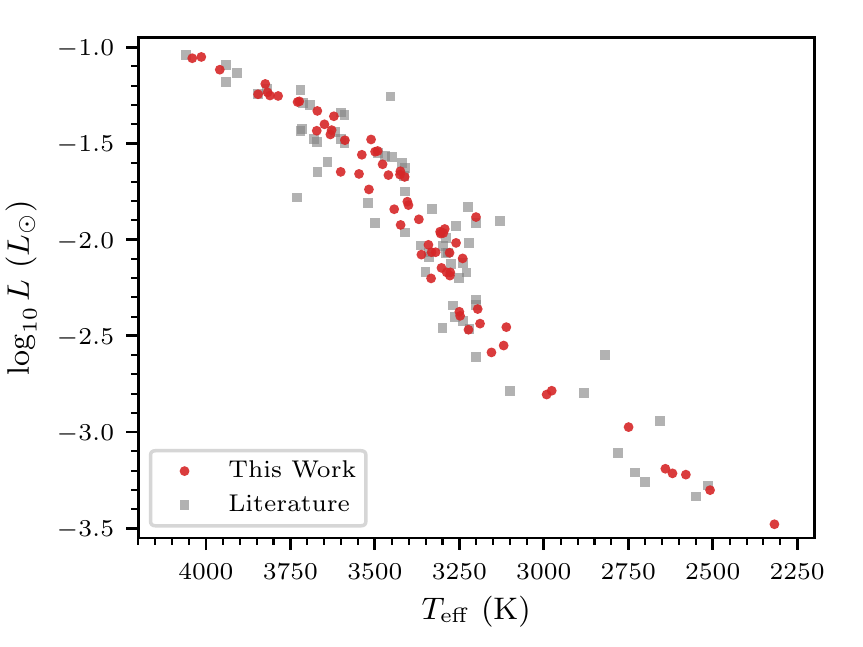} 
	\includegraphics{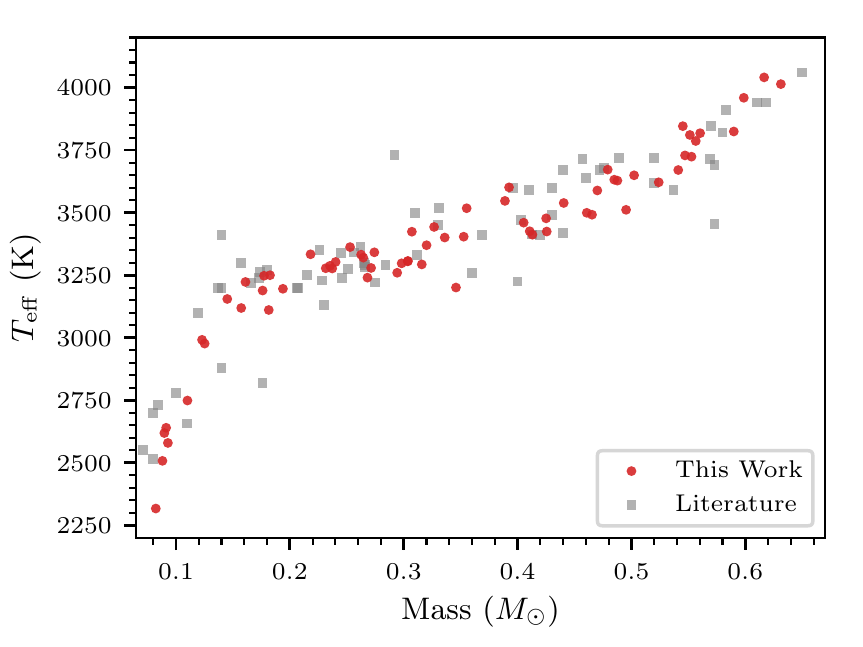} 
	\caption{Compared to the values found in the literature (squares), our median stellar parameter determinations (circles) define a sequence across the M-dwarf regime with less scatter in both the $L_{\mathrm{bol}}$-$T_{\mathrm{eff}}$ (\textit{Left}) and $\mathcal{M}$-$T_{\mathrm{eff}}$ (\textit{Right}) planes.}
	\label{fig:paramLit}
\end{figure*}

\subsection{Results and Applications}

\subsubsection{Comparison of Stellar Properties with Literature}\label{sec:litcomp}

By applying our methodology to the MUSS field stars, we have consistently determined their properties with well defined uncertainties and their correlations. To test how our methods compare against typical literature determinations we have compiled the stellar parameters for the MUSS stars found in eclectic assessments across numerous works. We typically utilized estimates from the same work, using the radii and effective temperatures to set the bolometric luminosity. These effective temperatures were often spectroscopically determined (atmospheric model based or otherwise), with radii coming from semi-empirical relations like that of \cite{Mann2015}. Literature masses were often determined from earlier iterations of the $\mathcal{M}$-$M_{K_{S}}$ relation. The literature sources are listed in Appendix~\ref{sec:ap_litcite}. 

We provide this comparison in Figure~\ref{fig:paramLit}, showing both luminosity vs.\ effective temperature, and effective temperature vs.\ mass. Often the estimates, like those of the interferometric objects, are nearly identical. Nevertheless, in both planes, our best fit stellar parameters provide ensemble M-dwarf sequences with reduced scatter. As not every work treated all these properties, the compilation as is typical in the literature can be inconsistent, leading to larger outliers.

\subsubsection{M-dwarf Main Sequences}\label{sec:GP}

\begin{deluxetable}{l c c c}
	\tablecaption{ GP M-dwarf Main Sequence Models\tablenotemark{a}
		\label{tab:GPresults} }
	\tablehead{
		\colhead{Variables\tablenotemark{b}}  & \colhead{$l$} & \colhead{$\eta$}  & \colhead{$\sigma$} 
	}
	\startdata
	$L$-$T_{\mathrm{eff}}$	& $\mathcal{HC}(100)$ K & $\mathcal{HC}(3)$ dex & $\mathcal{HC}(0.4)$ dex \\ 
&$2260 \pm ^{800}_{550} $ & $2.4 \pm ^{1.7}_{0.8} $& $0.096 \pm ^{0.014}_{0.012}$ \\		 $L$-$\mathcal{M}$	& $\mathcal{HC}(0.3)$ $M_{\odot}$ & $\mathcal{HC}(3)$ dex & $\mathcal{HC}(0.1)$ dex \tablenotemark{c} \\ 
	& $0.55 \pm ^{0.23}_{0.15} $ & $4.6 \pm ^{4.0}_{1.8} $& $<0.03$\\ 
	$T_{\mathrm{eff}}$-$\mathcal{M}$	& $\mathcal{HC}(0.1)$ $M_{\odot}$ & $\mathcal{HC}(50)$ K &  $\mathcal{HC}(50)$ K  \\
	& $0.57 \pm ^{0.21}_{0.14} $ $M_{\odot}$ & $3860 \pm ^{3300}_{1430} $  K & $57 \pm ^{10}_{9}$ K \\
	\enddata
	\tablenotetext{a}{The rows for each variable pair show first the model priors, and underneath the median and central 68\% confidence intervals of the fit parameters. $\mathcal{HC}(\beta)$ indicates a Half-Cauchy distribution with shape parameter $\beta$. }
	\tablenotetext{b}{In practice, for the luminosity sequences we fit in the space of $\log_{10}$ of the bolometric luminosity relative to solar.}
	\tablenotetext{c}{For the mass-luminosity sequence, we additionally bounded the scatter to be greater than 0.001 dex, to prevent the small scatter from causing the likelihood to diverge.}
\end{deluxetable}

The MUSS field targets span the entire M-dwarf regime, and with consistently determined properties we now examine the main sequence of these low-mass stars. Using our stellar parameters with well defined uncertainties, we applied Gaussian Process (GP) regression models to the ensemble to capture the shape of the stellar main sequence across different parameters: $L_{\mathrm{bol}}$ vs.\ $T_{\mathrm{eff}}$, $L_{\mathrm{bol}}$ vs.\ $\mathcal{M}$, and $T_{\mathrm{eff}}$ vs.\ $\mathcal{M}$. GP models enable a flexible means of fitting, and have been used extensively to fit data of unknown functional form \citep[e.g.,][]{Aigrain2016MNRAS.459.2408A}.   We used a Matern52 covariance kernel, $k$, and fit the hyper-parameters of scale length, $l$, covariance amplitude, $\eta$, and added an intrinsic noise, $\sigma$, at fixed abscissa to the dataset. Specifically, we used the covariance function

\begin{eqnarray}
k(x,x') = \eta^{2} \left( 1 +  \frac{\sqrt{5(x - x')^{2}}}{l} + \frac{5(x - x')^{2}}{3 l}  \right) \times \nonumber  \\
  \exp \left( - \frac{\sqrt{5(x - x')^{2}}}{l} \right) \; .
\end{eqnarray}

\noindent This choice of kernel is general purpose, flexible, and yields GP models that are smooth without being too smooth (i.e.\ infinitely differentiable).

\begin{figure*}[tbp]
	\centering
	\includegraphics{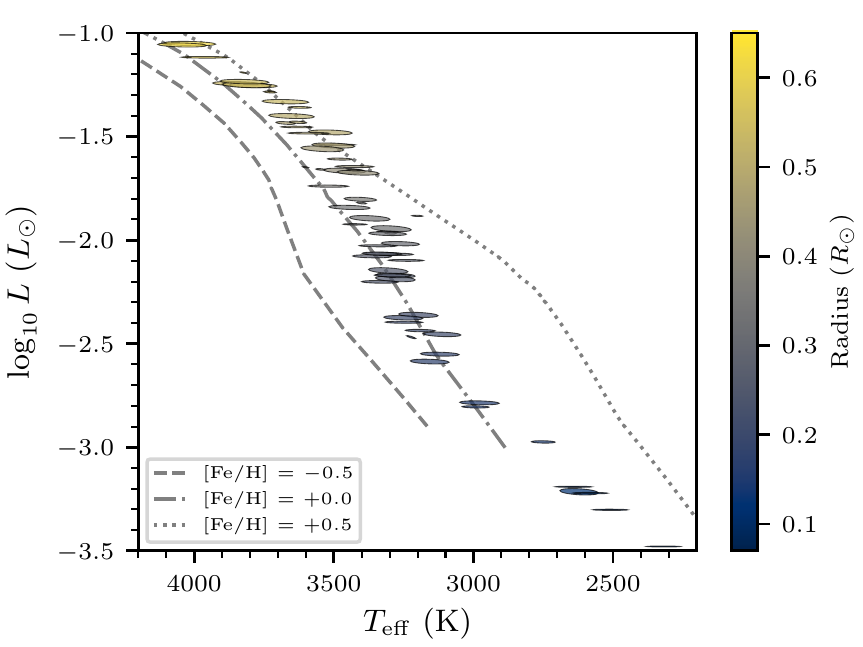} 
	\includegraphics{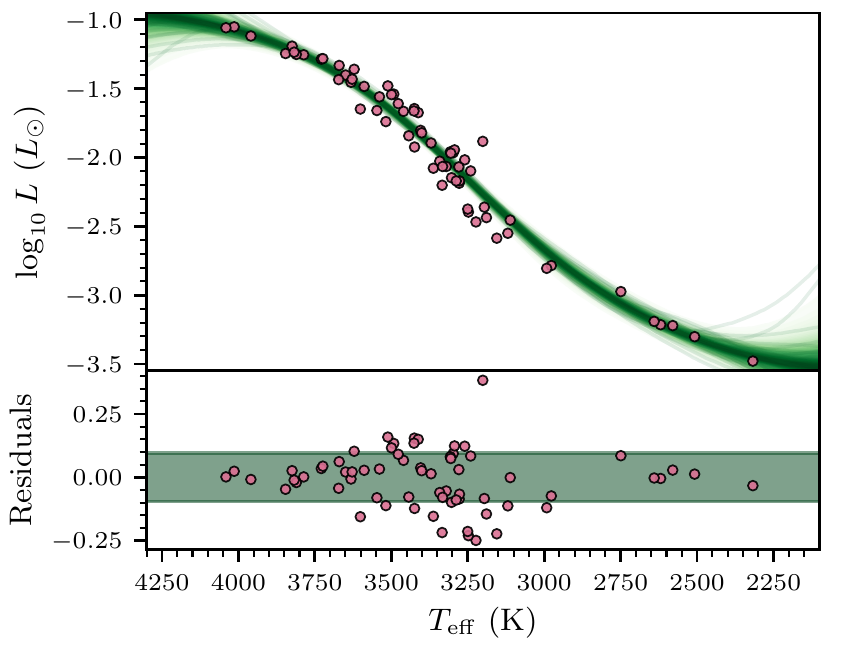} 
	\caption{\textit{Left} - The MUSS data set illustrates the M-dwarf main sequence in luminosity vs.\ effective temperature, with each data point shown by representative 1$\sigma$ error ellipses, with shading according to the stellar radius. Different metallicity lines correspond to MIST model 5~Gyr isochrones. \textit{Right} - The GP regression model (green lines) of the data set, shown as circles plotted at the median stellar properties, traces the data well, but reveals a $\sim$0.1 dex scatter of luminosity at fixed effective temperature. This scatter is shaded in the lower panel of the data residuals relative to the median GP solution.}
	\label{fig:LumTeffR}
\end{figure*}

\begin{figure*}[tbp]
	\centering
	\includegraphics{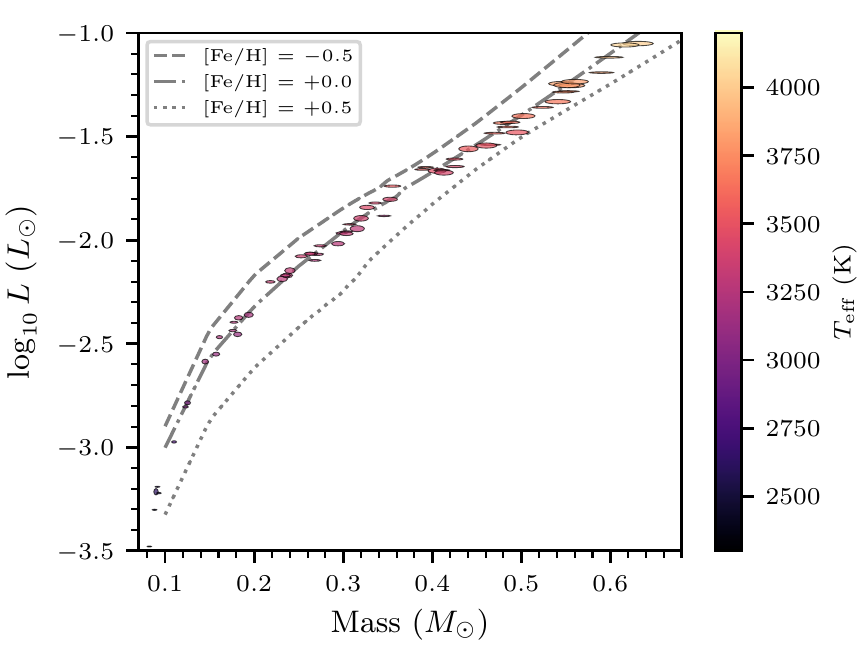} 
	\includegraphics{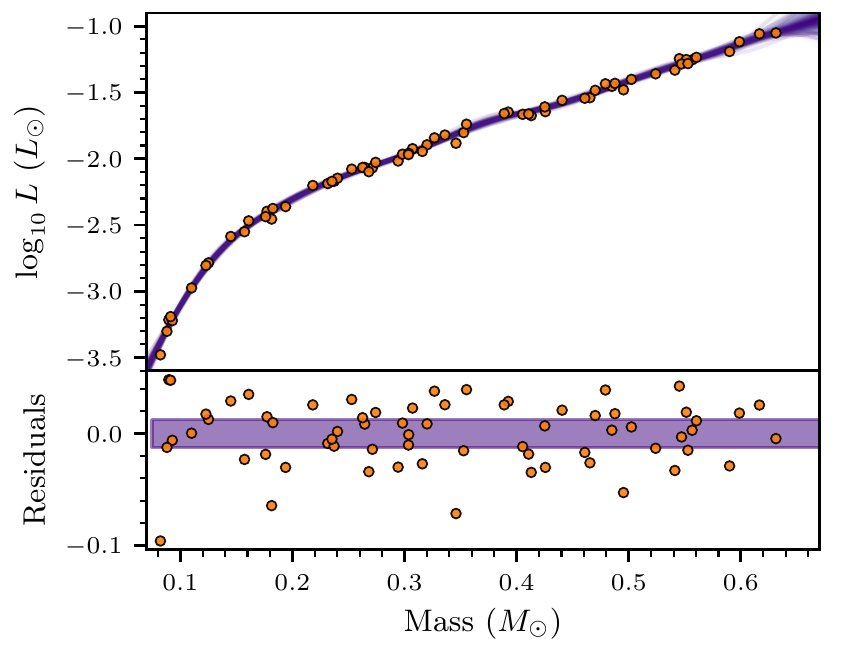} 
	\caption{Same as Figure~\ref{fig:LumTeffR}, but for luminosity vs.\ mass. In the left panel the error ellipses are shaded with the stellar effective temperature. The residuals of the right panel show an intrinsic scatter about the GP regression model of 0.012 dex.}
	\label{fig:MassLumT}
\end{figure*}

\begin{figure*}[tbp]
	\centering
	\includegraphics{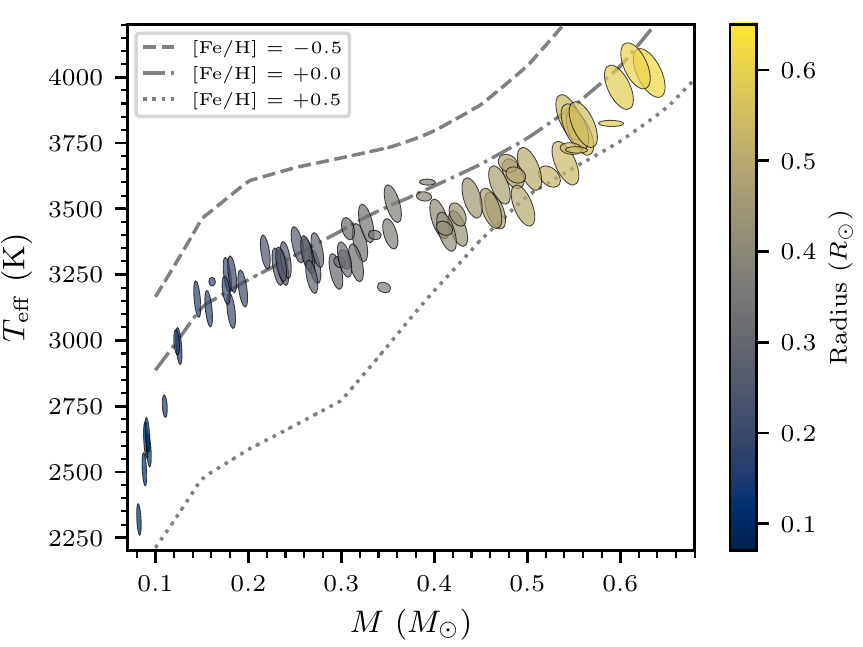} 
	\includegraphics{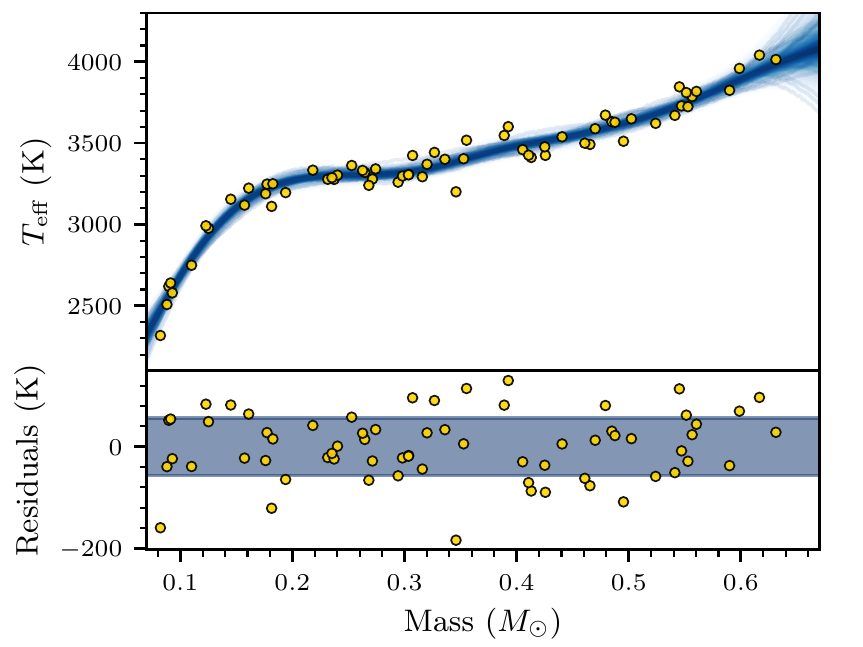} 
	\caption{Same as Figure~\ref{fig:LumTeffR}, but for effective temperature vs.\ mass. In the left panel the error ellipses are shaded with the stellar radius. The residuals of the right panel show an intrinsic scatter about the GP regression model of 57~K.}
	\label{fig:MassTeffR}
\end{figure*}

In fitting a particular sequence, for example $L_{\mathrm{bol}}$-$T_{\mathrm{eff}}$, the data set consisted of bivariate normal distributions for each star, determined by the covariance and medians of the joint posteriors for the parameters from the Monte Carlo sampling explained in Section~\ref{sec:field}. This allows our methods to both account for uncertainties in both dimensions in the fitting and include the correlations of those uncertainties. The fitting was done through Monte Carlo sampling within PyMC3. We present the results of these methods in Figures~\ref{fig:LumTeffR}-\ref{fig:MassTeffR}. 

We illustrate the traditional theoretical HR diagram with luminosity as a function of effective temperature in Figure~\ref{fig:LumTeffR}. In the left panel, the data ensemble shows the shape of this main sequence, with the parameter uncertainties illustrated as 1$\sigma$ error ellipses, conveying the size and correlation of the output parameters, with shading corresponding to the stellar radius. The bulk of the data pertains to early and mid-M dwarfs with a handful of late-M-dwarfs extending the sequence to effective temperatures as low as 2300 K. For comparison, we also include the 5 Gyr isochrones from the MESA Isochrones and Stellar Tracks \citep[MIST;][]{Dotter2016ApJS..222....8D}. We show the results of the GP regression of $L_{\mathrm{bol}}$-$T_{\mathrm{eff}}$ in the right panel of Figure~\ref{fig:LumTeffR}, with data points displaying the central medians, and the best fit GP curves sampled from the posterior distribution. The ensemble of lines presents the best GP solutions, with the residual plot showing the best fit scatter about that solution. The result is smooth and fits the data well with an intrinsic scatter of $\sim$0.1 dex. 

In contrast to $L_{\mathrm{bol}}$-$T_{\mathrm{eff}}$, the $L_{\mathrm{bol}}$-$\mathcal{M}$ sequence shown in Figure~\ref{fig:MassLumT} is very well defined with a tighter scatter about the mean GP regression model. Figure~\ref{fig:MassLumT} is similar to Figure~\ref{fig:LumTeffR}, but the shading of the data ellipses in the left panel now show the target effective temperatures. The sample posterior for the fit luminosity scatter peaks at $\sim$0.012 dex, but the distribution on the scatter extends down toward zero, because the data point uncertainties can largely account for the bulk of the scatter in the model regression. Accordingly, we describe the sampled scatter in luminosity at fixed mass with a 3$\sigma$ upper limit of 0.03 dex in Table~\ref{tab:GPresults}. The differences between the $L_{\mathrm{bol}}$-$T_{\mathrm{eff}}$ and the $L_{\mathrm{bol}}$-$\mathcal{M}$ sequences are largely a consequence of the mid-M dwarfs. Objects around 3300 K show a broad range of possible luminosities, whereas in $L_{\mathrm{bol}}$-$\mathcal{M}$ space, the relationship exhibits far less scatter.

We show this further in the $T_{\mathrm{eff}}$-$\mathcal{M}$ relationship of Figure~\ref{fig:MassTeffR}, with the shading of the error ellipses in the left panel indicative of the stellar radius. Among the very-low mass stars, the higher-mass M dwarfs exhibit much higher effective temperatures, with differences of $\sim$600~K between 0.1 and 0.2 $M_{\odot}$ stars. The effective temperature sequence then plateaus in the mid-M dwarfs before increasing again in the early-to-mid M-dwarf regime. For the mid-M dwarfs, on average across the sample between about 0.2 and 0.3 solar masses, the effective temperature is ostensibly constant. Furthermore, the GP regression model in $T_{\mathrm{eff}}$-$\mathcal{M}$ reveals an intrinsic scatter of only $\sim$60~K, at fixed mass. 
The results of all our GP regression models are summarized in Table~\ref{tab:GPresults} and we further publicly provide the GP solutions with 68\% confidence curves.\footnote{\url{https://github.com/jspineda/stellarprop}}

Collectively, these results show how in M-dwarfs, especially the mid-M dwarfs, effective temperature, when used to predict other stellar properties is subject to a great deal of scatter. Although $T_{\mathrm{eff}}$ is a useful quantity and directly related to the inputs to model spectra, its use when possible should be tied to empirically determined temperatures, like those from interferometric measurements. This is especially true if the metallicity is unknown. Mass and luminosity are both more fundamental and show a very tight relationship, and are thus better suited as the basis for inferring additional stellar properties. Our sample averages over the possible metallicity differences, and being composed largely of near solar metallicity objects, is representative of the nearby stellar neighborhood. It is likely that the intrinsic scatter in the $T_{\mathrm{eff}}$-$\mathcal{M}$ relation is defined by the distribution of metallicities in the MUSS field stars (see Figure~\ref{fig:fedist}).

In Figures~\ref{fig:LumTeffR},~\ref{fig:MassLumT}, and~\ref{fig:MassTeffR}, we also showed how the MIST isochrones at 5 Gyr compare to our empirical sequences. The $L_{\mathrm{bol}}$-$\mathcal{M}$ plane shows that the solar metallicity track matches the bulk of the data really well (the model does not extend to the lowest-mass stars in our sample). However, the model effective temperatures are systematically higher than the MUSS sample stars. With well matched bolometric luminosities, this likely reflects the long standing issue that the evolutionary models often under-predict M-dwarf stellar radii \citep[e.g.,][]{LopezMorales2007ApJ...660..732L,Kesseli2018}, and thus need higher $T_{\mathrm{eff}}$ to match the luminosity.

\subsubsection{An Empirical Boundary for Fully Convective Interiors?}\label{sec:fcbound}

\begin{figure}[tbp]
	\centering
	\includegraphics{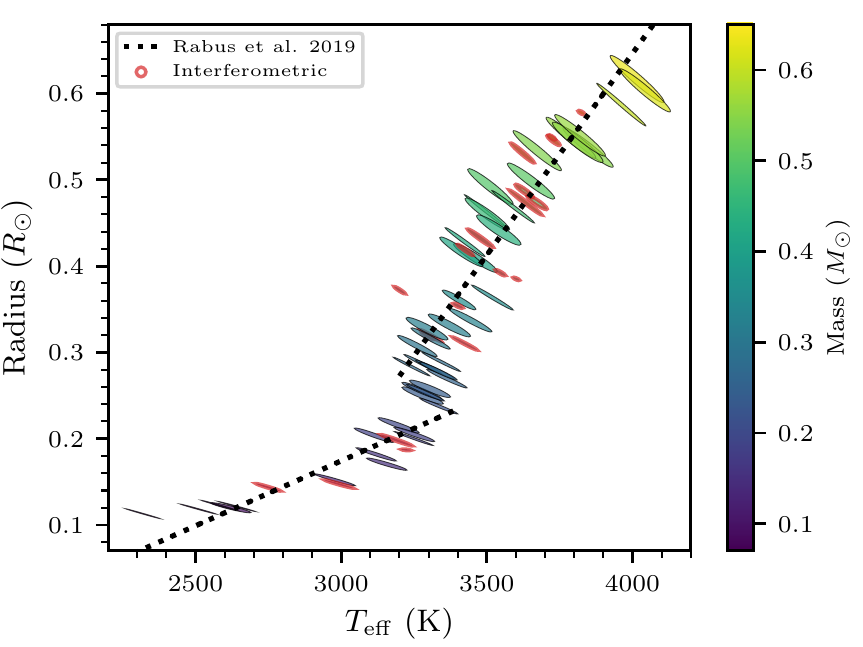} 
	\caption{The $\mathcal{R}$-$T_{\mathrm{eff}}$ data of the MUSS targets show the `discontinuity' evinced by \citet{Rabus2019}, including their linear fits as dashed lines, and MUSS objects with interferometric data highlighted. MUSS data are shown as 1$\sigma$ error ellipses, indicating how these properties are highly correlated, with shading indicative of the stellar mass.}
	\label{fig:TeffRadM}
\end{figure}

There is an apparent plateau at low-masses in the $T_{\mathrm{eff}}$-$\mathcal{M}$ relationship (see Figure~\ref{fig:MassTeffR}). \citet{Rabus2019} showed using an ensemble of M-dwarfs that the $\mathcal{R}$-$T_{\mathrm{eff}}$ relation appeared to have a `discontinuity' at effective temperatures between 3200 and 3340 K. Within this temperature regime their relation was effectively vertical; we illustrate this feature in Figure~\ref{fig:TeffRadM} using the MUSS sample, and including the linear fits from \cite{Rabus2019} (their Equation~7), not accounting for possible metallicity effects. The linear fits from \cite{Rabus2019} are consistent with our data.

Given the largely one-to-one relationship between stellar mass and radius (see Section~\ref{sec:relationMR}), recasting this relation as $T_{\mathrm{eff}}$-$\mathcal{M}$ yields the results we have shown in Figure~\ref{fig:MassTeffR}. The effective temperature plateau at low masses ($\sim$0.25 $M_{\odot}$) corresponds exactly to the this `discontinuity' in $\mathcal{R}$-$T_{\mathrm{eff}}$. \citet{Rabus2019} considered this feature within the M-dwarf population as indicative of the transition between partially and fully convective interiors, a claim we discuss further below. Firstly, we consider the question: at what mass does this feature occur?

\begin{figure}[tbp]
	\centering
	\includegraphics{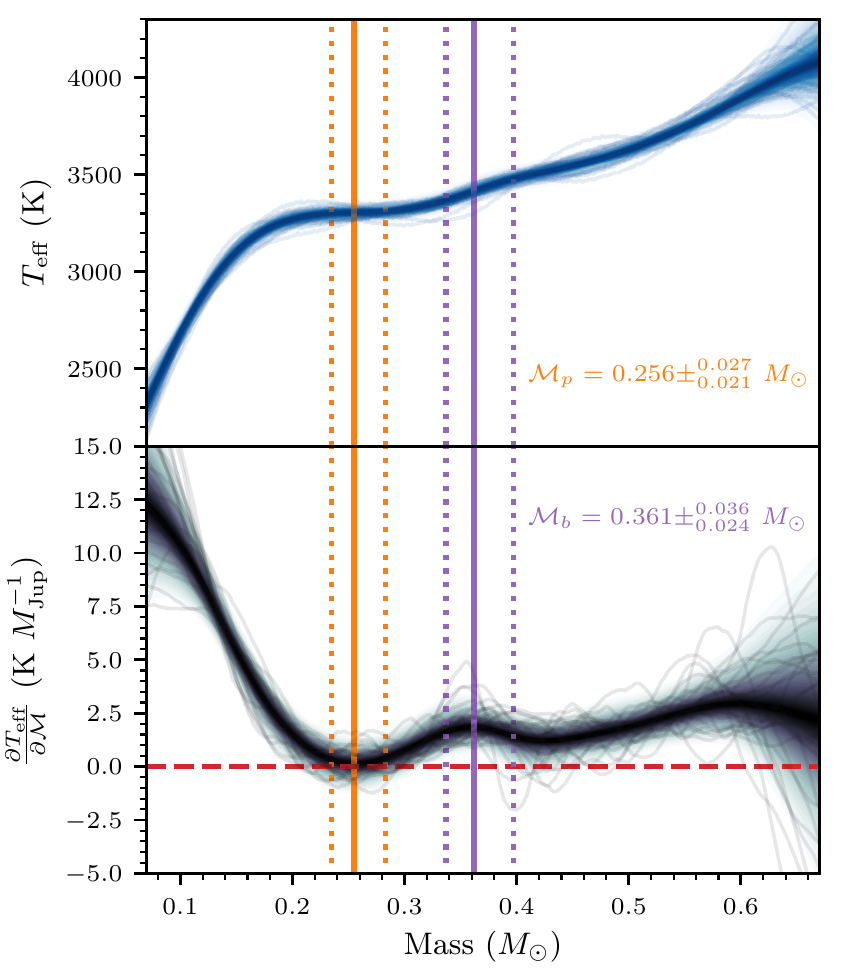} 
	\caption{Using a GP regression model for $T_{\mathrm{eff}}$-$\mathcal{M}$ we determined the derivative of the model function, and use it to determine the location of the effective temperature plateau for mid-M dwarfs at $\mathcal{M}_{p} \sim$ 0.26 $M_{\odot}$, a feature first highlighted by \citet{Rabus2019}. We separately measure a local maximum at $\mathcal{M}_{b} \sim$ 0.36 $M_{\odot}$ that is likely associated with the fully convective boundary, Section~\ref{sec:fcbound}. \textit{Top} - Our GP result for $T_{\mathrm{eff}}$-$\mathcal{M}$, as in Figure~\ref{fig:MassTeffR}. \textit{Bottom} - The derivative of effective temperature as a function of stellar mass, predicted from the GP model. Units are in degrees Kelvin per Jupiter mass, $M_{\mathrm{Jup}}\sim10^{-3} M_{\odot}$. For example, near 0.15 $M_{\odot}$, the addition of 1 Jupter's worth of mass would increase the stellar effective temperature by $\sim$5 K. }
	\label{fig:DerTM}
\end{figure}

To identify the plateau feature we determined where the derivative of the GP model approaches zero, or is otherwise minimal. One benefit of the GP model is that not only can it predict the values of the functional relationship, but the derivative of the GP is itself a GP, and can therefore be used similarly to define the derivative of that function \citep{Rasmussen2006gpml.book.....R}. Using the results of the regression model, we further calculate the derivative, $\partial T_{\mathrm{eff}} / \partial \mathcal{M}$, and show those results in Figure~\ref{fig:DerTM}, sampling at $\Delta \mathcal{M} = 0.001$  $M_{\odot}$.\footnote{This explicitly uses the GP derivative function, but we note that finite difference methods applied to the GP regression curves would yield similar results. Our chosen methods are more computationally intensive, but circumvent the finite difference approximations, fully accounting for all of the associated uncertainties, and can be calculated at arbitrary mass points.} The top panel re-plots the $T_{\mathrm{eff}}$-$\mathcal{M}$ relation, with the bottom one showing its derivative. At the location of the plateau, the derivative indeed appears to dip to zero, but is otherwise positive.

To define at which mass this feature exists, we took the ensemble of GP curves that represent the derivative, and defined a representative sequence from the median value of the derivative at each mass. The median derivative minimum occurs clearly at 0.256~$M_{\odot}$. To estimate a confidence interval, we restricted consideration of each individual GP derivative curve from the ensemble to masses below 0.35 $M_{\odot}$. Then for each curve we recorded within that range ($\mathcal{M}<$0.35 $M_{\odot}$) where the curve was at its lowest point. Although individual curves can be variable, the ensemble produced a distribution of masses reflective of the plateau feature. The result with 68\% confidence interval was $\mathcal{M}_{p} = 0.256 \pm ^{0.027}_{0.021}$ $M_{\odot}$. Along the median $T_{\mathrm{eff}}$-$\mathcal{M}$ relation, this mass corresponds to 3305~K.

\begin{figure}[tbp]
	\centering
	\includegraphics{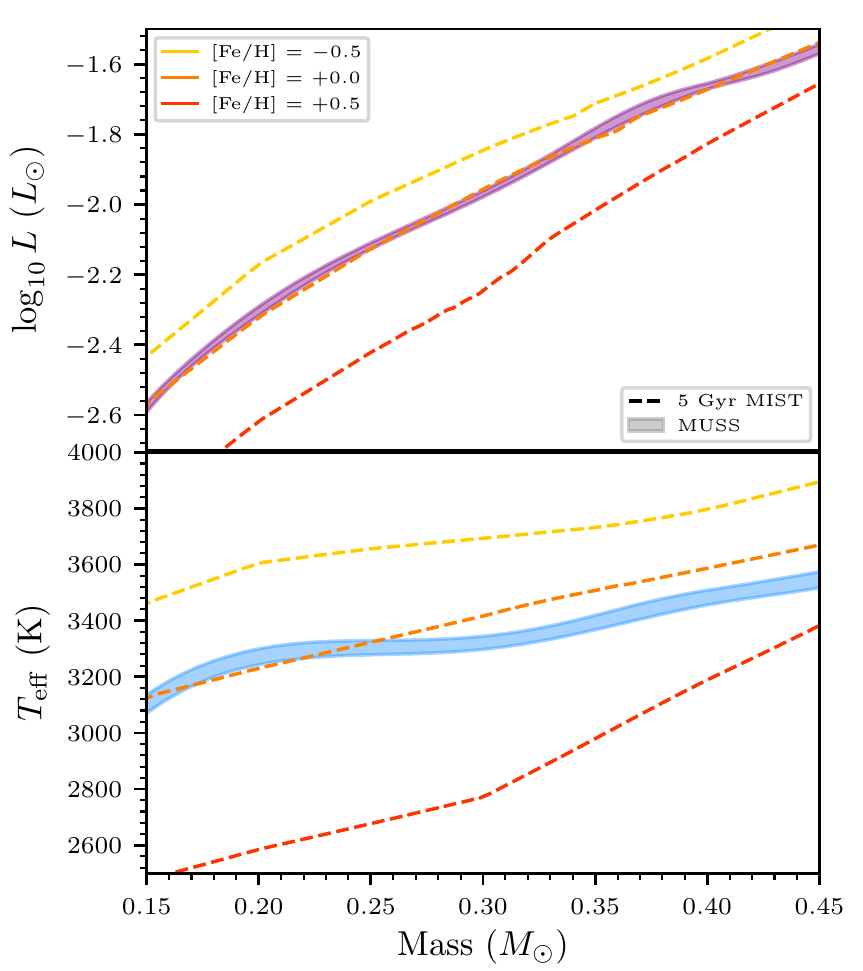} 
	\caption{We compare our empirical MUSS GP stellar sequences, as shaded central 68\% confidence intervals in $L_{\mathrm{bol}}$-$\mathcal{M}$ (\emph{top}) and $T_{\mathrm{eff}}$-$\mathcal{M}$ (\emph{bottom}), to the MIST model 5 Gyr isochrones. The different colored lines indicate distinct metallicities. The solar metallicity model can reproduce the stellar bolometric luminosities, but misses the shape of the effective temperature sequence.}
	\label{fig:MIST}
\end{figure}

\begin{figure}[tbp]
	\centering
	\includegraphics{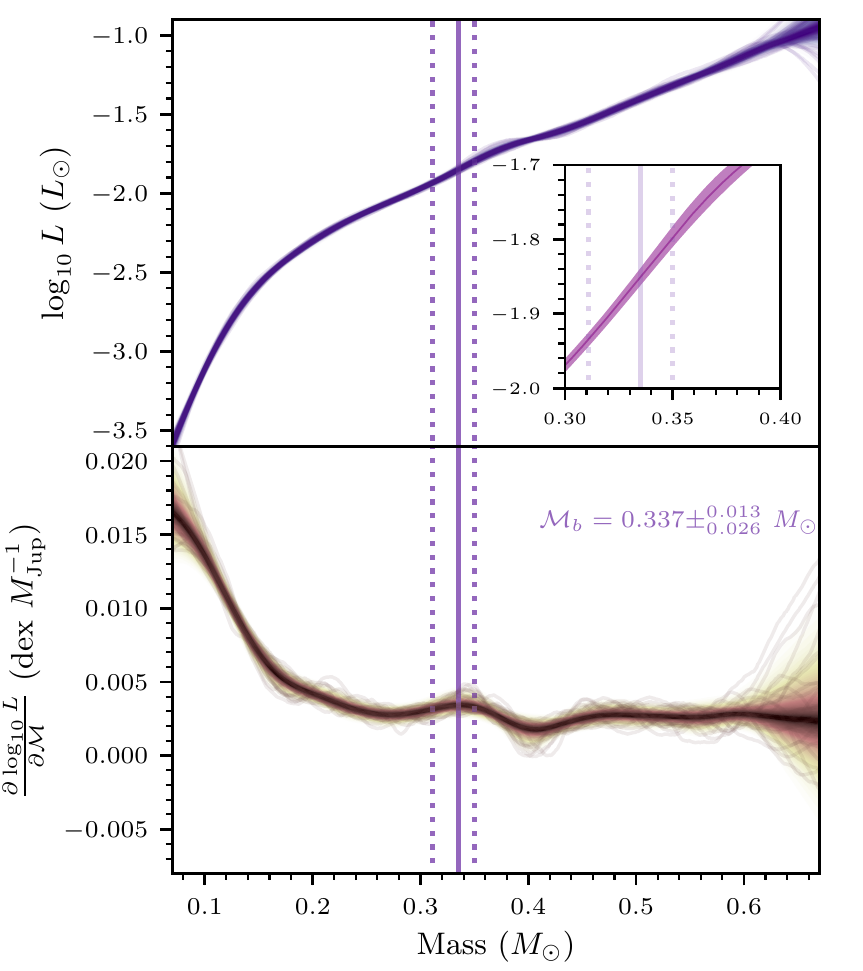} 
	\caption{Using a GP regression model for $L_{\mathrm{bol}}$-$\mathcal{M}$ (\emph{top}) we determined the derivative of the model function (\emph{bottom}). The sequence slope is plotted as the derivative of the log luminosity with respect to mass in units of dex luminosity relative to solar per unit Jupiter mass. For example, at around 0.1~$M_{\odot}$ the addition of 1 Jovian mass increases the stellar bolometric luminosity by almost 0.015 dex relative to solar. Changes in the slope are subtle, as shown in the inset of top panel with median and central 68\% confidence bounds on the GP curves, but are clear from the derivative curve ensemble. The first local maximum occurs at $\sim$0.34 $M_{\odot}$, likely a consequence of internal structure changes associated with the transition from fully to partially convective interiors.}
	\label{fig:DerLM}
\end{figure}

\citet{Rabus2019} suggested that this plateau was indicative of the transition between fully and partly convective interiors, with it taking place at masses just below that corresponding to the Gaia M-dwarf gap \citep[$\sim$0.35~$M_{\odot}$,][]{Jao2018ApJ...861L..11J,Feiden2021ApJ...907...53F}. However, modeling of the impact of the $^{3}$He instability \citep{vanSaders2012ApJ...751...98V}, the cause for the Jao Gap, indicates that the onset of this instability marks the transition between partly and fully convective interiors \citep{Baraffe2018A&A...619A.177B,MacDonald2018MNRAS.480.1711M,Feiden2021ApJ...907...53F}. The plateau mass ($\sim$0.26 $M_{\odot}$) is too far from the masses corresponding to the Gaia M-dwarf gap to likely be associated with the convective transition. 

It is unclear what effect may be leading to small or no changes in effective temperature between 0.2 and 0.3 $M_{\odot}$. For example, field age model solar metallicity isochrones do not exhibit such a shallow effective temperature derivative in this regime \citep[e.g.,][]{Baraffe2015,Feiden2016}. We illustrate this by plotting our median GP curves against 5 Gyr MIST isochrones \citep{Morton2015ascl.soft03010M, Dotter2016ApJS..222....8D} in Figure~\ref{fig:MIST}, zooming into this mass regime around 0.3~$M_{\odot}$. Despite matching the $L_{\mathrm{bol}}$-$\mathcal{M}$ sequence well, the model effective temperature sequence does not reproduce the extent of the plateau, and generally over estimates $T_{\mathrm{eff}}$. It is possible that the observed plateau feature may be an emergent effect of the stellar sample, which averages together a range of stellar metallicities (see Figure~\ref{fig:fedist}), however, the observed slope remains difficult to reproduce. Nevertheless, on average it appears observationally that the radial response to increasing luminosity over that mass range balances such that the effective temperature remains nearly constant.

The desire to measure the plateau feature in the $T_{\mathrm{eff}}$-$\mathcal{M}$ sequence prompted our derivative analysis. Interestingly, this derivative also reveals a local maximum at $\sim$0.35 $M_{\odot}$, which is easy to miss when simply examining the standard sequence (e.g., top panel of Figure~\ref{fig:DerTM}). To define at which mass this local maximum takes place we examined the median GP derivative, identifying the peak to be at 0.361~$M_{\odot}$. Considering the individual GP curves within the range 0.25-0.42 $M_{\odot}$ (roughly between the local derivative minima, see Figure~\ref{fig:DerTM}), we recorded the location of the maximal value of the derivative curves, finding $\mathcal{M}_{b} = 0.361 \pm ^{0.036}_{0.024}$ $M_{\odot}$. This coincides with theoretical predictions for the solar metallicity mass boundary between partly and fully convective interiors in low-mass stars \citep[e.g.,][]{Chabrier2000}. The corresponding temperature along the median $T_{\mathrm{eff}}$-$\mathcal{M}$ sequence is 3414~K. These mass values, ${M}_{p}$ and ${M}_{b}$, correspond to the temperatures between M3 and M4 spectral types \citep[e.g.,][]{Boyajian2012,Pecaut2013}, where the convective transition has been thought to occur \citep{Reid2005nlds.book.....R}. We consider only the higher mass feature (${M}_{b}$) as indicative of the convective transition, as it is consistent with theoretical predictions.

\begin{figure}[tbp]
	\centering
	\includegraphics{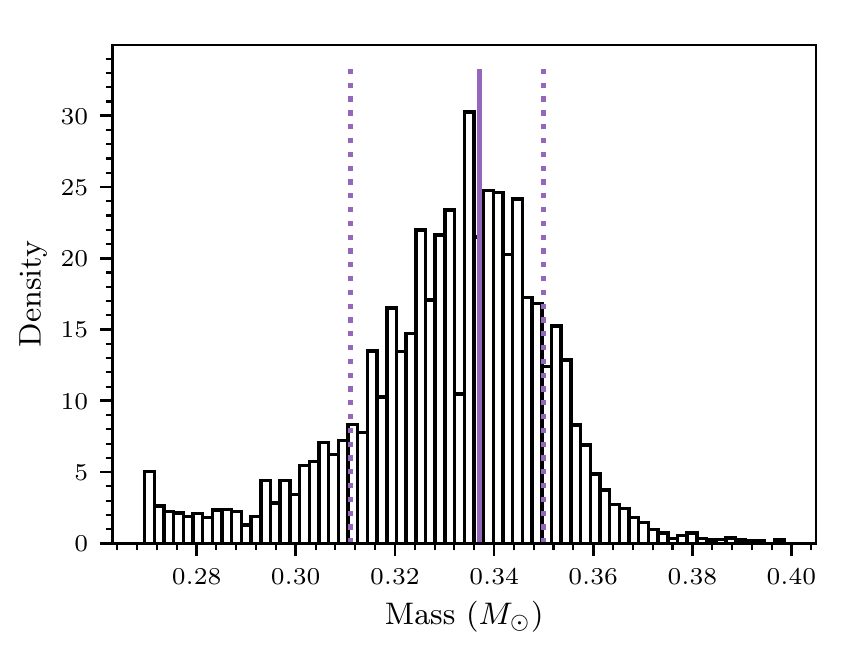} 
	\caption{This distribution for where the ensemble GP luminosity-mass derivatives peak within 0.27-0.40~$M_{\odot}$ defines the uncertainties for our location estimate of the transition mass between partially and fully convective dwarf star interiors: $\mathcal{M}_{b} = 0.337 \pm ^{0.013}_{0.026}$ $M_{\odot}$.}
	\label{fig:convdist}
\end{figure}

To corroborate this finding, we also examined the derivative of the GP $L_{\mathrm{bol}}$-$\mathcal{M}$ sequence. We show those results in Figure~\ref{fig:DerLM}. The median derivative sequence shows a well defined local maximum at 0.337~$M_{\odot}$, with the slope remaining positive throughout the masses corresponding to the temperature plateau. We defined the confidence interval for this feature from where the ensemble derivative curves were maximal on the interval 0.27-0.40~$M_{\odot}$, yielding a result of $\mathcal{M}_{b} = 0.337 \pm ^{0.013}_{0.026}$ $M_{\odot}$ (see Figure~\ref{fig:convdist}). Along the median $L_{\mathrm{bol}}$-$\mathcal{M}$ sequence this mass corresponds to a log luminosity of $-1.843$ relative to solar. This corresponds exactly to the stellar luminosity that near solar metallicity models predict exhibit the $^{3}$He instability \citep{Feiden2021ApJ...907...53F} --- the marker for the convective transition, and explanation for the Gaia M-dwarf gap.

Although within uncertainties, this mass value is slightly smaller than that found in the $T_{\mathrm{eff}}$-$\mathcal{M}$ sequence, possibly due to the effective temperature plane combining the effects on radius and luminosity across the convective transition \citep{Baraffe2018A&A...619A.177B}. In both luminosity and temperature planes we see derivative undulations with changing stellar mass indicating perturbations along the main stellar sequence. With these features occurring across the 0.3-0.4 $M_{\odot}$ range, we take the first feature ($\mathcal{M}_{b} = 0.337 \pm ^{0.013}_{0.026}$ $M_{\odot}$) in the derivative sequence with increasing mass as indicative of the transition onset.

We consider the feature in the luminosity-mass sequence to be the more accurate empirical measure of the transition from fully to partially convective interiors, as this stellar sequence shows less scatter than the effective temperature-mass sequence. This boundary measurement not only agrees with theoretical predictions, its determination from the derivative of the stellar sequence, agrees well with the physical expectations of the $^{3}$He instability. \cite{MacDonald2018MNRAS.480.1711M} note that the instability should coincide with local extremal changes in the luminosity, seen here as a local maximum in the derivative. \citet{Baraffe2018A&A...619A.177B} also showed model isochrones that illustrate this idea, with each isochrone curve exhibiting abrupt luminosity shifts at slightly different masses corresponding to the convective transition. These shifts can be difficult to discern and may not be totally evident in the models (see Figure~\ref{fig:MIST}). Our analysis illuminated these features as undulations in the sequence derivative. The exact location and width of the features we measured likely reflects the distribution of ages and metallicities of the MUSS field stars, and thus represents the solar neighborhood result for the empirical fully convective boundary mass of M-dwarf stars.

\section{Conclusions}\label{sec:conc}

Well defined empirical relationships for stellar properties are critical for many astrophysics investigations. The new mass-radius relationship presented in this article (Section~\ref{sec:relationMR}) is one such example, and can be readily used to empirically assess masses and/or radii in the low-mass star regime. This and the other empirical relationships used in characterizing the M-dwarf Ultraviolet Spectroscopic Sample will surely be supplanted as calibration data sets expand; however, the methods we have demonstrated (Section~\ref{sec:bayes}) will facilitate their future incorporation in Bayesian model fitting. 

To further illustrate how: imagine a hypothetical joint Bayesian regression analysis of stellar radial velocity data, and an exoplanet transit light curve for some system. The host star mass and radius are among the parameters of such a fit. Without any additional information, our formulation for an empirical mass-radius relationship (e.g., Equation~\ref{eq:fR}) can be multiplied to the model likelihood to give stellar mass and radius solutions (and thus the planet properties that depend on them) entirely consistent with the empirical relationship. This could properly and fully incorporate all the sources of uncertainty, and their correlations. Additional constraints (e.g., stellar angular diameters) could provide improved precision that would naturally propagate through to all of the parameter posteriors.

Often in the literature, the approach to this problem does not fully account for the sources of known uncertainty, or ignores their correlations, potentially biasing results or mischaracterizing the actual error distributions. In particular, ignoring the correlations can lead to overestimated errors. These effects could lead to potentially missing important astrophysical trends or overconfidence in possible results. More data is often the solution, but better statistical characterization can more accurately direct significant astrophysical findings.

Consistently applying our methods further enables a clear understanding of potential underlying systematics, and provides a platform for ensemble studies of stellar properties, as we have demonstrated with the MUSS stellar sequences (Section~\ref{sec:GP}). This Gaussian Process (GP) regression is flexible and provides a powerful tool for examining the stellar sequences. Features along these curves can be difficult to discern, and often elided through standard parametric regressions assuming a functional form, but by employing the GP derivative predictions we have revealed features that may otherwise have been missed. Using this analysis we were able to define the plateau in the effective temperature-mass relation to be at $\sim$0.26 $M_{\odot}$. Most importantly, the derivatives in $L_{\mathrm{bol}}$-$\mathcal{M}$ and $T_{\mathrm{eff}}$-$\mathcal{M}$ showed undulations that were predicted by models of stellar interiors as a consequence of the transition between partly and fully convective interiors. We thus empirically defined that boundary to be at $\sim$0.34 $M_{\odot}$ with the solar neighborhood MUSS stars.

The exact measures of these features reflect average sequence properties for solar neighborhood M-dwarfs, likely a consequence of age and metallicity distributions, but present a clear empirical result for the characteristics of nearby stars. These methods provide a powerful new way to examine empirical stellar sequences, for comparison to model predictions. Our results contribute to the empirical evidence of the convective transition in low-mass stellar interiors \citep[e.g.,][]{Jao2018ApJ...861L..11J}, and a first empirical measure of the corresponding mass. 

Larger stellar samples with narrower metallicity distributions, and accurately/consistently defined properties, as we have employed here with the MUSS stars, will enable more precise examinations of how ensemble stellar sequences can be used to probe the physics of stellar interiors.

\section{Summary}\label{sec:summary}

We summarize our key results as follows:

\begin{itemize}
	\item We have defined a new M-dwarf mass-radius relation that gives 3.1\% uncertainties at fixed mass across a range of 0.09-0.7 $M_{\odot}$, see Section~\ref{sec:relationMR}.
	\item We have developed a self-consistent Bayesian framework to simultaneously incorporate empirical relationships in Bayesian model fitting. Although general in nature, we illustrated its application in stellar parameter estimation, flexibly incorporating observational constraints, see Section~\ref{sec:bayes}. 
	\item Combining multiple photometric empirical relations and simultaneously accounting for the known uncertainty across relations, we applied these methods to the M-dwarf Ultraviolet Spectroscopic Sample (MUSS) field stars consistently defining their $\mathcal{M}$, $\mathcal{R}$, $T_{\mathrm{eff}}$, and $L_{\mathrm{bol}}$, see Section~\ref{sec:field}. These stars are of interest to the community and these properties have already been used by the analyses of \citet{Pineda2021arXiv210212485P}, and \citet{Youngblood2021arXiv210212504Y}.
	\item The MUSS results define stellar sequences, which we fit using Gaussian Process regression models, now publically available, see Section~\ref{sec:GP}.
	\item We empirically assessed the boundary mass between partly and fully convective M-dwarfs to be $\mathcal{M}_{b} = 0.337 \pm ^{0.013}_{0.026}$ $M_{\odot}$, see Section~\ref{sec:fcbound}.
	\item A skew normal distribution provides a useful way to approximate published results reporting asymmetric errorbars, that preserves the central confidence interval and the centroid, for use as prior information in additional studies, see Appendix~\ref{sec:ap_prior}. 
\end{itemize}

\section*{Acknowledgments}

The authors would like to thank Zachary Berta-Thompson, Elisabeth Newton, Girish Duvvuri, Aurora Kesseli, and Eileen Gonzales for useful discussions and commentary in the preparation of this work. The authors would also like to thanks the anonymous referee for their thoughtful comments and constructive feedback.

Support for Program numbers HST-GO 14640, 14633 and 15071 were provided by NASA
through grants from the Space Telescope Science Institute, which is
operated by the Association of Universities for Research in Astronomy,
Incorporated, under NASA contract NAS5-26555.

A.Y. acknowledges support by an appointment to the NASA Postdoctoral Program
at Goddard Space Flight Center, administered by USRA through a contract with NASA.

This publication makes use of data products from the Two Micron All Sky Survey, which is a joint project of the University of Massachusetts and the Infrared Processing and Analysis Center/California Institute of Technology, funded by the National Aeronautics and Space Administration and the National Science Foundation.

\section*{Appendix}
\renewcommand{\thesubsection}{\Alph{subsection}}

\subsection{Comments on Specific Targets}\label{sec:ap_starspec}

In Section~\ref{sec:empfield}, we discussed our methods for determining the stellar properties of the MUSS field stars. These methods relied on combining multiple empirical relations and available data for constraining the properties, usually photometry, and, whenever available, measured bolometric fluxes or interferometric angular diameters. We show the specific data used in the parameter estimates of all field stars in Table~\ref{tab:field_obsdata}. A few of these stars warranted additional commentary, which we provide below.

 \textit{Kepler-138}: For Kepler-138 there have been some inconsistent parameter estimations published in the literature. \cite{Almenara2018} discuss the history of these estimates and possible implications for the planetary system properties. Our new values are consistent with those used by \citet{Kipping2014} from \citet{Pineda2013a}, and show agreement with the analysis of \cite{Berger2020}. For comparison with transit data, our inferred stellar density for Kepler-138 is $4.95 \pm ^{0.61}_{0.54}$ g cm$^{-3}$, whereas \cite{Almenara2018} report a value of $3.92 \pm^{0.81}_{0.66}$ g cm$^{-3}$ from light-curve modeling, which is consistent within uncertainties. Their lower density value, however, led them to infer a larger stellar radius than we report, and in spectral energy distribution fitting thus infer a distance of $74.3 \pm 5.8$ pc, which is larger than the Gaia DR2 parallax inferred distance of $66.99 \pm 0.11$ pc. The smaller radius (higher density) value would move that estimate toward agreement. We additionally note that the spectrophotometric distance from \cite{Pineda2013a} of $66.5 \pm 7.3$ pc is surprisingly close to the Gaia value, lending additional validation to those methods, and studies of Kepler-138 that have since used their parameter estimates. 

 \textit{TOI-1235}: TOI-1235 (TYC 4384-1735-1) has recently been shown to host a transiting exoplanet near the gap in the exoplanet radius distribution \citep{Cloutier2020AJ....159..211C, Bluhm2020}. Our analysis of the stellar properties of this star ($M = 0.617 \pm^{0.016}_{0.017} M_{\odot}$, $R = 0.604 \pm 0.027$) give slightly smaller values in mass and radius than that used in those exoplanet papers, although generally within uncertainties. Comparing these values, the \cite{Cloutier2020AJ....159..211C} work relied on \cite{Benedict2016}, which gives systematically higher masses than our methods (see discussion within \cite{Mann2019}). Both \cite{Cloutier2020AJ....159..211C} and \cite{Bluhm2020} rely on spectroscopic effective temperatures tied to model atmospheres, and determine $T_{\mathrm{eff}}$s that are cooler than our more empirical approach. Given that our bolometric luminosity estimates are in strong agreement, this difference in assumed temperatures implies larger stellar radii. In general, our stellar parameters are in better agreement with those used by \cite{Bluhm2020} than \cite{Cloutier2020AJ....159..211C}, with our implied stellar density $3.95\pm^{0.51}_{0.44}$ g cm$^{-3}$ in good agreement with the density from transit light curve modeling \citep[$3.74 \pm^{0.30}_{0.31}$ g cm$^{-3}$;][]{Bluhm2020}. Our smaller stellar radius implies a smaller planet, and systematically shifts the planet away from the gap in the exoplanet radius distribution. 

 \textit{TRAPPIST-1}: Our properties for TRAPPIST-1 may be of particular interest. The bolometric flux is taken from \cite{Gonzales2019}. Although not listed in that work explicitly, the value and uncertainties quoted in Table~\ref{tab:field_obsdata} were provided by them (private communication, Gonzales, E.). Our stellar parameter values are empirical ($M =94.1 \pm 2.6$ $M_{\mathrm{Jup}}$, $R = 1.09 \pm 0.06$ $R_{\mathrm{Jup}}$), and agree remarkably well with their evolutionary model dependent approach ($M = 90 \pm 8$ $M_{\mathrm{Jup}}$, $R = 1.16 \pm 0.03$ $R_{\mathrm{Jup}}$ ).

\begin{figure*}[tbp]
	\centering
	\includegraphics{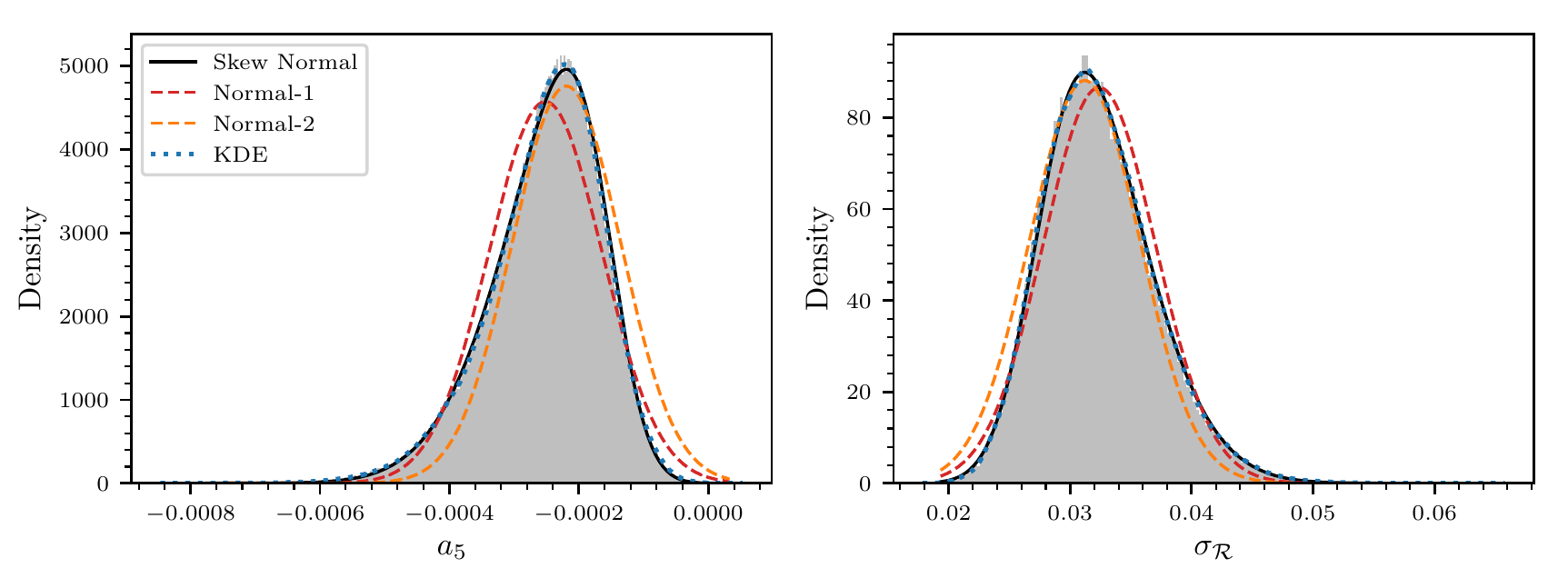} 
	\caption{A skew normal distribution can provide a good approximation to asymmetric posterior distributions from complex statistical models, for use in subsequent studies. The skew normal approximation (black) closely matches the best kernel density estimate (dotted line) for the histogram (gray), and provides a better match than typical normally distributed approximations (dashed lines), see Appendix~\ref{sec:ap_prior}. The two examples here are $a_{5}$ (\textit{left}) from the marginalized posterior distribution of \cite{Mann2019} for their $M_{K_{s}}$-$M$ polynomial fit, and $\sigma_{\mathcal{R}}$ in our fractional scatter model for $\mathcal{M}$-$\mathcal{R}$ relation in Section~\ref{sec:relationMR}. The displayed skew normal distributions have location, scale and shape parameters, respectively, of $\{-1.556 \times 10^{-4},  1.285\times 10^{-4}, -2.62\}$ for $a_{5}$ and $\{  0.0277, 0.00656, 1.96 \}$ for $\sigma_{\mathcal{R}}$. }
	\label{fig:ex_asym}
\end{figure*}

\subsection{On Using Asymmetric Priors}\label{sec:ap_prior}

Probabilistic modeling of astronomical phenomena has become a widespread practice across a variety of subfields, enabled by increased computational ability and efficient Monte Carlo methods for sampling likelihood functions. An outcome of this work is often then probability distributions describing parameters of interest, usually quoted as a `best' value and 1$\sigma$ errorbars corresponding to the middle $\sim$68\% confidence interval. These properties are then typically assumed to describe a normal distribution with mean at the `best' value and variance $\sigma^2$. In most instances, the probability distribution for the parameter of interest can indeed be approximated by a normal distribution and these results can be readily incorporated into subsequent studies as prior information. However, with typically non-linear models and transformations between random variables of interest, results often do not yield symmetric, normally distributed posterior distributions. The 68\% confidence interval is then denoted with unequal values, $\sigma_{-}$ and $\sigma_{+}$, below and above the `best' estimate, respectively. These results can then become problematic to use in new analyses. A common and unjustified practice, is to average the uncertainties above and below separately when combining these data with other estimates \citep{Barlow2003}. Another typical approach is to approximate the probability distribution as Gaussian using an estimate for the mean and averaging $\sigma_{-}$ and $\sigma_{+}$ to approximate the distribution standard deviation, giving an easy to use informative prior on the physical quantity of interest \citep{DAgostini2004}. This is reasonable when the posterior distribution is well constrained, such that the log of the distribution around its peak can be approximated by a parabola. Furthermore, central limit theorem arguments make the normal distribution a good limiting function for use in many cases. As we show below, this practice, while giving reasonable results can shift the central 68\% confidence interval, as well as the mode of the distribution, potentially biasing the analysis, when an informative prior is incorporated within Bayesian inference modeling.

As an illustration of this effect, we consider two clearly asymmetric posterior distributions, one from the literature, the parameter $a_{5}$ corresponding to the coefficient of the highest order term in the fifth order polynomial fit for the mass-luminosity relationship of \cite{Mann2019}, and the second as the marginalized posterior for the fractional scatter, $\sigma_{\mathcal{R}}$, for the mass-radius relation defined in this work (Section~\ref{sec:relationMR}). In Figure~\ref{fig:ex_asym}, we plot the histograms of these distributions for $a_{5}$ (\textit{left}) and $\sigma_{\mathcal{R}}$ (\textit{right}) in gray, with two normal approximations as dashed lines. The line `Normal-1' plots a Gaussian distribution centered at the mean and standard deviation matching that of the posterior samples. The line `Normal-2' indicates a similar Gaussian distribution, but now centered at the peak of the posterior with standard deviation given by the average of $\sigma_{+}$ and $\sigma_{-}$ defined by the central 68\% confidence interval of the posterior samples relative to the peak.

Because the distributions are asymmetric, neither `Normal' approximation is ideal for conveying the probability density function of these variables. `Normal-1' shifts the peak of the distributions away from the most-probable value, whereas `Normal-2' retains that peak but places much greater probability mass away from the actual skew in the posterior distribution. Both approximations bias subsequent results away from the most-likely confidence interval defined by the actual posterior samples. If the posterior distribution for a given quantity of interest is available, it can be sampled directly, avoiding such issues. However, in practice, this is rarely the case, and published values generally give only a `best' value with a 1$\sigma$ confidence interval. 

We consider here a method to estimate a useful probability density function from these `best' values that is Gaussian-like in shape but preserves the asymmetry of the quoted result. In effect, we draw upon the class of three parameter probability density functions that are perturbations to a standard normal distribution. In particular, we employ the skew normal distribution,

\begin{equation}
\mathcal{S}(x) = \frac{2}{\sigma} \phi \left( \frac{x - \mu}{\sigma} \right) \Phi \left(\alpha \frac{x - \mu}{\sigma} \right) \; ,
\end{equation}

\noindent where $\phi$ is the standard normal distribution, $\Phi$ is the cumulative distribution function for $\phi$, $\mu$ is a location parameter, $\sigma$ sets the scale, and $\alpha$ is the shape parameter. In the limit of $\alpha \rightarrow 0$, the distribution matches that of a Gaussian and becomes increasingly asymmetric for increasing $|\alpha|$, with the limit of $|\alpha| \rightarrow \infty$ yielding a Half-Normal distribution. The sign of $\alpha$ is also indicative of the direction of skew, e.g., for $\alpha >0$ the skew normal distribution exhibits more probability mass at values greater than the mean, i.e.\ to the right, and opposite for $\alpha < 0$.

In Figure~\ref{fig:ex_asym}, we show that a skew normal approximation (solid line) does a remarkably good job of matching the shape of the posterior distributions for $a_{5}$ and $\sigma_{\mathcal{R}}$: the kernel density estimate (KDE, dotted) and the skew normal curve for both parameters largely overlap, matching the envelope of the histogram. 

The skew normal approximation is determined by numerically solving the system of equations for the 3 variables $\mu$, $\sigma$, and $\alpha$, such that the resulting skew normal distribution yields a central 68\% confidence interval, and median (or mode) that match the desired input `best' value with its asymmetric error bars. Using the cumulative function for the skew normal distribution the system can be expresses as 

\begin{eqnarray}
	\int_{-\infty}^{x_{u}} \mathcal{S}(x; \mu, \sigma, \alpha) \, dx  - 0.84 &=& 0 \; , \\
	\int_{-\infty}^{x_{l}} \mathcal{S}(x; \mu, \sigma, \alpha) \, dx - 0.16 &=& 0 \; , \\ \label{eq:sys_med}
	\int_{-\infty}^{x_{m}} \mathcal{S}(x; \mu, \sigma, \alpha) \, dx - 0.50 &=& 0 \;  , \\ 
	\alpha \phi\left(\alpha \, \frac{x_{p} - \mu}{\sigma}\right) - \Phi \left(\alpha \frac{x_{p} - \mu}{\sigma}\right)\frac{x_{p} - \mu}{\sigma} &=& 0 \; , \label{eq:sys_mode}
\end{eqnarray}

\noindent where we use only one of either Equation~\ref{eq:sys_med} or Equation~\ref{eq:sys_mode}, depending on whether the central value in question is the median or mode respectively, $x_{m}$ is that input median value, $x_{p}$ is the desired peak of the output skew normal distribution, $x_{u}$ is the upper value on the interval bound (e.g., $x_{m} + \sigma_{+}$), and $x_{l}$ is the lower value on the interval bound (e.g., $x_{m} - \sigma_{-}$).

We provide code for determining this numerical approximation in an online repository.\footnote{ \url{https://github.com/jspineda/stellarprop}} This approach has the advantages of formally matching the central confidence interval for the quantity of interest, gives a probability density function skewed in the appropriate direction, and can be applied to any result publishing asymmetric errorbars as long as it is clear what defines the central quantity (either a median or mode). The skew normal distribution is already implemented in commonly used software, including Python's \texttt{scipy}.

This approach remains an approximation and should not replace directly sampling the posterior when available, although there may be use cases for which having a well defined functional form may be advantageous. In practice, while we only demonstrate its applicability for two example posteriors, this method should work well for distributions with Gaussian tails and only slight deviations from symmetric. A skew normal distribution constructed in this way will preserve the range of the central confidence interval and likely match the proper median (or mode) of the parameter in question. By definition this will provide better approximations to asymmetric distributions that are assumed to be Gaussian-like rather than assuming a symmetric Gaussian distribution. Sample distributions with long tails, that depart more significantly from a normal distribution will consequently not be well approximated with a skew normal distribution, although a similar approach could be taken with an exponential Gaussian distribution or similar 3 parameter probability density functions.

\subsection{On Correlated Uncertainties}\label{sec:ap_corr }

We note throughout this article that ignoring the correlations in the uncertainties enlarges the resulting error estimates when using empirical fit relationships, for example, from published mass-radius relations for M-dwarfs. Although the correlations are not often published, typical polynomial fits of any kind when applied to data will yield parameters that have correlated uncertainties. In general, the severity of ignoring the correlations when using such relationships will depend on the details of the polynomial relation and its parameter uncertainties. Nevertheless, that process will always yield errors larger than implied by the underlying data. This is because, by ignoring the correlations, one allows the fit parameter uncertainties to include parameter space that is actually strongly disfavored by the regression analysis. Here we provide a specific example using the results from Section~\ref{sec:relationMR}, for the magnitude of this effect.

If we consider a 0.6~$M_{\odot}$ star, its implied radius using our fractional scatter $\mathcal{M}$-$\mathcal{R}$ relation (Section~\ref{sec:relationMR}) is $0.588\pm^{0.018}_{0.019}$~$R_{\odot}$. If we ignore the correlations of the fit parameter uncertainties, the implied result becomes $0.588\pm^{0.021}_{0.020}$~$R_{\odot}$. That corresponds to $\sim$$10$\% increase in the standard deviation. Similarly for a 0.1~$M_{\odot}$ star, the corresponding results would be $0.129\pm0.005$~$R_{\odot}$, and $0.129\pm0.006$~$R_{\odot}$, for including and excluding the uncertainty correlations. That corresponds to $\sim$$20$\% increase in the standard deviation. In these cases the difference is not especially dramatic because the uncertainties are dominated by the intrinsic scatter term. However, when just considering the contribution from the polynomial parameter uncertainties, the implied errors are at least twice as large if the correlations are ignored.

Whether these effects are very significant depends on the science question. Nevertheless, it is straightforward to account for uncertainty correlations, and doing so enables more accurate results and reliable probabilistic calculations.

\subsection{Literature Stellar Parameters}\label{sec:ap_litcite}

In Section~\ref{sec:litcomp}, we compared our new stellar parameter determinations with typical values found in the literature. We collected the masses, radii and effective temperatures for that comparison from the following list of references: \citet{Bonfils2005A&A...443L..15B,Demory2009,Houdebine2010MNRAS.407.1657H,Berta2011ApJ...736...12B,Kraus2011ApJ...728...48K,vonBraun2011,Boyajian2012,RojasAyala2012,vonBraun2012,AngladaEscude2013A&A...556A.126A,Lepine2013AJ....145..102L,Kordopatis2013AJ....146..134K,Pecaut2013,Pineda2013a,Rajpurohit2013A&A...556A..15R,Santos2013A&A...556A.150S,Tuomi2013A&A...556A.111T,Gaidos2014,Neves2014,vonBraun2014,BertaThompson2015Natur.527..204B,Mann2015,Terrien2015,Awiphan2016MNRAS.463.2574A,Gillon2016,Houdebine2016ApJ...822...97H,Kopytova2016A&A...585A...7K,AstudilloDefru2017,Dressing2017ApJ...836..167D, Newton2017,Houdebine2017,Berger2018ApJ...866...99B,Damasso2018A&A...615A..69D,vanGrootel2018ApJ...853...30V,Houdebine2019AJ....158...56H,Rabus2019,Jeffers2020Sci...368.1477J} .


\startlongtable
\centerwidetable
\begin{deluxetable*}{l c c c c c c c c c}
	\tablecaption{ Data Constraints for Field Stars \label{tab:field_obsdata} }
	\tablehead{
		\colhead{Name} &\colhead{$\pi$ \tablenotemark{a}} & \colhead{$J$\tablenotemark{b}} & \colhead{$K_{S}$\tablenotemark{b}} & \colhead{$V-J$\tablenotemark{c}}& \colhead{$r-J$\tablenotemark{c}}  & \colhead{[Fe/H]} & \colhead{$F_{\mathrm{bol}}$} & \colhead{$\theta$} \\
		& \colhead{(mas)} &  & & &   & & \colhead{($10^{-8}$ erg s$^{-1}$ cm$^{-2}$)} & ($\mu$as)    
	}
	\startdata
	\multicolumn{9}{c}{MUSCLES} \\
	\hline
	GJ 436  & 102.5015 $\pm$ 0.0936 & --- &  6.073$\pm$0.016 &---& --- &---& 0.8277$\pm$0.0093 & 417$\pm$13  \\
	GJ 832  & 201.4073 $\pm$ 0.0429 & 5.349$\pm$0.032 &  4.501$\pm$0.018 & 3.323$\pm$0.033& --- & $-0.11 \pm$0.09 & ---  &--- \\	
	GJ 876 & 213.8669 $\pm$ 0.0758 & ---  & 5.01$\pm$0.021& --- & --- & --- & 1.9160$\pm$0.015 &  746$\pm$9 \\
	GJ 581  & 158.7492 $\pm$ 0.0523  &--- & 5.837$\pm$0.023  & ---  & --- & --- & 0.9609$\pm$0.0077 & 446$\pm$14\\
	GJ 176 & 105.5627 $\pm$ 0.07 & --- & 5.607$\pm$0.034  & --- & --- & --- & 1.254$\pm$0.011 & 448$\pm$21\\
	GJ 1214  & 68.2653 $\pm$ 0.1723  &9.75$\pm$0.024 &  8.782 $\pm$0.02 & --- & 4.2714$\pm$0.024 & $+0.11\pm0.09$ & --- & ---\\
	GJ 667C & 138.0171 $\pm$ 0.0918  &6.848 $\pm$0.021 & 6.036 $\pm$0.02 & 3.438$\pm$0.045& --- & $-0.30\pm0.08$ & --- & --- \\
	Prox. Cen & 768.5004 $\pm$ 0.2030  & --- & 4.384 $\pm$0.033 & ---& --- & --- & 2.961$\pm$0.037 & 1020.8$\pm$43.6\\
	Gl 887 & 304.219 $\pm$ 0.0451  & --- & 3.36$\pm$0.02 & --- & --- & --- &10.89$\pm$0.17  & 1337$\pm$25\\
	GJ 628  & 232.2095 $\pm$ 0.063 &  --- & 5.075$\pm$0.024 & --- &--- & --- & 1.897$\pm$0.017&695$\pm$18  \\
	GJ 1061 & 272.2446 $\pm$ 0.0661 & 7.523$\pm$0.02& 6.61$\pm$0.021 & 5.514 $\pm$0.077 & --- & $-0.03\pm0.09$ & --- & ---\\   
	GJ 725A & $283.9489\pm0.0624$ & --- & $4.432\pm0.020$& --- & --- & --- & $3.898 \pm 0.045$ & $937 \pm 8$ \\ 
	\hline
	\multicolumn{9}{c}{Living with Red Dwarf} \\
	\hline
	GJ 213  & 172.7068 $\pm$ 0.0788& --- & 6.389$\pm$0.016 & --- & --- & --- & 0.600$\pm$0.009 & --- \\
	GJ 821  & 84.7044 $\pm$ 0.0685  & --- & 6.909$\pm$0.029 & ---& --- &--- & 0.4185$\pm$0.0051 &---  \\
	\hline
	\multicolumn{9}{c}{FUMES} \\
	\hline
	GJ 4334  & 39.4732 $\pm$ 0.0766 & 9.878$\pm$0.022 & 8.980$\pm$0.020 &  ---  & 4.151$\pm$0.022& $+0.52\pm$0.11 & --- & ---  \\
	GJ 49 & 101.465 $\pm$ 0.0335 & 6.230$\pm$0.021 & 5.371$\pm$0.020& --- & 3.026$\pm$0.064  &$+0.15\pm$0.08& --- &--- \\	
	G 249-11 & 34.3159  $\pm$ 0.0451 & 10.552 $\pm$0.024 & 9.729 $\pm$0.023 & 4.578$\pm$0.055 & --- & --- & --- & ---  \\	
	LP 55-41 & 27.0006 $\pm$ 0.035 & 9.907 $\pm$0.023 & 9.031 $\pm$0.023 & 3.951$\pm$0.064 &--- &  --- & --- & --- \\
	LP 247-13  & 28.5412 $\pm$ 0.1102 & 9.317$\pm$0.023 & $8.410\pm0.018$ & --- & 3.204$\pm$0.025 &  ---&--- &---\\
	GJ 410 & 83.7765 $\pm$ 0.0573 & 6.522$\pm$0.020& 5.688$\pm$0.021 & ---& 2.552$\pm$0.028  & --- & --- & --- \\
	\hline
	\multicolumn{9}{c}{Mega-MUSCLES} \\
	\hline
	GJ 676A &  	$62.3650 \pm0.0391$   & $6.711\pm	0.02$ & $5.825	\pm 0.029$ & $2.874 \pm 0.021 $ & --- & --- &--- & ---  \\
	GJ 15A &  $	280.6902 \pm 0.0429 $& ---  &  $4.039 \pm 0.020$& --- & --- & --- & $5.669 \pm 0.045$ & $1005 \pm 5$ \\
	GJ 649 & $96.3141 \pm 0.0311 $ & --- & $5.624	\pm 0.016$& ---& ---& ---&$1.298\pm0.013$& $484 \pm 12 $ \\
	GJ 163 & $ 	66.0714 \pm 0.0314 $ & $7.948	\pm 0.026$&  $7.135	\pm 0.021$ & $3.886\pm0.037$& --- & --- & --- & ---  \\
	GJ 849 &  $	113.6000 \pm 0.0463$ & --- & $ 5.594	\pm 0.017$& ---&---&---&$1.192 \pm0.010$ & --- \\
	GJ 674 &$ 	219.8012 \pm 0.0487 $ & $5.711 \pm 0.019$ & $ 4.855	\pm 0.018$& $3.696 \pm  0.022$ & --- & --- & --- &  $ 737 \pm 37 $\\ 
	GJ 699 &  	$547.4506 \pm0.2899 $& --- & $ 4.524	\pm 0.02$& --- &---&---& $3.263 \pm 0.057$& $952.8 \pm 5.0 $ \\
	L 980-5 &$74.5644 \pm0.0808$&  $ 8.948	\pm 0.03$& $8.093\pm0.026$& $4.223 \pm 0.036$&---&---&---&--- \\ 
	LP 756-18 & $ 	80.2319 \pm0.0881$ & $9.717 \pm 0.024$& $ 8.915 \pm 0.021$& $4.923 \pm 0.028$& --- &---&--- &---\\
	LHS 2686 &$  	82.0336 \pm 0.0641$ & $9.584	\pm0.02$& $8.687 \pm 0.016$& $5.006 \pm 0.034$&---& $+0.2\pm0.11$&---& --- \\ 
	GJ 729 &$ 	336.1228 \pm 0.0641$ &  --- & $5.37	\pm 0.016$&---&---&---&$1.451 \pm 0.017$&--- \\
	GJ 1132 & 79.2543 $\pm$ 0.0438  &9.245$\pm$0.026 & 8.322$\pm$0.027  & 4.259$\pm$0.034 &--- & --- &--- &--- \\  
	TRAPPIST-1 &$ 	80.4512 \pm 0.1211$&  --- & $10.296\pm0.023$ &---&---&---& $ (1.266 \pm 0.044) \times 10^{-2} $& --- \\
	\hline
	\multicolumn{9}{c}{Miscellaneous} \\
	\hline
	AD Leo & 201.3683 $\pm$ 0.0679  &--- & 4.593$\pm$0.017 & --- & ---& --- & 2.940$\pm$0.035 & --- \\ 
	EV Lac & 198.0112 $\pm$ 0.038 & 6.106$\pm$0.030 & 5.299$\pm$0.024 & 4.154$\pm$0.067 & --- & $-0.03\pm0.08$ & --- &--- \\
	YZ CMi  & 167.0186 $\pm$ 0.0592  &$6.581\pm0.024$ & $5.698\pm0.017$ & $4.438\pm0.22$ & --- & $0.23\pm0.08$ & ---& ---\\
	Kepler-138 &  	$14.9272 \pm0.0238$&  $10.293 \pm 0.022$ & $ 9.506  \pm 0.011$ & $2.875 \pm 0.040$ &--- & $-0.16 \pm 0.11 $ & --- & --- \\	
	TOI-1235 &  	$25.2016 \pm 0.0302$ &  $8.711 \pm 0.020$ & $ 7.893 \pm 0.023$ & $2.784 \pm 0.059$ &--- & --- & $0.1780\pm0.0041$ & --- \\	
	GJ~3470 &  	$33.9601 \pm0.0582$ & $8.794\pm0.026$ & $7.989 \pm0.023$ & --- & $2.936 \pm 0.028$ & $0.14 \pm 0.10$ & --- & --- \\
	GJ~1243 & $83.4814\pm0.0366$ & $8.586\pm0.023$ & $7.773\pm0.016 $ & $4.375 \pm 0.189$ & --- & --- & --- & --- \\
	LHS~2065 & $115.3036\pm0.1132$ & --- & $9.942\pm0.024$ & --- & --- & --- &  $ (1.412 \pm 0.007) \times 10^{-2} $ & --- \\
	LHS~3003 & $141.6865\pm0.1063$ & --- & $8.928 \pm 0.027$ & --- & --- & --- & $(3.86 \pm 0.03) \times 10^{-2} $ & --- \\
	VB~10  & $168.9620\pm0.1299$  & --- & $ 8.765 \pm 0.022$  & --- & --- & --- &  $(4.56 \pm 0.03) \times 10^{-2} $ & --- \\
	VB~8  & $153.8139 \pm 0.1148$ & --- & $8.816\pm 0.023$ & --- & --- & --- &  $(4.88 \pm 0.03) \times 10^{-2} $ & --- \\
	LHS~1140 & $66.6996 \pm 0.0674$ & $9.612 \pm 0.023$ & $8.821\pm 0.024$ & $4.538 \pm 0.066$ & --- & $-0.30 \pm 0.08$ & --- & --- \\
	K2-3 & $22.6610 \pm 0.0553$ & $9.421 \pm 0.027$ & $8.561 \pm 0.023$ & $2.747 \pm 0.028$ & --- & $-0.32 \pm 0.13$ & --- & --- \\
	GJ~273 & $262.961\pm1.387$ & --- & $4.857\pm0.023$ & --- & --- & --- & $2.395 \pm 0.021 $ & $783 \pm 10$ \\ 
	GJ~205 & $175.4287 \pm 0.0672$ & --- & $3.856 \pm 0.019$ &  --- & --- & --- & $6.340 \pm 0.054$ & $943 \pm 4$ \\ 
	GJ~588 & $169.0074\pm0.0580$ & $5.647\pm0.021$ & $4.759\pm0.024$ &$3.664 \pm 0.024$ & --- &$ 0.12 \pm 0.09$ & --- &--- \\
	GJ~338~A & $157.8796\pm0.0366$ & --- & $4.032\pm0.020$ & --- & --- & --- & $6.087\pm0.057$ & --- \\
	GJ~411 & $392.64\pm0.67$hipp & --- & $3.361\pm0.020$ & --- & --- & --- & $10.821 \pm0.096$ &  $1432 \pm 13$\\ 
	GJ~526 &  $183.9836 \pm 0.0509$ & --- &  $4.415 \pm 0.017$ &  --- & --- & --- & $4.013\pm 0.053 $ &  $835\pm 14$ \\
	GJ~687 & $219.7807 \pm 0.0324$ & --- &  $4.504,\pm0.02$ & --- & --- & ---& $3.363 \pm0.028$  & $859 \pm 14$ \\
	GJ~809 & $142.0341 \pm 0.0305$ & --- & $4.618 \pm 0.024$ & --- & --- & --- & $3.348\pm 0.036$ & $722 \pm 8 $ \\
	GJ~880 & $145.6107 \pm 0.0388$ & --- & $4.523 \pm 0.016$ &  --- & --- & --- & $3.545\pm 0.027 $ & $744 \pm 4$ \\
	GJ~406 & $ 413.13 \pm 1.27$ & --- & $6.084 \pm 0.017 $ & --- & --- & --- & $ 0.5798 \pm 0.0069 $ & $582 \pm 20 $  \\
	GJ~447 & $ 296.3073 \pm 0.0699$ & --- & $ 5.654 \pm0.024 $ & --- & --- & --- & $ 1.028 \pm 0.014$ & $ 540 \pm 29 $ \\
	GJ~4367 & $ 58.9687 \pm 0.1169$ & --- & $ 8.043 \pm 0.024$ & --- & --- & --- & $0.1196 \pm 0.0025$ & ---  \\
	GJ~3991 & $134.5971 \pm 0.4894 $ & --- & $6.485 \pm 0.016$ & --- & --- & --- & $0.4951\pm0.0053$ & ---  \\
	GJ~403 & $75.4576 \pm 0.0579 $ & --- & $7.795  \pm 0.026$ & --- & --- & --- & $0.1567 \pm 0.0024 $ & ---  \\
	GJ~3325 & $108.3263  \pm 0.0500$ & --- & $ 6.936 \pm 0.021 $ & --- & --- & --- & $0.3526 \pm 0.0041$ & ---  \\
	GJ~545 & $72.4350 \pm 0.0724$ & --- & $ 7.976 \pm 0.031  $ & --- & --- & --- & $ 0.1403 \pm  0.0024$ & ---  \\
	GJ~3378 & $129.4589 \pm 0.0799 $ & --- & $ 6.639 \pm 0.018$ & --- & --- & --- & $0.4613 \pm 0.0085 $ & ---  \\
	GJ~402 & $ 143.4971 \pm 0.0626 $ & --- &  $ 6.371 \pm 0.016 $ & --- & --- & --- & $ 0.5261 \pm 0.0055$ & ---  \\
	GJ~1207 & $ 	115.0339 \pm 0.0823	  $ & --- &  $ 7.120 \pm 0.021  $ & --- & --- & --- & $ 0.2861  \pm 0.0046 $ & ---  \\
	\hline
	\multicolumn{9}{c}{HAZMAT} \\ 
	\hline
	G~75-55 & $ 41.8484 \pm 0.0884$ & $8.021 \pm 0.021 $ & $7.171\pm0.027 $ & $2.937 \pm 0.056$ & --- & $-0.10 \pm 0.12$ & --- & --- \\
	GJ~173 & $ 89.1708 \pm 0.0313	$ & --- &  $ 6.091 \pm 0.021$ & --- & --- & --- & $0.8354 \pm 0.008$ & ---  \\	
	\enddata
	\tablenotetext{a}{Parallaxes are from Gaia DR2 \citep{Gaia2018}, except for GJ~273 \citep{Rabus2019}, GJ~406 \citep{Weinberger2016}, and GJ~411 \citep[Hipparcos;][]{vanLeeuwen2007}.}
	\tablenotetext{b}{$J$ and $K_{S}$ are 2MASS photometry \citep{Cutri2003}, except for Gl 205, GJ 338 A, GJ 411, and GJ 687, for which we use the synthetic $K_{S}$ values of \citet{Mann2015}.}
	\tablenotetext{c}{When bolometric luminosity is unavailable the optical/IR color (either $V-J$ or $r-J$, with possibly [Fe/H], see Section~\ref{sec:bayes}) is used to set the bolometric correction when estimating the stellar luminosity. Photometry taken from \cite{Koen2010MNRAS.403.1949K}, \cite{Zacharias2012yCat.1322....0Z}, \cite{Hosey2015AJ....150....6H}, \cite{Henden2016}, and \cite{Chambers2016arXiv161205560C}. }
	\tablenotetext{}{A\textsc{dditional} R\textsc{eferences.} -- \emph{Metallicity}: \cite{Demory2013ApJ...768..154D}, \cite{Gaidos2014}, \cite{Neves2014}, \cite{Terrien2015}, \cite{Crossfield2015ApJ...804...10C}, and \cite{Dittmann2017Natur.544..333D}  ; \emph{Bolometric Luminosity}: \cite{Boyajian2012}, \cite{Mann2015}, \cite{Filippazzo2015},\cite{Terrien2015}, and \cite{Gonzales2019} ; \emph{Angular Diameter}: \cite{Segransan2003}, \cite{Demory2009},\cite{vonBraun2011}, \cite{Boyajian2012}, \cite{vonBraun2014}, \cite{Kane2017}, and \cite{Rabus2019}.}
	
\end{deluxetable*}

\bibliographystyle{aasjournal}
\bibliography{muss1_arxiv}

\begin{thebibliography}{}
\expandafter\ifx\csname natexlab\endcsname\relax\def\natexlab#1{#1}\fi
\providecommand{\url}[1]{\href{#1}{#1}}
\providecommand{\dodoi}[1]{doi:~\href{http://doi.org/#1}{\nolinkurl{#1}}}
\providecommand{\doeprint}[1]{\href{http://ascl.net/#1}{\nolinkurl{http://ascl.net/#1}}}
\providecommand{\doarXiv}[1]{\href{https://arxiv.org/abs/#1}{\nolinkurl{https://arxiv.org/abs/#1}}}

\bibitem[{{Aigrain} {et~al.}(2016){Aigrain}, {Parviainen}, \&
  {Pope}}]{Aigrain2016MNRAS.459.2408A}
{Aigrain}, S., {Parviainen}, H., \& {Pope}, B.~J.~S. 2016, \mnras, 459, 2408,
  \dodoi{10.1093/mnras/stw706}

\bibitem[{{Allard} {et~al.}(2000){Allard}, {Hauschildt}, \&
  {Schwenke}}]{Allard2000ApJ...540.1005A}
{Allard}, F., {Hauschildt}, P.~H., \& {Schwenke}, D. 2000, \apj, 540, 1005,
  \dodoi{10.1086/309366}

\bibitem[{{Allard} {et~al.}(2012){Allard}, {Homeier}, \&
  {Freytag}}]{Allard2012}
{Allard}, F., {Homeier}, D., \& {Freytag}, B. 2012, Philosophical Transactions
  of the Royal Society of London Series A, 370, 2765,
  \dodoi{10.1098/rsta.2011.0269}

\bibitem[{{Almenara} {et~al.}(2018){Almenara}, {D{\'\i}az}, {Dorn}, {Bonfils},
  \& {Udry}}]{Almenara2018}
{Almenara}, J.~M., {D{\'\i}az}, R.~F., {Dorn}, C., {Bonfils}, X., \& {Udry}, S.
  2018, \mnras, 478, 460, \dodoi{10.1093/mnras/sty1050}

\bibitem[{{Anglada-Escud{\'e}} {et~al.}(2013){Anglada-Escud{\'e}}, {Tuomi},
  {Gerlach}, {Barnes}, {Heller}, {Jenkins}, {Wende}, {Vogt}, {Butler},
  {Reiners}, \& {Jones}}]{AngladaEscude2013A&A...556A.126A}
{Anglada-Escud{\'e}}, G., {Tuomi}, M., {Gerlach}, E., {et~al.} 2013, \aap, 556,
  A126, \dodoi{10.1051/0004-6361/201321331}

\bibitem[{{Astudillo-Defru} {et~al.}(2017){Astudillo-Defru}, {Delfosse},
  {Bonfils}, {Forveille}, {Lovis}, \& {Rameau}}]{AstudilloDefru2017}
{Astudillo-Defru}, N., {Delfosse}, X., {Bonfils}, X., {et~al.} 2017, \aap, 600,
  A13, \dodoi{10.1051/0004-6361/201527078}

\bibitem[{{Awiphan} {et~al.}(2016){Awiphan}, {Kerins}, {Pichadee},
  {Komonjinda}, {Dhillon}, {Rujopakarn}, {Poshyachinda}, {Marsh}, {Reichart},
  {Ivarsen}, \& {Haislip}}]{Awiphan2016MNRAS.463.2574A}
{Awiphan}, S., {Kerins}, E., {Pichadee}, S., {et~al.} 2016, \mnras, 463, 2574,
  \dodoi{10.1093/mnras/stw2148}

\bibitem[{{Baraffe} \& {Chabrier}(2018)}]{Baraffe2018A&A...619A.177B}
{Baraffe}, I., \& {Chabrier}, G. 2018, \aap, 619, A177,
  \dodoi{10.1051/0004-6361/201834062}

\bibitem[{{Baraffe} {et~al.}(2015){Baraffe}, {Homeier}, {Allard}, \&
  {Chabrier}}]{Baraffe2015}
{Baraffe}, I., {Homeier}, D., {Allard}, F., \& {Chabrier}, G. 2015, \aap, 577,
  A42, \dodoi{10.1051/0004-6361/201425481}

\bibitem[{{Barlow}(2003)}]{Barlow2003}
{Barlow}, R. 2003, arXiv e-prints, physics/0306138.
\newblock \doarXiv{physics/0306138}

\bibitem[{{Benedict} {et~al.}(2016){Benedict}, {Henry}, {Franz}, {McArthur},
  {Wasserman}, {Jao}, {Cargile}, {Dieterich}, {Bradley}, \&
  {Nelan}}]{Benedict2016}
{Benedict}, G.~F., {Henry}, T.~J., {Franz}, O.~G., {et~al.} 2016, \aj, 152,
  141, \dodoi{10.3847/0004-6256/152/5/141}

\bibitem[{{Berger} {et~al.}(2018){Berger}, {Huber}, {Gaidos}, \& {van
  Saders}}]{Berger2018ApJ...866...99B}
{Berger}, T.~A., {Huber}, D., {Gaidos}, E., \& {van Saders}, J.~L. 2018, \apj,
  866, 99, \dodoi{10.3847/1538-4357/aada83}

\bibitem[{{Berger} {et~al.}(2020){Berger}, {Huber}, {van Saders}, {Gaidos},
  {Tayar}, \& {Kraus}}]{Berger2020}
{Berger}, T.~A., {Huber}, D., {van Saders}, J.~L., {et~al.} 2020, arXiv
  e-prints, arXiv:2001.07737.
\newblock \doarXiv{2001.07737}

\bibitem[{{Berta} {et~al.}(2011){Berta}, {Charbonneau}, {Bean}, {Irwin},
  {Burke}, {D{\'e}sert}, {Nutzman}, \& {Falco}}]{Berta2011ApJ...736...12B}
{Berta}, Z.~K., {Charbonneau}, D., {Bean}, J., {et~al.} 2011, \apj, 736, 12,
  \dodoi{10.1088/0004-637X/736/1/12}

\bibitem[{{Berta-Thompson} {et~al.}(2015){Berta-Thompson}, {Irwin},
  {Charbonneau}, {Newton}, {Dittmann}, {Astudillo-Defru}, {Bonfils}, {Gillon},
  {Jehin}, {Stark}, {Stalder}, {Bouchy}, {Delfosse}, {Forveille}, {Lovis},
  {Mayor}, {Neves}, {Pepe}, {Santos}, {Udry}, \&
  {W{\"u}nsche}}]{BertaThompson2015Natur.527..204B}
{Berta-Thompson}, Z.~K., {Irwin}, J., {Charbonneau}, D., {et~al.} 2015, \nat,
  527, 204, \dodoi{10.1038/nature15762}

\bibitem[{{Bluhm} {et~al.}(2020){Bluhm}, {Luque}, {Espinoza}, {Palle},
  {Caballero}, {Dreizler}, {Livingston}, {Mathur}, {Quirrenbach}, {Stock}, {Van
  Eylen}, {Nowak}, {Lopez}, {Csizmadia}, {Zapatero Osorio}, {Schoefer},
  {Lillo-Box}, {Oshagh}, {Amado}, {Barrado}, {Bejar}, {Cale}, {Chaturvedi},
  {Cifuentes}, {Cochran}, {Collins}, {Collins}, {Cortes-Contreras}, {Diez
  Alonso}, {El Mufti}, {Ercolino}, {Fridlund}, {Gaidos}, {Garcia},
  {Gonzalez-Alvarez}, {Gonzalez-Cuesta}, {Guerra}, {Hatzes}, {Henning},
  {Herrero}, {Hidalgo}, {Isopi}, {Jeffers}, {Jenkins}, {Jensen}, {Kabath},
  {Kemmer}, {Korth}, {Kossakowski}, {Kuerster}, {Lafarga}, {Mallia}, {Montes},
  {Morales}, {Morales-Calderon}, {Murgas}, {Narita}, {Plavchan}, {Passegger},
  {Pedraz}, {Rauer}, {Redfield}, {Reffert}, {Reiners}, {Ribas}, {Ricker},
  {Rodriguez-Lopez}, {Santos}, {Seager}, {Shan}, {Schlecker}, {Schweitzer},
  {Soto}, {Subjak}, {Tal-Or}, {Trifonov}, {Vanaverbeke}, {Vanderspek},
  {Wittrock}, {Zechmeister}, \& {Zohrabi}}]{Bluhm2020}
{Bluhm}, P., {Luque}, R., {Espinoza}, N., {et~al.} 2020, arXiv e-prints,
  arXiv:2004.06218.
\newblock \doarXiv{2004.06218}

\bibitem[{{Bochanski} {et~al.}(2010){Bochanski}, {Hawley}, {Covey}, {West},
  {Reid}, {Golimowski}, \& {Ivezi{\'c}}}]{Bochanski2010}
{Bochanski}, J.~J., {Hawley}, S.~L., {Covey}, K.~R., {et~al.} 2010, \aj, 139,
  2679, \dodoi{10.1088/0004-6256/139/6/2679}

\bibitem[{{Bonfils} {et~al.}(2005){Bonfils}, {Forveille}, {Delfosse}, {Udry},
  {Mayor}, {Perrier}, {Bouchy}, {Pepe}, {Queloz}, \&
  {Bertaux}}]{Bonfils2005A&A...443L..15B}
{Bonfils}, X., {Forveille}, T., {Delfosse}, X., {et~al.} 2005, \aap, 443, L15,
  \dodoi{10.1051/0004-6361:200500193}

\bibitem[{{Bourrier} {et~al.}(2017{\natexlab{a}}){Bourrier}, {de Wit},
  {Bolmont}, {Stamenkovi{\'c}}, {Wheatley}, {Burgasser}, {Delrez}, {Demory},
  {Ehrenreich}, {Gillon}, {Jehin}, {Leconte}, {Lederer}, {Lewis}, {Triaud}, \&
  {Van Grootel}}]{Bourrier2017b}
{Bourrier}, V., {de Wit}, J., {Bolmont}, E., {et~al.} 2017{\natexlab{a}}, \aj,
  154, 121, \dodoi{10.3847/1538-3881/aa859c}

\bibitem[{{Bourrier} {et~al.}(2017{\natexlab{b}}){Bourrier}, {Ehrenreich},
  {Wheatley}, {Bolmont}, {Gillon}, {de Wit}, {Burgasser}, {Jehin}, {Queloz}, \&
  {Triaud}}]{Bourrier2017a}
{Bourrier}, V., {Ehrenreich}, D., {Wheatley}, P.~J., {et~al.}
  2017{\natexlab{b}}, \aap, 599, L3, \dodoi{10.1051/0004-6361/201630238}

\bibitem[{{Bourrier} {et~al.}(2018){Bourrier}, {Lecavelier des Etangs},
  {Ehrenreich}, {Sanz-Forcada}, {Allart}, {Ballester}, {Buchhave}, {Cohen},
  {Deming}, {Evans}, {Garc{\'\i}a Mu{\~n}oz}, {Henry}, {Kataria}, {Lavvas},
  {Lewis}, {L{\'o}pez-Morales}, {Marley}, {Sing}, \&
  {Wakeford}}]{Bourrier2018AA...620A.147B}
{Bourrier}, V., {Lecavelier des Etangs}, A., {Ehrenreich}, D., {et~al.} 2018,
  \aap, 620, A147, \dodoi{10.1051/0004-6361/201833675}

\bibitem[{{Boyajian} {et~al.}(2012){Boyajian}, {von Braun}, {van Belle},
  {McAlister}, {ten Brummelaar}, {Kane}, {Muirhead}, {Jones}, {White},
  {Schaefer}, {Ciardi}, {Henry}, {L{\'o}pez-Morales}, {Ridgway}, {Gies}, {Jao},
  {Rojas-Ayala}, {Parks}, {Sturmann}, {Sturmann}, {Turner}, {Farrington},
  {Goldfinger}, \& {Berger}}]{Boyajian2012}
{Boyajian}, T.~S., {von Braun}, K., {van Belle}, G., {et~al.} 2012, \apj, 757,
  112, \dodoi{10.1088/0004-637X/757/2/112}

\bibitem[{{Chabrier} \& {Baraffe}(2000)}]{Chabrier2000}
{Chabrier}, G., \& {Baraffe}, I. 2000, \araa, 38, 337,
  \dodoi{10.1146/annurev.astro.38.1.337}

\bibitem[{{Chambers} {et~al.}(2016){Chambers}, {Magnier}, {Metcalfe},
  {Flewelling}, {Huber}, {Waters}, {Denneau}, {Draper}, {Farrow}, {Finkbeiner},
  {Holmberg}, {Koppenhoefer}, {Price}, {Rest}, {Saglia}, {Schlafly}, {Smartt},
  {Sweeney}, {Wainscoat}, {Burgett}, {Chastel}, {Grav}, {Heasley}, {Hodapp},
  {Jedicke}, {Kaiser}, {Kudritzki}, {Luppino}, {Lupton}, {Monet}, {Morgan},
  {Onaka}, {Shiao}, {Stubbs}, {Tonry}, {White}, {Ba{\~n}ados}, {Bell},
  {Bender}, {Bernard}, {Boegner}, {Boffi}, {Botticella}, {Calamida},
  {Casertano}, {Chen}, {Chen}, {Cole}, {Deacon}, {Frenk}, {Fitzsimmons},
  {Gezari}, {Gibbs}, {Goessl}, {Goggia}, {Gourgue}, {Goldman}, {Grant},
  {Grebel}, {Hambly}, {Hasinger}, {Heavens}, {Heckman}, {Henderson}, {Henning},
  {Holman}, {Hopp}, {Ip}, {Isani}, {Jackson}, {Keyes}, {Koekemoer}, {Kotak},
  {Le}, {Liska}, {Long}, {Lucey}, {Liu}, {Martin}, {Masci}, {McLean}, {Mindel},
  {Misra}, {Morganson}, {Murphy}, {Obaika}, {Narayan}, {Nieto-Santisteban},
  {Norberg}, {Peacock}, {Pier}, {Postman}, {Primak}, {Rae}, {Rai}, {Riess},
  {Riffeser}, {Rix}, {R{\"o}ser}, {Russel}, {Rutz}, {Schilbach}, {Schultz},
  {Scolnic}, {Strolger}, {Szalay}, {Seitz}, {Small}, {Smith}, {Soderblom},
  {Taylor}, {Thomson}, {Taylor}, {Thakar}, {Thiel}, {Thilker}, {Unger},
  {Urata}, {Valenti}, {Wagner}, {Walder}, {Walter}, {Watters}, {Werner},
  {Wood-Vasey}, \& {Wyse}}]{Chambers2016arXiv161205560C}
{Chambers}, K.~C., {Magnier}, E.~A., {Metcalfe}, N., {et~al.} 2016, arXiv
  e-prints, arXiv:1612.05560.
\newblock \doarXiv{1612.05560}

\bibitem[{{Chen} {et~al.}(2014){Chen}, {Girardi}, {Bressan}, {Marigo},
  {Barbieri}, \& {Kong}}]{Chen2014}
{Chen}, Y., {Girardi}, L., {Bressan}, A., {et~al.} 2014, \mnras, 444, 2525,
  \dodoi{10.1093/mnras/stu1605}

\bibitem[{{Cloutier} \& {Menou}(2020)}]{Cloutier2020AJ....159..211C}
{Cloutier}, R., \& {Menou}, K. 2020, \aj, 159, 211,
  \dodoi{10.3847/1538-3881/ab8237}

\bibitem[{{Crossfield} {et~al.}(2015){Crossfield}, {Petigura}, {Schlieder},
  {Howard}, {Fulton}, {Aller}, {Ciardi}, {L{\'e}pine}, {Barclay}, {de Pater},
  {de Kleer}, {Quintana}, {Christiansen}, {Schlafly}, {Kaltenegger}, {Crepp},
  {Henning}, {Obermeier}, {Deacon}, {Weiss}, {Isaacson}, {Hansen}, {Liu},
  {Greene}, {Howell}, {Barman}, \& {Mordasini}}]{Crossfield2015ApJ...804...10C}
{Crossfield}, I. J.~M., {Petigura}, E., {Schlieder}, J.~E., {et~al.} 2015,
  \apj, 804, 10, \dodoi{10.1088/0004-637X/804/1/10}

\bibitem[{{Cutri} {et~al.}(2003){Cutri}, {Skrutskie}, {van Dyk}, {Beichman},
  {Carpenter}, {Chester}, {Cambresy}, {Evans}, {Fowler}, \&
  {Gizis}}]{Cutri2003}
{Cutri}, R.~M., {Skrutskie}, M.~F., {van Dyk}, S., {et~al.} 2003, VizieR Online
  Data Catalog, II/246

\bibitem[{{D'Agostini}(2004)}]{DAgostini2004}
{D'Agostini}, G. 2004, arXiv e-prints, physics/0403086.
\newblock \doarXiv{physics/0403086}

\bibitem[{{Damasso} {et~al.}(2018){Damasso}, {Bonomo}, {Astudillo-Defru},
  {Bonfils}, {Malavolta}, {Sozzetti}, {Lopez}, {Zeng}, {Haywood}, {Irwin},
  {Mortier}, {Vanderburg}, {Maldonado}, {Lanza}, {Affer}, {Almenara},
  {Benatti}, {Biazzo}, {Bignamini}, {Borsa}, {Bouchy}, {Buchhave}, {Cameron},
  {Carleo}, {Charbonneau}, {Claudi}, {Cosentino}, {Covino}, {Delfosse},
  {Desidera}, {Di Fabrizio}, {Dressing}, {Esposito}, {Fares}, {Figueira},
  {Fiorenzano}, {Forveille}, {Giacobbe}, {Gonz{\'a}lez-{\'A}lvarez}, {Gratton},
  {Harutyunyan}, {Johnson}, {Latham}, {Leto}, {Lopez-Morales}, {Lovis},
  {Maggio}, {Mancini}, {Masiero}, {Mayor}, {Micela}, {Molinari}, {Motalebi},
  {Murgas}, {Nascimbeni}, {Pagano}, {Pepe}, {Phillips}, {Piotto}, {Poretti},
  {Rainer}, {Rice}, {Santos}, {Sasselov}, {Scandariato}, {S{\'e}gransan},
  {Smareglia}, {Udry}, {Watson}, \& {W{\"u}nsche}}]{Damasso2018A&A...615A..69D}
{Damasso}, M., {Bonomo}, A.~S., {Astudillo-Defru}, N., {et~al.} 2018, \aap,
  615, A69, \dodoi{10.1051/0004-6361/201732459}

\bibitem[{{Delfosse} {et~al.}(2000){Delfosse}, {Forveille}, {S{\'e}gransan},
  {Beuzit}, {Udry}, {Perrier}, \& {Mayor}}]{Delfosse2000}
{Delfosse}, X., {Forveille}, T., {S{\'e}gransan}, D., {et~al.} 2000, \aap, 364,
  217.
\newblock \doarXiv{astro-ph/0010586}

\bibitem[{{Demory} {et~al.}(2009){Demory}, {S{\'e}gransan}, {Forveille},
  {Queloz}, {Beuzit}, {Delfosse}, {di Folco}, {Kervella}, {Le Bouquin}, \&
  {Perrier}}]{Demory2009}
{Demory}, B.~O., {S{\'e}gransan}, D., {Forveille}, T., {et~al.} 2009, \aap,
  505, 205, \dodoi{10.1051/0004-6361/200911976}

\bibitem[{{Demory} {et~al.}(2013){Demory}, {Torres}, {Neves}, {Rogers},
  {Gillon}, {Horch}, {Sullivan}, {Bonfils}, {Delfosse}, {Forveille}, {Lovis},
  {Mayor}, {Santos}, {Seager}, {Smalley}, \&
  {Udry}}]{Demory2013ApJ...768..154D}
{Demory}, B.-O., {Torres}, G., {Neves}, V., {et~al.} 2013, \apj, 768, 154,
  \dodoi{10.1088/0004-637X/768/2/154}

\bibitem[{{Dittmann} {et~al.}(2017){Dittmann}, {Irwin}, {Charbonneau},
  {Bonfils}, {Astudillo-Defru}, {Haywood}, {Berta-Thompson}, {Newton},
  {Rodriguez}, {Winters}, {Tan}, {Almenara}, {Bouchy}, {Delfosse}, {Forveille},
  {Lovis}, {Murgas}, {Pepe}, {Santos}, {Udry}, {W{\"u}nsche}, {Esquerdo},
  {Latham}, \& {Dressing}}]{Dittmann2017Natur.544..333D}
{Dittmann}, J.~A., {Irwin}, J.~M., {Charbonneau}, D., {et~al.} 2017, \nat, 544,
  333, \dodoi{10.1038/nature22055}

\bibitem[{{Dotter}(2016)}]{Dotter2016ApJS..222....8D}
{Dotter}, A. 2016, \apjs, 222, 8, \dodoi{10.3847/0067-0049/222/1/8}

\bibitem[{{Dressing} \& {Charbonneau}(2015)}]{Dressing2015}
{Dressing}, C.~D., \& {Charbonneau}, D. 2015, \apj, 807, 45,
  \dodoi{10.1088/0004-637X/807/1/45}

\bibitem[{{Dressing} {et~al.}(2017){Dressing}, {Newton}, {Schlieder},
  {Charbonneau}, {Knutson}, {Vanderburg}, \&
  {Sinukoff}}]{Dressing2017ApJ...836..167D}
{Dressing}, C.~D., {Newton}, E.~R., {Schlieder}, J.~E., {et~al.} 2017, \apj,
  836, 167, \dodoi{10.3847/1538-4357/836/2/167}

\bibitem[{{Feiden}(2016)}]{Feiden2016}
{Feiden}, G.~A. 2016, Astronomy and Astrophysics, 593, A99,
  \dodoi{10.1051/0004-6361/201527613}

\bibitem[{{Feiden} {et~al.}(2021){Feiden}, {Skidmore}, \&
  {Jao}}]{Feiden2021ApJ...907...53F}
{Feiden}, G.~A., {Skidmore}, K., \& {Jao}, W.-C. 2021, \apj, 907, 53,
  \dodoi{10.3847/1538-4357/abcc03}

\bibitem[{{Feinstein} {et~al.}(2019){Feinstein}, {Schlieder}, {Livingston},
  {Ciardi}, {Howard}, {Arnold}, {Barentsen}, {Bristow}, {Christiansen},
  {Crossfield}, {Dressing}, {Gonzales}, {Kosiarek}, {Lintott}, {Miller},
  {Morales}, {Petigura}, {Thackeray}, {Ault}, {Baeten}, {Jonkeren}, {Langley},
  {Moshinaly}, {Pearson}, {Tanner}, \&
  {Treasure}}]{Feinstein2019AJ....157...40F}
{Feinstein}, A.~D., {Schlieder}, J.~E., {Livingston}, J.~H., {et~al.} 2019,
  \aj, 157, 40, \dodoi{10.3847/1538-3881/aafa70}

\bibitem[{{Filippazzo} {et~al.}(2015){Filippazzo}, {Rice}, {Faherty}, {Cruz},
  {Van Gordon}, \& {Looper}}]{Filippazzo2015}
{Filippazzo}, J.~C., {Rice}, E.~L., {Faherty}, J., {et~al.} 2015, \apj, 810,
  158, \dodoi{10.1088/0004-637X/810/2/158}

\bibitem[{{France} {et~al.}(2012){France}, {Schindhelm}, {Herczeg}, {Brown},
  {Abgrall}, {Alexander}, {Bergin}, {Brown}, {Linsky}, {Roueff}, \&
  {Yang}}]{France2012ApJ...756..171F}
{France}, K., {Schindhelm}, E., {Herczeg}, G.~J., {et~al.} 2012, \apj, 756,
  171, \dodoi{10.1088/0004-637X/756/2/171}

\bibitem[{{France} {et~al.}(2013){France}, {Froning}, {Linsky}, {Roberge},
  {Stocke}, {Tian}, {Bushinsky}, {D{\'e}sert}, {Mauas}, {Vieytes}, \&
  {Walkowicz}}]{France2013}
{France}, K., {Froning}, C.~S., {Linsky}, J.~L., {et~al.} 2013, \apj, 763, 149,
  \dodoi{10.1088/0004-637X/763/2/149}

\bibitem[{{France} {et~al.}(2016){France}, {Parke Loyd}, {Youngblood}, {Brown},
  {Schneider}, {Hawley}, {Froning}, {Linsky}, {Roberge}, {Buccino},
  {Davenport}, {Fontenla}, {Kaltenegger}, {Kowalski}, {Mauas}, {Miguel},
  {Redfield}, {Rugheimer}, {Tian}, {Vieytes}, {Walkowicz}, \&
  {Weisenburger}}]{France2016}
{France}, K., {Parke Loyd}, R.~O., {Youngblood}, A., {et~al.} 2016, \apj, 820,
  89, \dodoi{10.3847/0004-637X/820/2/89}

\bibitem[{{France} {et~al.}(2020){France}, {Duvvuri}, {Egan}, {Koskinen},
  {Wilson}, {Youngblood}, {Froning}, {Brown}, {Alvarado-G{\'o}mez},
  {Berta-Thompson}, {Drake}, {Garraffo}, {Kaltenegger}, {Kowalski}, {Linsky},
  {Loyd}, {Mauas}, {Miguel}, {Pineda}, {Rugheimer}, {Schneider}, {Tian}, \&
  {Vieytes}}]{France2020AJ....160..237F}
{France}, K., {Duvvuri}, G., {Egan}, H., {et~al.} 2020, \aj, 160, 237,
  \dodoi{10.3847/1538-3881/abb465}

\bibitem[{{Froning} {et~al.}(2019){Froning}, {Kowalski}, {France}, {Loyd},
  {Schneider}, {Youngblood}, {Wilson}, {Brown}, {Berta-Thompson}, {Pineda},
  {Linsky}, {Rugheimer}, \& {Miguel}}]{Froning2019}
{Froning}, C.~S., {Kowalski}, A., {France}, K., {et~al.} 2019, \apjl, 871, L26,
  \dodoi{10.3847/2041-8213/aaffcd}

\bibitem[{{Gaia Collaboration}(2018)}]{Gaia2018}
{Gaia Collaboration}. 2018, VizieR Online Data Catalog, I/345

\bibitem[{{Gaidos} \& {Mann}(2014)}]{Gaidos2014ApJ...791...54G}
{Gaidos}, E., \& {Mann}, A.~W. 2014, \apj, 791, 54,
  \dodoi{10.1088/0004-637X/791/1/54}

\bibitem[{{Gaidos} {et~al.}(2014){Gaidos}, {Mann}, {L{\'e}pine}, {Buccino},
  {James}, {Ansdell}, {Petrucci}, {Mauas}, \& {Hilton}}]{Gaidos2014}
{Gaidos}, E., {Mann}, A.~W., {L{\'e}pine}, S., {et~al.} 2014, Monthly Notices
  of the Royal Astronomical Society, 443, 2561, \dodoi{10.1093/mnras/stu1313}

\bibitem[{Gelman {et~al.}(2014)Gelman, Hwang, \& Vehtari}]{Gelman2014}
Gelman, A., Hwang, J., \& Vehtari, A. 2014, Statistics and Computing, 24, 997,
  \dodoi{10.1007/s11222-013-9416-2}

\bibitem[{{Gilbert} {et~al.}(2020){Gilbert}, {Barclay}, {Schlieder},
  {Quintana}, {Hord}, {Kostov}, {Lopez}, {Rowe}, {Hoffman}, {Walkowicz},
  {Silverstein}, {Rodriguez}, {Vanderburg}, {Suissa}, {Airapetian}, {Clement},
  {Raymond}, {Mann}, {Kruse}, {Lissauer}, {Col{\'o}n}, {Kopparapu},
  {Kreidberg}, {Zieba}, {Collins}, {Quinn}, {Howell}, {Ziegler}, {Vrijmoet},
  {Adams}, {Arney}, {Boyd}, {Brande}, {Burke}, {Cacciapuoti}, {Chance},
  {Christiansen}, {Covone}, {Daylan}, {Dineen}, {Dressing}, {Essack},
  {Fauchez}, {Galgano}, {Howe}, {Kaltenegger}, {Kane}, {Lam}, {Lee}, {Lewis},
  {Logsdon}, {Mandell}, {Monsue}, {Mullally}, {Mullally}, {Paudel},
  {Pidhorodetska}, {Plavchan}, {Reyes}, {Rinehart}, {Rojas-Ayala}, {Smith},
  {Stassun}, {Tenenbaum}, {Vega}, {Villanueva}, {Wolf}, {Youngblood}, {Ricker},
  {Vanderspek}, {Latham}, {Seager}, {Winn}, {Jenkins}, {Bakos}, {Brice{\~n}o},
  {Ciardi}, {Cloutier}, {Conti}, {Couperus}, {Di Sora}, {Eisner}, {Everett},
  {Gan}, {Hartman}, {Henry}, {Isopi}, {Jao}, {Jensen}, {Law}, {Mallia},
  {Matson}, {Shappee}, {Le Wood}, \& {Winters}}]{Gilbert2020AJ....160..116G}
{Gilbert}, E.~A., {Barclay}, T., {Schlieder}, J.~E., {et~al.} 2020, \aj, 160,
  116, \dodoi{10.3847/1538-3881/aba4b2}

\bibitem[{{Gillon} {et~al.}(2016){Gillon}, {Jehin}, {Lederer}, {Delrez}, {de
  Wit}, {Burdanov}, {Van Grootel}, {Burgasser}, {Triaud}, {Opitom}, {Demory},
  {Sahu}, {Bardalez Gagliuffi}, {Magain}, \& {Queloz}}]{Gillon2016}
{Gillon}, M., {Jehin}, E., {Lederer}, S.~M., {et~al.} 2016, \nat, 533, 221,
  \dodoi{10.1038/nature17448}

\bibitem[{{Gonzales} {et~al.}(2019){Gonzales}, {Faherty}, {Gagn{\'e}}, {Teske},
  {McWilliam}, \& {Cruz}}]{Gonzales2019}
{Gonzales}, E.~C., {Faherty}, J.~K., {Gagn{\'e}}, J., {et~al.} 2019, \apj, 886,
  131, \dodoi{10.3847/1538-4357/ab48fc}

\bibitem[{{Guinan} {et~al.}(2016){Guinan}, {Engle}, \& {Durbin}}]{Guinan2016}
{Guinan}, E.~F., {Engle}, S.~G., \& {Durbin}, A. 2016, \apj, 821, 81,
  \dodoi{10.3847/0004-637X/821/2/81}

\bibitem[{{Hawley} \& {Johns-Krull}(2003)}]{Hawley2003}
{Hawley}, S.~L., \& {Johns-Krull}, C.~M. 2003, \apjl, 588, L109,
  \dodoi{10.1086/375630}

\bibitem[{{Hawley} {et~al.}(2007){Hawley}, {Walkowicz}, {Allred}, \&
  {Valenti}}]{Hawley2007}
{Hawley}, S.~L., {Walkowicz}, L.~M., {Allred}, J.~C., \& {Valenti}, J.~A. 2007,
  \pasp, 119, 67, \dodoi{10.1086/510561}

\bibitem[{{Hawley} {et~al.}(2003){Hawley}, {Allred}, {Johns-Krull}, {Fisher},
  {Abbett}, {Alekseev}, {Avgoloupis}, {Deustua}, {Gunn}, {Seiradakis}, {Sirk},
  \& {Valenti}}]{Hawley2003b}
{Hawley}, S.~L., {Allred}, J.~C., {Johns-Krull}, C.~M., {et~al.} 2003, \apj,
  597, 535, \dodoi{10.1086/378351}

\bibitem[{Henden~A.A.(2016)}]{Henden2016}
Henden~A.A., Templeton~M., T. D. S. T. L. S. W.~D. 2016, {VizieR Online Data
  Catalog: AAVSO Photometric All Sky Survey (APASS) DR9}

\bibitem[{{Henry} {et~al.}(2018){Henry}, {Jao}, {Winters}, {Dieterich},
  {Finch}, {Ianna}, {Riedel}, {Silverstein}, {Subasavage}, \&
  {Vrijmoet}}]{Henry2018AJ....155..265H}
{Henry}, T.~J., {Jao}, W.-C., {Winters}, J.~G., {et~al.} 2018, \aj, 155, 265,
  \dodoi{10.3847/1538-3881/aac262}

\bibitem[{{Hosey} {et~al.}(2015){Hosey}, {Henry}, {Jao}, {Dieterich},
  {Winters}, {Lurie}, {Riedel}, \& {Subasavage}}]{Hosey2015AJ....150....6H}
{Hosey}, A.~D., {Henry}, T.~J., {Jao}, W.-C., {et~al.} 2015, \aj, 150, 6,
  \dodoi{10.1088/0004-6256/150/1/6}

\bibitem[{{Houdebine}(2010)}]{Houdebine2010MNRAS.407.1657H}
{Houdebine}, E.~R. 2010, \mnras, 407, 1657,
  \dodoi{10.1111/j.1365-2966.2010.16827.x}

\bibitem[{{Houdebine} {et~al.}(2017){Houdebine}, {Mullan}, {Bercu}, {Paletou},
  \& {Gebran}}]{Houdebine2017}
{Houdebine}, E.~R., {Mullan}, D.~J., {Bercu}, B., {Paletou}, F., \& {Gebran},
  M. 2017, \apj, 837, 96, \dodoi{10.3847/1538-4357/aa5cad}

\bibitem[{{Houdebine} {et~al.}(2019){Houdebine}, {Mullan}, {Doyle}, {de La
  Vieuville}, {Butler}, \& {Paletou}}]{Houdebine2019AJ....158...56H}
{Houdebine}, {\'E}.~R., {Mullan}, D.~J., {Doyle}, J.~G., {et~al.} 2019, \aj,
  158, 56, \dodoi{10.3847/1538-3881/ab23fe}

\bibitem[{{Houdebine} {et~al.}(2016){Houdebine}, {Mullan}, {Paletou}, \&
  {Gebran}}]{Houdebine2016ApJ...822...97H}
{Houdebine}, E.~R., {Mullan}, D.~J., {Paletou}, F., \& {Gebran}, M. 2016, \apj,
  822, 97, \dodoi{10.3847/0004-637X/822/2/97}

\bibitem[{{Jackson} {et~al.}(2018){Jackson}, {Deliyannis}, \&
  {Jeffries}}]{Jackson2018MNRAS.476.3245J}
{Jackson}, R.~J., {Deliyannis}, C.~P., \& {Jeffries}, R.~D. 2018, \mnras, 476,
  3245, \dodoi{10.1093/mnras/sty374}

\bibitem[{{Jao} {et~al.}(2018){Jao}, {Henry}, {Gies}, \&
  {Hambly}}]{Jao2018ApJ...861L..11J}
{Jao}, W.-C., {Henry}, T.~J., {Gies}, D.~R., \& {Hambly}, N.~C. 2018, \apjl,
  861, L11, \dodoi{10.3847/2041-8213/aacdf6}

\bibitem[{{Jeffers} {et~al.}(2020){Jeffers}, {Dreizler}, {Barnes}, {Haswell},
  {Nelson}, {Rodr{\'\i}guez}, {L{\'o}pez-Gonz‧lez}, {Morales}, {Luque},
  {Zechmeister}, {Vogt}, {Jenkins}, {Palle}, {Berdi {\~n}as}, {Coleman},
  {D{\'\i}az}, {Ribas}, {Jones}, {Butler}, {Tinney}, {Bailey}, {Carter},
  {O'Toole}, {Wittenmyer}, {Crane}, {Feng}, {Shectman}, {Teske}, {Reiners},
  {Amado}, \& {Anglada-Escud{\'e}}}]{Jeffers2020Sci...368.1477J}
{Jeffers}, S.~V., {Dreizler}, S., {Barnes}, J.~R., {et~al.} 2020, Science, 368,
  1477, \dodoi{10.1126/science.aaz0795}

\bibitem[{{Kane} {et~al.}(2017){Kane}, {von Braun}, {Henry}, {Waters},
  {Boyajian}, \& {Mann}}]{Kane2017}
{Kane}, S.~R., {von Braun}, K., {Henry}, G.~W., {et~al.} 2017, \apj, 835, 200,
  \dodoi{10.3847/1538-4357/835/2/200}

\bibitem[{{Kelly}(2007)}]{Kelly2007}
{Kelly}, B.~C. 2007, \apj, 665, 1489, \dodoi{10.1086/519947}

\bibitem[{{Kervella} {et~al.}(2008){Kervella}, {M{\'e}rand}, {Pichon},
  {Th{\'e}venin}, {Heiter}, {Bigot}, {ten Brummelaar}, {McAlister}, {Ridgway},
  \& {Turner}}]{Kervella2008}
{Kervella}, P., {M{\'e}rand}, A., {Pichon}, B., {et~al.} 2008, \aap, 488, 667,
  \dodoi{10.1051/0004-6361:200810080}

\bibitem[{{Kesseli} {et~al.}(2018){Kesseli}, {Muirhead}, {Mann}, \&
  {Mace}}]{Kesseli2018}
{Kesseli}, A.~Y., {Muirhead}, P.~S., {Mann}, A.~W., \& {Mace}, G. 2018, \aj,
  155, 225, \dodoi{10.3847/1538-3881/aabccb}

\bibitem[{{Kipping} {et~al.}(2014){Kipping}, {Nesvorn{\'y}}, {Buchhave},
  {Hartman}, {Bakos}, \& {Schmitt}}]{Kipping2014}
{Kipping}, D.~M., {Nesvorn{\'y}}, D., {Buchhave}, L.~A., {et~al.} 2014, \apj,
  784, 28, \dodoi{10.1088/0004-637X/784/1/28}

\bibitem[{{Koen} {et~al.}(2010){Koen}, {Kilkenny}, {van Wyk}, \&
  {Marang}}]{Koen2010MNRAS.403.1949K}
{Koen}, C., {Kilkenny}, D., {van Wyk}, F., \& {Marang}, F. 2010, \mnras, 403,
  1949, \dodoi{10.1111/j.1365-2966.2009.16182.x}

\bibitem[{{Kopytova} {et~al.}(2016){Kopytova}, {Brandner}, {Tognelli}, {Prada
  Moroni}, {Da Rio}, {R{\"o}ser}, \& {Schilbach}}]{Kopytova2016A&A...585A...7K}
{Kopytova}, T.~G., {Brandner}, W., {Tognelli}, E., {et~al.} 2016, \aap, 585,
  A7, \dodoi{10.1051/0004-6361/201527044}

\bibitem[{{Kordopatis} {et~al.}(2013){Kordopatis}, {Gilmore}, {Steinmetz},
  {Boeche}, {Seabroke}, {Siebert}, {Zwitter}, {Binney}, {de Laverny},
  {Recio-Blanco}, {Williams}, {Piffl}, {Enke}, {Roeser}, {Bijaoui}, {Wyse},
  {Freeman}, {Munari}, {Carrillo}, {Anguiano}, {Burton}, {Campbell}, {Cass},
  {Fiegert}, {Hartley}, {Parker}, {Reid}, {Ritter}, {Russell}, {Stupar},
  {Watson}, {Bienaym{\'e}}, {Bland-Hawthorn}, {Gerhard}, {Gibson}, {Grebel},
  {Helmi}, {Navarro}, {Conrad}, {Famaey}, {Faure}, {Just}, {Kos},
  {Matijevi{\v{c}}}, {McMillan}, {Minchev}, {Scholz}, {Sharma}, {Siviero}, {de
  Boer}, \& {{\v{Z}}erjal}}]{Kordopatis2013AJ....146..134K}
{Kordopatis}, G., {Gilmore}, G., {Steinmetz}, M., {et~al.} 2013, \aj, 146, 134,
  \dodoi{10.1088/0004-6256/146/5/134}

\bibitem[{{Kowalski} {et~al.}(2019){Kowalski}, {Wisniewski}, {Hawley}, {Osten},
  {Brown}, {Fari{\~n}a}, {Valenti}, {Brown}, {Xilouris}, {Schmidt}, \&
  {Johns-Krull}}]{Kowalski2019ApJ...871..167K}
{Kowalski}, A.~F., {Wisniewski}, J.~P., {Hawley}, S.~L., {et~al.} 2019, \apj,
  871, 167, \dodoi{10.3847/1538-4357/aaf058}

\bibitem[{{Kraus} {et~al.}(2011){Kraus}, {Tucker}, {Thompson}, {Craine}, \&
  {Hillenbrand}}]{Kraus2011ApJ...728...48K}
{Kraus}, A.~L., {Tucker}, R.~A., {Thompson}, M.~I., {Craine}, E.~R., \&
  {Hillenbrand}, L.~A. 2011, \apj, 728, 48, \dodoi{10.1088/0004-637X/728/1/48}

\bibitem[{{L{\'e}pine} {et~al.}(2013){L{\'e}pine}, {Hilton}, {Mann}, {Wilde},
  {Rojas-Ayala}, {Cruz}, \& {Gaidos}}]{Lepine2013AJ....145..102L}
{L{\'e}pine}, S., {Hilton}, E.~J., {Mann}, A.~W., {et~al.} 2013, \aj, 145, 102,
  \dodoi{10.1088/0004-6256/145/4/102}

\bibitem[{{L{\'o}pez-Morales}(2007)}]{LopezMorales2007ApJ...660..732L}
{L{\'o}pez-Morales}, M. 2007, \apj, 660, 732, \dodoi{10.1086/513142}

\bibitem[{{Loyd} {et~al.}(2018{\natexlab{a}}){Loyd}, {Shkolnik}, {Schneider},
  {Barman}, {Meadows}, {Pagano}, \& {Peacock}}]{Loyd2018ApJ...867...70L}
{Loyd}, R.~O.~P., {Shkolnik}, E.~L., {Schneider}, A.~C., {et~al.}
  2018{\natexlab{a}}, \apj, 867, 70, \dodoi{10.3847/1538-4357/aae2ae}

\bibitem[{{Loyd} {et~al.}(2018{\natexlab{b}}){Loyd}, {France}, {Youngblood},
  {Schneider}, {Brown}, {Hu}, {Segura}, {Linsky}, {Redfield}, {Tian},
  {Rugheimer}, {Miguel}, \& {Froning}}]{Loyd2018}
{Loyd}, R.~O.~P., {France}, K., {Youngblood}, A., {et~al.} 2018{\natexlab{b}},
  \apj, 867, 71, \dodoi{10.3847/1538-4357/aae2bd}

\bibitem[{{Loyd} {et~al.}(2021){Loyd}, {Shkolnik}, {Schneider},
  {Richey-Yowell}, {Jackman}, {Peacock}, {Barman}, {Pagano}, \&
  {Meadows}}]{Loyd2021ApJ...907...91L}
{Loyd}, R.~O.~P., {Shkolnik}, E.~L., {Schneider}, A.~C., {et~al.} 2021, \apj,
  907, 91, \dodoi{10.3847/1538-4357/abd0f0}

\bibitem[{{MacDonald} \& {Gizis}(2018)}]{MacDonald2018MNRAS.480.1711M}
{MacDonald}, J., \& {Gizis}, J. 2018, \mnras, 480, 1711,
  \dodoi{10.1093/mnras/sty1888}

\bibitem[{{Mann} {et~al.}(2013){Mann}, {Brewer}, {Gaidos}, {L{\'e}pine}, \&
  {Hilton}}]{Mann2013AJ....145...52M}
{Mann}, A.~W., {Brewer}, J.~M., {Gaidos}, E., {L{\'e}pine}, S., \& {Hilton},
  E.~J. 2013, \aj, 145, 52, \dodoi{10.1088/0004-6256/145/2/52}

\bibitem[{{Mann} {et~al.}(2015){Mann}, {Feiden}, {Gaidos}, {Boyajian}, \& {von
  Braun}}]{Mann2015}
{Mann}, A.~W., {Feiden}, G.~A., {Gaidos}, E., {Boyajian}, T., \& {von Braun},
  K. 2015, \apj, 804, 64, \dodoi{10.1088/0004-637X/804/1/64}

\bibitem[{{Mann} {et~al.}(2016){Mann}, {Feiden}, {Gaidos}, {Boyajian}, \& {von
  Braun}}]{Mann2016}
---. 2016, \apj, 819, 87, \dodoi{10.3847/0004-637X/819/1/87}

\bibitem[{{Mann} {et~al.}(2019){Mann}, {Dupuy}, {Kraus}, {Gaidos}, {Ansdell},
  {Ireland}, {Rizzuto}, {Hung}, {Dittmann}, \& {Factor}}]{Mann2019}
{Mann}, A.~W., {Dupuy}, T., {Kraus}, A.~L., {et~al.} 2019, \apj, 871, 63,
  \dodoi{10.3847/1538-4357/aaf3bc}

\bibitem[{{Melbourne} {et~al.}(2020){Melbourne}, {Youngblood}, {France},
  {Froning}, {Pineda}, {Shkolnik}, {Wilson}, {Wood}, {Basu}, {Roberge},
  {Schlieder}, {Cauley}, {Loyd}, {Newton}, {Schneider}, {Arulanantham},
  {Berta-Thompson}, {Brown}, {Buccino}, {Kempton}, {Linsky}, {Logsdon},
  {Mauas}, {Pagano}, {Peacock}, {Redfield}, {Rugheimer}, {Schneider}, {Teal},
  {Tian}, {Tilipman}, \& {Vieytes}}]{Melbourne2020AJ....160..269M}
{Melbourne}, K., {Youngblood}, A., {France}, K., {et~al.} 2020, \aj, 160, 269,
  \dodoi{10.3847/1538-3881/abbf5c}

\bibitem[{{Morton}(2015)}]{Morton2015ascl.soft03010M}
{Morton}, T.~D. 2015, {isochrones: Stellar model grid package}.
\newblock \doeprint{1503.010}

\bibitem[{{Neves} {et~al.}(2014){Neves}, {Bonfils}, {Santos}, {Delfosse},
  {Forveille}, {Allard}, \& {Udry}}]{Neves2014}
{Neves}, V., {Bonfils}, X., {Santos}, N.~C., {et~al.} 2014, \aap, 568, A121,
  \dodoi{10.1051/0004-6361/201424139}

\bibitem[{{Newton} {et~al.}(2015){Newton}, {Charbonneau}, {Irwin}, \&
  {Mann}}]{Newton2015}
{Newton}, E.~R., {Charbonneau}, D., {Irwin}, J., \& {Mann}, A.~W. 2015, \apj,
  800, 85, \dodoi{10.1088/0004-637X/800/2/85}

\bibitem[{{Newton} {et~al.}(2017){Newton}, {Irwin}, {Charbonneau}, {Berlind},
  {Calkins}, \& {Mink}}]{Newton2017}
{Newton}, E.~R., {Irwin}, J., {Charbonneau}, D., {et~al.} 2017, \apj, 834, 85,
  \dodoi{10.3847/1538-4357/834/1/85}

\bibitem[{{Osten} {et~al.}(2006){Osten}, {Hawley}, {Allred}, {Johns-Krull},
  {Brown}, \& {Harper}}]{Osten2006ApJ...647.1349O}
{Osten}, R.~A., {Hawley}, S.~L., {Allred}, J., {et~al.} 2006, \apj, 647, 1349,
  \dodoi{10.1086/504889}

\bibitem[{{Pagano} {et~al.}(2000){Pagano}, {Linsky}, {Carkner}, {Robinson},
  {Woodgate}, \& {Timothy}}]{Pagano2000ApJ...532..497P}
{Pagano}, I., {Linsky}, J.~L., {Carkner}, L., {et~al.} 2000, \apj, 532, 497,
  \dodoi{10.1086/308559}

\bibitem[{{Pecaut} \& {Mamajek}(2013)}]{Pecaut2013}
{Pecaut}, M.~J., \& {Mamajek}, E.~E. 2013, \apjs, 208, 9,
  \dodoi{10.1088/0067-0049/208/1/9}

\bibitem[{{Pineda} {et~al.}(2013){Pineda}, {Bottom}, \&
  {Johnson}}]{Pineda2013a}
{Pineda}, J.~S., {Bottom}, M., \& {Johnson}, J.~A. 2013, \apj, 767, 28,
  \dodoi{10.1088/0004-637X/767/1/28}

\bibitem[{{Pineda} {et~al.}(2021){Pineda}, {Youngblood}, \&
  {France}}]{Pineda2021arXiv210212485P}
{Pineda}, J.~S., {Youngblood}, A., \& {France}, K. 2021, arXiv e-prints,
  arXiv:2102.12485.
\newblock \doarXiv{2102.12485}

\bibitem[{{Rabus} {et~al.}(2019){Rabus}, {Lachaume}, {Jord{\'a}n}, {Brahm},
  {Boyajian}, {von Braun}, {Espinoza}, {Berger}, {Le Bouquin}, \&
  {Absil}}]{Rabus2019}
{Rabus}, M., {Lachaume}, R., {Jord{\'a}n}, A., {et~al.} 2019, \mnras, 484,
  2674, \dodoi{10.1093/mnras/sty3430}

\bibitem[{{Rajpurohit} {et~al.}(2013){Rajpurohit}, {Reyl{\'e}}, {Allard},
  {Homeier}, {Schultheis}, {Bessell}, \&
  {Robin}}]{Rajpurohit2013A&A...556A..15R}
{Rajpurohit}, A.~S., {Reyl{\'e}}, C., {Allard}, F., {et~al.} 2013, \aap, 556,
  A15, \dodoi{10.1051/0004-6361/201321346}

\bibitem[{{Rasmussen} \& {Williams}(2006)}]{Rasmussen2006gpml.book.....R}
{Rasmussen}, C.~E., \& {Williams}, C. K.~I. 2006, {Gaussian Processes for
  Machine Learning}

\bibitem[{{Reid} \& {Hawley}(2005)}]{Reid2005nlds.book.....R}
{Reid}, I.~N., \& {Hawley}, S.~L. 2005, {New light on dark stars : red dwarfs,
  low-mass stars, brown dwarfs}, \dodoi{10.1007/3-540-27610-6}

\bibitem[{{Reiners} {et~al.}(2018){Reiners}, {Zechmeister}, {Caballero},
  {Ribas}, {Morales}, {Jeffers}, {Sch{\"o}fer}, {Tal-Or}, {Quirrenbach},
  {Amado}, {Kaminski}, {Seifert}, {Abril}, {Aceituno}, {Alonso-Floriano},
  {Ammler-von Eiff}, {Antona}, {Anglada-Escud{\'e}}, {Anwand-Heerwart},
  {Arroyo-Torres}, {Azzaro}, {Baroch}, {Barrado}, {Bauer}, {Becerril},
  {B{\'e}jar}, {Ben{\'\i}tez}, {Berdinas̃}, {Bergond}, {Bl{\"u}mcke},
  {Brinkm{\"o}ller}, {del Burgo}, {Cano}, {C{\'a}rdenas V{\'a}zquez}, {Casal},
  {Cifuentes}, {Claret}, {Colom{\'e}}, {Cort{\'e}s-Contreras}, {Czesla},
  {D{\'\i}ez-Alonso}, {Dreizler}, {Feiz}, {Fern{\'a}ndez}, {Ferro},
  {Fuhrmeister}, {Galad{\'\i}-Enr{\'\i}quez}, {Garcia-Piquer}, {Garc{\'\i}a
  Vargas}, {Gesa}, {G{\'o}mez Galera}, {Gonz{\'a}lez Hern{\'a}ndez},
  {Gonz{\'a}lez-Peinado}, {Gr{\"o}zinger}, {Grohnert}, {Gu{\`a}rdia},
  {Guenther}, {Guijarro}, {de Guindos}, {Guti{\'e}rrez-Soto}, {Hagen},
  {Hatzes}, {Hauschildt}, {Hedrosa}, {Helmling}, {Henning}, {Hermelo},
  {Hern{\'a}ndez Arab{\'\i}}, {Hern{\'a}ndez Casta{\~n}o}, {Hern{\'a}ndez
  Hernando}, {Herrero}, {Huber}, {Huke}, {Johnson}, {de Juan}, {Kim}, {Klein},
  {Kl{\"u}ter}, {Klutsch}, {K{\"u}rster}, {Lafarga}, {Lamert}, {Lamp{\'o}n},
  {Lara}, {Laun}, {Lemke}, {Lenzen}, {Launhardt}, {L{\'o}pez del Fresno},
  {L{\'o}pez-Gonz{\'a}lez}, {L{\'o}pez-Puertas}, {L{\'o}pez Salas},
  {L{\'o}pez-Santiago}, {Luque}, {Mag{\'a}n Madinabeitia}, {Mall}, {Mancini},
  {Mand el}, {Marfil}, {Mar{\'\i}n Molina}, {Maroto Fern{\'a}ndez},
  {Mart{\'\i}n}, {Mart{\'\i}n-Ruiz}, {Marvin}, {Mathar}, {Mirabet}, {Montes},
  {Moreno-Raya}, {Moya}, {Mundt}, {Nagel}, {Naranjo}, {Nortmann}, {Nowak},
  {Ofir}, {Oreiro}, {Pall{\'e}}, {Pand uro}, {Pascual}, {Passegger}, {Pavlov},
  {Pedraz}, {P{\'e}rez-Calpena}, {P{\'e}rez Medialdea}, {Perger}, {Perryman},
  {Pluto}, {Rabaza}, {Ram{\'o}n}, {Rebolo}, {Redondo}, {Reffert}, {Reinhart},
  {Rhode}, {Rix}, {Rodler}, {Rodr{\'\i}guez}, {Rodr{\'\i}guez-L{\'o}pez},
  {Rodr{\'\i}guez Trinidad}, {Rohloff}, {Rosich}, {Sadegi},
  {S{\'a}nchez-Blanco}, {S{\'a}nchez Carrasco}, {S{\'a}nchez-L{\'o}pez},
  {Sanz-Forcada}, {Sarkis}, {Sarmiento}, {Sch{\"a}fer}, {Schmitt}, {Schiller},
  {Schweitzer}, {Solano}, {Stahl}, {Strachan}, {St{\"u}rmer}, {Su{\'a}rez},
  {Tabernero}, {Tala}, {Trifonov}, {Tulloch}, {Ulbrich}, {Veredas}, {Vico
  Linares}, {Vilardell}, {Wagner}, {Winkler}, {Wolthoff}, {Xu}, {Yan}, \&
  {Zapatero Osorio}}]{Reiners2018A&A...612A..49R}
{Reiners}, A., {Zechmeister}, M., {Caballero}, J.~A., {et~al.} 2018, \aap, 612,
  A49, \dodoi{10.1051/0004-6361/201732054}

\bibitem[{{Ricker} {et~al.}(2015){Ricker}, {Winn}, {Vanderspek}, {Latham},
  {Bakos}, {Bean}, {Berta-Thompson}, {Brown}, {Buchhave}, {Butler}, {Butler},
  {Chaplin}, {Charbonneau}, {Christensen-Dalsgaard}, {Clampin}, {Deming},
  {Doty}, {De Lee}, {Dressing}, {Dunham}, {Endl}, {Fressin}, {Ge}, {Henning},
  {Holman}, {Howard}, {Ida}, {Jenkins}, {Jernigan}, {Johnson}, {Kaltenegger},
  {Kawai}, {Kjeldsen}, {Laughlin}, {Levine}, {Lin}, {Lissauer}, {MacQueen},
  {Marcy}, {McCullough}, {Morton}, {Narita}, {Paegert}, {Palle}, {Pepe},
  {Pepper}, {Quirrenbach}, {Rinehart}, {Sasselov}, {Sato}, {Seager},
  {Sozzetti}, {Stassun}, {Sullivan}, {Szentgyorgyi}, {Torres}, {Udry}, \&
  {Villasenor}}]{Ricker2015}
{Ricker}, G.~R., {Winn}, J.~N., {Vanderspek}, R., {et~al.} 2015, Journal of
  Astronomical Telescopes, Instruments, and Systems, 1, 014003,
  \dodoi{10.1117/1.JATIS.1.1.014003}

\bibitem[{{Rojas-Ayala} {et~al.}(2012){Rojas-Ayala}, {Covey}, {Muirhead}, \&
  {Lloyd}}]{RojasAyala2012}
{Rojas-Ayala}, B., {Covey}, K.~R., {Muirhead}, P.~S., \& {Lloyd}, J.~P. 2012,
  \apj, 748, 93, \dodoi{10.1088/0004-637X/748/2/93}

\bibitem[{Salvatier {et~al.}(2016)Salvatier, Wiecki, \&
  Fonnesbeck}]{Salvatier2016}
Salvatier, J., Wiecki, T.~V., \& Fonnesbeck, C. 2016, {PeerJ} Computer Science,
  2, e55, \dodoi{10.7717/peerj-cs.55}

\bibitem[{{Santos} {et~al.}(2013){Santos}, {Sousa}, {Mortier}, {Neves},
  {Adibekyan}, {Tsantaki}, {Delgado Mena}, {Bonfils}, {Israelian}, {Mayor}, \&
  {Udry}}]{Santos2013A&A...556A.150S}
{Santos}, N.~C., {Sousa}, S.~G., {Mortier}, A., {et~al.} 2013, \aap, 556, A150,
  \dodoi{10.1051/0004-6361/201321286}

\bibitem[{{Schweitzer} {et~al.}(2019){Schweitzer}, {Passegger}, {Cifuentes},
  {B{\'e}jar}, {Cort{\'e}s-Contreras}, {Caballero}, {del Burgo}, {Czesla},
  {K{\"u}rster}, \& {Montes}}]{Schweitzer2019}
{Schweitzer}, A., {Passegger}, V.~M., {Cifuentes}, C., {et~al.} 2019, \aap,
  625, A68, \dodoi{10.1051/0004-6361/201834965}

\bibitem[{{S{\'e}gransan} {et~al.}(2003){S{\'e}gransan}, {Kervella},
  {Forveille}, \& {Queloz}}]{Segransan2003}
{S{\'e}gransan}, D., {Kervella}, P., {Forveille}, T., \& {Queloz}, D. 2003,
  \aap, 397, L5, \dodoi{10.1051/0004-6361:20021714}

\bibitem[{{Terrien} {et~al.}(2015){Terrien}, {Mahadevan}, {Deshpande}, \&
  {Bender}}]{Terrien2015}
{Terrien}, R.~C., {Mahadevan}, S., {Deshpande}, R., \& {Bender}, C.~F. 2015,
  \apjs, 220, 16, \dodoi{10.1088/0067-0049/220/1/16}

\bibitem[{{Tinney} {et~al.}(1993){Tinney}, {Mould}, \&
  {Reid}}]{Tinney1993AJ....105.1045T}
{Tinney}, C.~G., {Mould}, J.~R., \& {Reid}, I.~N. 1993, \aj, 105, 1045,
  \dodoi{10.1086/116492}

\bibitem[{{Tuomi} \& {Anglada-Escud{\'e}}(2013)}]{Tuomi2013A&A...556A.111T}
{Tuomi}, M., \& {Anglada-Escud{\'e}}, G. 2013, \aap, 556, A111,
  \dodoi{10.1051/0004-6361/201321174}

\bibitem[{{Van Grootel} {et~al.}(2018){Van Grootel}, {Fernandes}, {Gillon},
  {Jehin}, {Manfroid}, {Scuflaire}, {Burgasser}, {Barkaoui}, {Benkhaldoun},
  {Burdanov}, {Delrez}, {Demory}, {de Wit}, {Queloz}, \&
  {Triaud}}]{vanGrootel2018ApJ...853...30V}
{Van Grootel}, V., {Fernandes}, C.~S., {Gillon}, M., {et~al.} 2018, \apj, 853,
  30, \dodoi{10.3847/1538-4357/aaa023}

\bibitem[{{van Leeuwen}(2007)}]{vanLeeuwen2007}
{van Leeuwen}, F. 2007, \aap, 474, 653, \dodoi{10.1051/0004-6361:20078357}

\bibitem[{{van Saders} \& {Pinsonneault}(2012)}]{vanSaders2012ApJ...751...98V}
{van Saders}, J.~L., \& {Pinsonneault}, M.~H. 2012, \apj, 751, 98,
  \dodoi{10.1088/0004-637X/751/2/98}

\bibitem[{{Veyette} {et~al.}(2016){Veyette}, {Muirhead}, {Mann}, \&
  {Allard}}]{Veyette2016ApJ...828...95V}
{Veyette}, M.~J., {Muirhead}, P.~S., {Mann}, A.~W., \& {Allard}, F. 2016, \apj,
  828, 95, \dodoi{10.3847/0004-637X/828/2/95}

\bibitem[{{von Boetticher} {et~al.}(2019){von Boetticher}, {Triaud}, {Queloz},
  {Gill}, {Maxted}, {Almleaky}, {Anderson}, {Bouchy}, {Burdanov}, \& {Collier
  Cameron}}]{vonBoetticher2019}
{von Boetticher}, A., {Triaud}, A. H.~M.~J., {Queloz}, D., {et~al.} 2019, \aap,
  625, A150, \dodoi{10.1051/0004-6361/201834539}

\bibitem[{{von Braun} {et~al.}(2011){von Braun}, {Boyajian}, {Kane}, {van
  Belle}, {Ciardi}, {L{\'o}pez-Morales}, {McAlister}, {Henry}, {Jao}, {Riedel},
  {Subasavage}, {Schaefer}, {ten Brummelaar}, {Ridgway}, {Sturmann},
  {Sturmann}, {Mazingue}, {Turner}, {Farrington}, {Goldfinger}, \&
  {Boden}}]{vonBraun2011}
{von Braun}, K., {Boyajian}, T.~S., {Kane}, S.~R., {et~al.} 2011, \apjl, 729,
  L26, \dodoi{10.1088/2041-8205/729/2/L26}

\bibitem[{{von Braun} {et~al.}(2012){von Braun}, {Boyajian}, {Kane}, {Hebb},
  {van Belle}, {Farrington}, {Ciardi}, {Knutson}, {ten Brummelaar}, \&
  {L{\'o}pez-Morales}}]{vonBraun2012}
---. 2012, \apj, 753, 171, \dodoi{10.1088/0004-637X/753/2/171}

\bibitem[{{von Braun} {et~al.}(2014){von Braun}, {Boyajian}, {van Belle},
  {Kane}, {Jones}, {Farrington}, {Schaefer}, {Vargas}, {Scott}, \& {ten
  Brummelaar}}]{vonBraun2014}
{von Braun}, K., {Boyajian}, T.~S., {van Belle}, G.~T., {et~al.} 2014, \mnras,
  438, 2413, \dodoi{10.1093/mnras/stt2360}

\bibitem[{{Waalkes} {et~al.}(2019){Waalkes}, {Berta-Thompson}, {Bourrier},
  {Newton}, {Ehrenreich}, {Kempton}, {Charbonneau}, {Irwin}, \&
  {Dittmann}}]{Waalkes2019AJ....158...50W}
{Waalkes}, W.~C., {Berta-Thompson}, Z., {Bourrier}, V., {et~al.} 2019, \aj,
  158, 50, \dodoi{10.3847/1538-3881/ab24c2}

\bibitem[{{Waalkes} {et~al.}(2021){Waalkes}, {Berta-Thompson}, {Collins},
  {Feinstein}, {Tofflemire}, {Rojas-Ayala}, {Silverstein}, {Newton}, {Ricker},
  {Vanderspek}, {Latham}, {Seager}, {Winn}, {Jenkins}, {Christiansen}, {Goeke},
  {Levine}, {Osborn}, {Rinehart}, {Rose}, {Ting}, {Twicken}, {Barkaoui},
  {Bean}, {Brice{\~n}o}, {Ciardi}, {Collins}, {Conti}, {Gan}, {Gillon},
  {Isopi}, {Jehin}, {Jensen}, {Kielkopf}, {Law}, {Mallia}, {Mann}, {Montet},
  {Pozuelos}, {Relles}, {Libby-Roberts}, \&
  {Ziegler}}]{Waalkes2021AJ....161...13W}
{Waalkes}, W.~C., {Berta-Thompson}, Z.~K., {Collins}, K.~A., {et~al.} 2021,
  \aj, 161, 13, \dodoi{10.3847/1538-3881/abc3b9}

\bibitem[{Watanabe(2010)}]{Watanabe2010}
Watanabe, S. 2010, CoRR, abs/1004.2316.
\newblock \doarXiv{1004.2316}

\bibitem[{{Weinberger} {et~al.}(2016){Weinberger}, {Boss}, {Keiser},
  {Anglada-Escud{\'e}}, {Thompson}, \& {Burley}}]{Weinberger2016}
{Weinberger}, A.~J., {Boss}, A.~P., {Keiser}, S.~A., {et~al.} 2016, \aj, 152,
  24, \dodoi{10.3847/0004-6256/152/1/24}

\bibitem[{{Wood} {et~al.}(2001){Wood}, {Linsky}, {M{\"u}ller}, \&
  {Zank}}]{Wood2001ApJ...547L..49W}
{Wood}, B.~E., {Linsky}, J.~L., {M{\"u}ller}, H.-R., \& {Zank}, G.~P. 2001,
  \apjl, 547, L49, \dodoi{10.1086/318888}

\bibitem[{{Youngblood} {et~al.}(2021){Youngblood}, {Pineda}, \&
  {France}}]{Youngblood2021arXiv210212504Y}
{Youngblood}, A., {Pineda}, J.~S., \& {France}, K. 2021, arXiv e-prints,
  arXiv:2102.12504.
\newblock \doarXiv{2102.12504}

\bibitem[{{Zacharias} {et~al.}(2012){Zacharias}, {Finch}, {Girard}, {Henden},
  {Bartlett}, {Monet}, \& {Zacharias}}]{Zacharias2012yCat.1322....0Z}
{Zacharias}, N., {Finch}, C.~T., {Girard}, T.~M., {et~al.} 2012, VizieR Online
  Data Catalog, I/322A

\end{thebibliography}

\end{document}